\documentclass[prd,aps,twocolumn,a4paper,showkeys,nofootinbib,showpacs]{revtex4-1}

\usepackage[colorlinks=true,linkcolor=blue,linktocpage=true,anchorcolor=black,citecolor=blue,filecolor=black,menucolor=black,urlcolor=blue]{hyperref}

\usepackage{graphicx,psfrag}
\usepackage{mathrsfs}
\usepackage{amsmath,amsfonts,amssymb}
\usepackage{comment}
\usepackage{bm}
\usepackage{ulem}
\usepackage[utf8]{inputenc}
\usepackage[top=2cm,bottom=2cm,left=1.4cm,right=1.4cm]{geometry}

\newcommand{\be}{\begin{equation}}
\newcommand{\ee}{\end{equation}}
\newcommand{\bea}{\begin{eqnarray}}
\newcommand{\eea}{\end{eqnarray}}
\newcommand{\bel}{\begin{align}}
\newcommand{\eel}{\end{align}}

\def\p{\partial}

\def\GMc2{G M_{\odot} c^{-2}}

\def\O{\mathcal{O}}

\def\F{{\cal F}}

\def\p{\partial}

\def\de{\partial}

\def\F{{\cal F}}

\def\O{{\cal O}}

\def\TEOBResumS{\texttt{TEOBResumS}}

\def\SEOBNRvq{{\texttt{SEOBNRv4}}}
\def\SEOBNRvqT{{\texttt{SEOBNRv4T}}}

\DeclareSymbolFontAlphabet{\mathrsfs}{rsfs}
\DeclareMathAlphabet{\mathcal}{OMS}{cmsy}{m}{n}
\DeclareSymbolFontAlphabet{\mathrsfs}{rsfs}
\DeclareMathAlphabet\mathbfcal{OMS}{cmsy}{b}{n}

\usepackage{color}
\definecolor{cyan}{rgb}{0,0.9,0.9}
\definecolor{orange}{rgb}{0.9,0.5,0}
\definecolor{magenta}{rgb}{1,0,1}
\definecolor{purple}{rgb}{0.8,0.4,0.8}
\definecolor{gray}{rgb}{0.8242,0.8242,0.8242}
\definecolor{dodgerblue}{rgb}{0.12, 0.56, 1.0}

\def\AA{{\mathbb A}}
\def\BB{{\mathbb B}}
\def\QQ{{\mathbb Q}}
\def\DD{{\mathbb D}}
\def\GG{{\mathbb G}}
\def\qrq{{\sqrt{\QQ}}}
\def\HH{{\mathbb H}}

\def\pphih{{\left( \bm{p} \cdot \bm{\xi} r \right)}}
\def\prh{{\left( \bm{p} \cdot \bm{n} \right)}}
\def\pthh{{\left( \bm{p} \cdot \bm{v} r \right)}}

\def\Sn{{\left( \vec{\mathbb S}_{*} \cdot \bm{n} \right)}}
\def\Sv{{\left( \vec{\mathbb S}_{*} \cdot \bm{v} \right)}}
\def\Sxi{{\left( \vec{\mathbb S}_{*} \cdot \bm{\xi} \right)}}
\def\Skerr{{\left( \vec{\mathbb S}_{*} \cdot \bm{s} \right)}}

\def\mut{{\tilde{\mu}}}

\begin{document}

\title{Comparing Effective One Body Hamiltonians for spin-aligned coalescing binaries}

\author{Piero \surname{Rettegno}${}^{1,2}$}
\author{Fabio \surname{Martinetti}${}^{1,2}$}
\author{Alessandro \surname{Nagar}${}^{1,3}$}
\author{Donato \surname{Bini}${}^{4,5}$}
\author{Gunnar \surname{Riemenschneider}${}^{1,2}$}
\author{Thibault \surname{Damour}${}^{3}$}

\affiliation{${}^1$INFN Sezione di Torino, Via P. Giuria 1, 10125 Torino, Italy}
\affiliation{${}^2$Dipartimento di Fisica, Universit\`a di Torino, via P. Giuria 1, 10125 Torino, Italy}
\affiliation{${}^3$Institut des Hautes Etudes Scientifiques, 91440 Bures-sur-Yvette, France}
\affiliation{${}^4$Istituto per le Applicazioni del Calcolo `` M.~Picone", CNR, I-00185 Rome, Italy}
\affiliation{${}^5$INFN, Sezione di Roma Tre, I-00146 Roma, Italy}

\begin{abstract}
\TEOBResumS{} and \SEOBNRvq{} are the two existing semi-analytical gravitational waveform 
models for spin-aligned coalescing black hole binaries based on the effective-one-body approach.
They are informed by numerical relativity simulations and provide the relative dynamics
and waveforms from early inspiral to plunge, merger and ringdown
The central building block of each model is the EOB resummed Hamiltonian.
The two models implement different Hamiltonians that are both
deformations of the Hamiltonian of a test spinning black hole moving around a Kerr black hole.
Here we analytically compare, element by element, the two Hamiltonians. 
In particular: we illustrate that one can introduce a {\it centrifugal radius} in \SEOBNRvq{},
so to rewrite the Hamiltonian in a more compact form that is analogous to the one of \TEOBResumS{}.
The latter centrifugal radius cannot, however, be identified with the one used in \TEOBResumS{}
because the two models differ in their ways of incorporating spin effects in their respective deformations
of the background Kerr Hamiltonian.
We performed extensive comparisons between the energetics corresponding to the
two Hamiltonians using gauge-invariant quantities.
Finally, as an exploratory investigation, we apply the post-adiabatic approximation
to the newly rewritten \SEOBNRvq{} Hamiltonian, illustrating that it is possible
to generate long-inspiral waveforms with negligible computational cost.
\end{abstract}

\maketitle

\section{Introduction}
Gravitational wave (GW) astronomy with coalescing compact binaries (CBC), either binary neutron stars (BNS)
or binary black holes (BBH), needs accurate waveform modeling. Efficient and accurate, analytical,
waveform models were essential for the analysis of the so-far published 11 GW transients
(10~BBHs and 1~BNS ~\cite{LIGOScientific:2018mvr}) discovered by the LIGO and Virgo collaboration.
The effective-one-body (EOB) approach to the general relativistic two-body problem~\cite{Buonanno:1998gg,Buonanno:2000ef,Damour:2000we,Damour:2008gu} represents one of the most complete, accurate and flexible analytical formalism able to describe the binary
dynamics and waveform from coalescing compact binaries from the early inspiral, through plunge,
merger, postmerger and ringdown. The crucial improvement brought by the EOB formalism is that
it {\it resums}, in special ways, state-of-the art post-Newtonian (PN)
results~\cite{Blanchet:2013haa,Schafer:2018kuf} so to improve their predictive power and robustness in
the strong-field, fast-velocity regime. Once complemented by strong-field, non-perturbative,
information provided by Numerical Relativity (NR) simulations to improve the behavior
of the EOB dynamics and waveform in the last few orbits (as well as in the merger-ringdown phase)
one has access to EOB-NR theoretical waveform models, routinely used on GW transient
data by the LIGO-Virgo collaboration.
The construction of such models has been a long process, with progressive steps forward
related to both: (i) new theoretical ideas and developments and (ii) the improved accuracy
of NR simulations. Nowadays, \SEOBNRvq{}~\cite{Bohe:2016gbl},
{\tt SEOBNRv4HM}~\cite{Cotesta:2018fcv} (its version with higher modes)
and \TEOBResumS{}~\cite{Nagar:2018zoe,Nagar:2018plt} are the two
state-of-the-art EOB models, improved (or calibrated) by NR simulations
for spin-aligned compact binaries. 
For generic, precessing, spins currently only the {\tt SEOBNRv3} model is
available~\cite{Pan:2013rra,Babak:2016tgq}, while {\tt SEOBNRv4PHM}
(that incorporates both higher modes and precession effects)
is in progress~\cite{Ossokine:2019}.
The model \TEOBResumS{} also incorporates matter effects [that is, tidal interactions
as well as EOS dependent spin-square effects~\cite{Nagar:2018plt} up to next-to-next-to-leading order (NNLO)]
so to provide BNS waveforms and has been recently extended to also have a phenomenological
description of the BNS postmerger phase~\cite{Breschi:2019srl}. 
By contrast \SEOBNRvqT{}~\cite{Hinderer:2016eia,Steinhoff:2016rfi,Lackey:2018zvw}
is the tidal extension of \SEOBNRvq{} that incorporates ``dynamical tides''~\cite{Steinhoff:2016rfi},
that is a special enhancement of the tidal interaction due to the coupling between the orbital
motion and the $f$-mode oscillation of each neutron star.

The EOB model is characterized by three building blocks: (i) a Hamiltonian that describes
the conservative part of the relative dynamics; (ii) a flux, that accounts for the energy and
angular momentum losses through GW emission; (iii) a prescription for computing the waveform from
this dynamics.
Although \SEOBNRvq{} and \TEOBResumS{} are {\it both} EOB-based waveform models and the waveform they provide are both considered ``faithful'', by current standards, when compared
to NR simulations, they were developed independently and are structurally different,
especially for what concerns the conservative dynamics. The aim of this paper is to
compare and contrast among the two Hamiltonians, so to highlight the analytical similarities
and differences between them. For \SEOBNRvq{} we will condense here, and suitably rewrite,
the information distributed in Refs.~\cite{Barausse:2009xi,Barausse:2009aa,Barausse:2011ys,Taracchini:2012ig,Taracchini:2013rva,Hinderer:2013uwa,Bohe:2016gbl}.
For \TEOBResumS{} we will refer to~\cite{Damour:2014sva,Nagar:2015xqa,Nagar:2017jdw,Nagar:2018zoe,Nagar:2018gnk}.
Loosely speaking, an EOB Hamiltonian that describes the relative dynamics of a two-body system,
with comparable masses $(m_1,m_2)$, is always constructed as a {\it deformation} of the Hamiltonian of
a (spinning) particle moving on a Schwarzschild (for nonspinning bodies) or Kerr (for spinning bodies)
spacetime. The deformation parameter is the symmetric mass ratio $\nu\equiv m_1 m_2/(m_1+m_2)^2$.
There might be a certain degree of arbitrariness in this process, provided that two constraints
are preserved: (i) on the one hand, one must recover the known, $\nu$-dependent, PN results;
(ii) on the other hand, one wants the exact test-particle Hamiltonian to be correctly recovered
when $\nu \to 0$. Such process of deformation crucially regards also the spins. For example,
starting from the Hamiltonian of a spinning particle on a rotating black hole (BH), there are several
ways of defining ``effective spins'' that are combinations of the spins of two objects when the
masses are comparable. 
In simple words, the main differences between the \TEOBResumS{} and \SEOBNRvq{} Hamiltonians
is that the $\nu$-deformation of the limiting Kerr one was implemented in different ways. 
In particular, the constructions of the two models were rooted in different
ways of thinking about the extreme-mass-ratio limit ($m_2/m_1 \to 0$) of a system of two
spinning BHs (see Sec.~\ref{sec:kerr} below). 
This conceptual difference then entailed
distinct approaches to defining the respective $\nu$-deformations, with emphases
on different parts of the spin-dependent dynamics, as well as different ways to incorporate the non-perturbative information extracted from NR simulations.
As a consequence, one cannot define
a one-to-one map between the various building blocks of the two EOB models.
However, both models incorporate (modulo the choice of different gauges) the same PN information.
On top of such structural difference, the two Hamiltonians also
differ because of: (i) the choice of the effective-spin variable, which entails a different
EOB representation of spin-spin effects, that are only partially resummed, already at
leading order (LO), in \SEOBNRvq{}; (ii) the spin-orbit sector, both because of the different
$\nu$-deformation and because of the contribution of the Hamiltonian of a spinning particle
is only partially included in \TEOBResumS{}, while it is fully incorporated in \SEOBNRvq{}:
(iii) the gauge choice for the spin-orbit sector of above; (iv) the resummation of the
various potentials entering the Hamiltonian. 
To facilitate the comparison, we found that
it is possible to modify the original notation of the \SEOBNRvq{}
Hamiltonian~\cite{Barausse:2009xi}, by introducing a suitable, new, centrifugal radius function. 
This allows us to bring the structure of \SEOBNRvq{} Hamiltonian much closer
to the one of \TEOBResumS{}, providing easier comparisons between the various functions.

The paper is organized as follows: in Sec.~\ref{sec:kerr} we review
the Hamiltonian of a spinning test-body
(a particle) on a Kerr BH in equatorial motion, and introduce several
useful analytic structures. In Sec.~\ref{sec:TEOB}
we review the \TEOBResumS{} Hamiltonian. 
The \SEOBNRvq{} one is studied
in Sec.~\ref{sec:seobNRv4}, where various subsection are dedicated to a substantial
rewriting of the expressions of Ref.~\cite{Barausse:2009xi} (specified to spin-aligned binaries) so to put them in a notation similar to the \TEOBResumS{} one.
In particular, Sec.~\ref{sec:HNS} illustrates that a new
centrifugal radius function can be introduced so that (part of) the even-in-spin
 sector can be recasted in a way formally analogous,
though quantitatively different, to the corresponding
one of \TEOBResumS{}, see Eq.~\eqref{eq:HseobEff}.
Similarly, Sec.~\ref{spin_orbit_seob} performs a similar operation on the spin-orbit
sector of the effective Hamiltonian and eventually illustrates that it is formally
analogous to the one of \TEOBResumS{}, though with different gyro-gravitomagnetic
functions. 
Finally, Sec.~\ref{sec:seob_SS} applies the same procedure also on the remaining
part of the spin-spin contribution, that is written in compact form in
Eq.~\eqref{eq:Hss_seob}. We also highlight, in the same section, how the LO
spin-spin term is recovered from either model.Section \ref{sec:comparisons}
is dedicated to select comparisons between the energetics of the two
Hamiltonians, so to especially point out differences in the spin-spin
sector. We also analyse, in Sec.~\ref{sec:exploringA} the performance of
a new nonspinning EOB model that is based on the \TEOBResumS{} framework
but where the $A$ potential is resummed like in \SEOBNRvq{}.
Finally, to put our results in perspective for GW data analysis applications,
in Sec.~\ref{pa} we construct a EOB model that uses the \SEOBNRvq{}
Hamiltonian, with a certain radiation reaction force and waveform, 
and apply to it the post-adiabatic approximation of Ref.~\cite{Nagar:2018gnk}
to solve the inspiral dynamics. This allows us to compute long-inspiral
waveforms using the \SEOBNRvq{} Hamiltonian, as previously done
for \TEOBResumS~\cite{Nagar:2018gnk,Akcay:2018yyh,Nagar:2018plt}.
Throughout this work we use geometric units with $G=c=1$.

\section{Hamiltonian of a spinning particle on a Kerr background}
\label{sec:kerr}

Both EOB models use as starting point the Hamiltonian of a test spinning object
around a background spinning BH (described by a Kerr metric). 
However, the constructions of the two spinning EOB models were rooted in different ways
of considering the extreme-mass-ratio ($m_2 \ll m_1$, i.e. $\nu \to 0$) limit defining the
undeformed, background Kerr Hamiltonian.
The \TEOBResumS{} construction was initially based on Ref.~\cite{Damour:2008qf}
which considered the extreme-mass-ratio limit of a system of two spinning BHs,
while the \SEOBNRvq{} construction was initially based on Refs.~\cite{Barausse:2009aa,Barausse:2009xi,Barausse:2011ys},
which considered the extreme-mass-ratio limit describing a test particle
endowed with an {\it unlimited} spin moving in a Kerr metric.
The difference between the two ways of considering the limit is that,
as the spin of a small BH of mass $m_2 \ll m_1$ is physically bounded by the inequality
$\chi_2 \leq 1$, i.e. $S_2 \leq m_2^2$, the former way of thinking about the limit leads to
an Hamiltonian describing a nonspinning test particle around a Kerr BH.
Technically, when considering the general Hamiltonian described in Eq.\eqref{eq:HeffTEOB} below,
the spin combination $\hat{S}_*$ defined in Eq.\eqref{eq:Sstar} goes to zero proportionally to $\nu$
in this way of considering the extreme-mass-ratio limit. 
By contrast, in the second
way of considering the extreme-mass-ratio limit, in which one formally considers
overspinning test objects having $\chi_2 \gg 1$, but a fixed value of $\tilde{a}_2 = S_2/(m_2 M)$,
the spin combination $\hat{S}_*$ does not go to zero as $\nu \to 0$.
This motivated Ref.\cite{Barausse:2009xi,Barausse:2011ys} to pay particular attention
to the part of the spin-orbit sector linked to the coupling of $\hat{S}_*$, i.e. to
the second gyro-gravitomagnetic factor $G_{S_*}$ in Eq.\eqref{eq:GSs_T} below.
On the other hand, the construction of the \TEOBResumS{} model paid particular attention
to the first gyro-gravitomagnetic factor $G_S$, and,
following the construction of the first spinning EOB model~\cite{Damour:2001tu}, to ways
of incorporating spin-spin effects through the definition of a suitable background Kerr
spin variable $\tilde{a}_0$ (see below).

In this Section, we indicate with $M$ the mass of the Kerr BH and with $\mu$
the mass of the particle. Their spins are addressed as $S_{\rm Kerr}$ and $S_*$
respectively, with dimensionless spin variables that read $\hat{a}\equiv S_{\rm Kerr}/M^2$
and $\tilde{a}_*\equiv S_*/(\mu M)$. We restrict ourselves to equatorial
orbits ($\theta = \pi/2$) and parallel spins, using dimensionless phase space variables
defined as ($r \equiv R/M$, $t \equiv T/M$, $p_r \equiv P_r/ \mu$, $p_\varphi\equiv P_\varphi/(\mu M)$). 

The Kerr metric in Boyer-Lindquist coordinates and restricted to the equatorial plane reads
\begin{align}
ds^2 =& - \dfrac{\Lambda}{\Delta^K \Sigma} dt^2 + \dfrac{\Delta^K}{\Sigma} dr^2 \nonumber\\
&+ \dfrac{1}{\Lambda} \left( -\dfrac{4 r^2 \hat{a}^2}{\Delta^K \Sigma} + \Sigma \right) d\varphi^2 - \dfrac{2 r \hat{a}}{\Delta^K \Sigma} dt d\varphi,
\label{kerr_metric_inv}
\end{align}
where
\begin{align}
\Sigma &\equiv r^2,\\ 
\Delta^K &\equiv r^2 \left(1 - \frac{2}{r}\right) + \hat{a}^2, \\
\label{eq:Lambda}
\Lambda&\equiv (r^2+\hat{a}^2)^2-\hat{a}^2 \Delta^K.
\end{align} 

From the relativistic mass-shell condition \hbox{$g^{\mu \nu} p_\mu p_\nu = -1$},
one obtains the Hamiltonian of a nonspinning particle on a Kerr background, $\hat{H}^K_0\equiv - p_0$, as 
\begin{equation}
\hat{H}^K_0 = \alpha \sqrt{1+\gamma^{ij}\, p_i\, p_j} + \beta^i\, p_i,
\end{equation}
with standard lapse-shift decomposition of the metric
\begin{align}
\label{eq:alpha}
\alpha & = \dfrac{1}{\sqrt{-g^{tt}}}, \\
\beta^i& = \dfrac{g^{ti}}{g^{tt}}, \\
\gamma^{ij} &= g^{ij}-\dfrac{g^{ti}g^{tj}}{g^{tt}}.
\label{eq:gamma}
\end{align}

The same Hamiltonian can be written equivalently as
\begin{equation}
\hat{H}^K_0 = \sqrt{ A^{K} \left( 1+\dfrac{p_\varphi^2}{(r_c^{K})^2 } + \dfrac{p^2_r}{B^{K}} \right)} + G_S^{K} \hat{a}\, p_\varphi,
\label{kerrhnsp}
\end{equation}
where we introduced the centrifugal radius
\begin{align}
\label{eq:kerr_rc}
\left(r^{K}_c\right)^2 = \dfrac{\Lambda}{\Sigma} =\dfrac{(r^2+\hat{a}^2)^2 - \hat{a}^2 \Delta^K}{r^2}
= r^2+\hat{a}^2\left(1+\dfrac{2}{r}\right),
\end{align}
and the functions $(A^K,B^K,r_c^K,G_S^K)$ are expressed in terms
of the Kerr metric functions as
\begin{align}
\label{eq:kerr_A}
A^{K} &= \dfrac{\Delta^K \Sigma}{\Lambda} = \left( 1 - 2 u^K_c \right) \dfrac{1 + 2 u^K_c}{1+2u},
 \\
\label{eq:kerr_B}
B^{K} &= \dfrac{\Sigma}{\Delta^K} = \dfrac{(u_c^{K})^2}{u^2} \dfrac{1}{A^{K}},
 \\
\label{eq:kerr_GS}
G_S^{K} &= \dfrac{2 r \hat{a}}{\Lambda} = 2\, u\, (u_c^{K})^2,
\end{align}
where $u^K_c \equiv 1/r^K_c$.
These two different formulations of the Kerr Hamiltonian are
at the core of the differences between the two EOB-NR models dynamics.
We will expand our discussion on this topic in the following sections.

When we consider a (over)spinning particle, an additional spin-orbit coupling
term $G^K_{S_*}\tilde{a}_* p_\varphi$ is present, so that the Kerr
Hamiltonian in the extreme mass-ratio limit~\cite{Barausse:2009aa,Damour:2014sva} reads
\begin{equation}
\hat{H}^{K} = \sqrt{ A^{K} \left( 1+\dfrac{p_\varphi^2}{(r^{K}_c)^2 } + \dfrac{p^2_r}{B^{K}} \right)} + \left( G_S^{K} \hat{a} + G_{S_*}^{K} \tilde{a}_* \right) p_\varphi.
\label{kerrsph}
\end{equation}
The expression of $G^K_{S_*}$ is not trivial (see Ref.~\cite{Barausse:2009xi}).
The re-derivation of Ref.~\cite{Bini:2015xua} showed that for the equatorial,
parallel-spin, case it can be written as (see Eq.~(2.21) therein)
\begin{align}
\label{kerrGss}
{G}_{S_{*}}^{K} = \frac{1}{\left(r_c^K\right)^2} \Bigg\{&~ \dfrac{\sqrt{A^K}}{\sqrt{Q^K}} \left[ 1 - \frac{\left(r_c^K\right)'}{\sqrt{B^K}} \right] + \nonumber\\
&+ \dfrac{r^K_c}{2\left(1+\sqrt{Q^K}\right)}\dfrac{\left(A^K\right)'}{\sqrt{A^K B^K}} \Bigg\},
\end{align}
where the radial derivatives are indicated as $(\cdot)^{'} \equiv \partial_r(\cdot)$ and
\begin{align}
Q^K \equiv&~ 1+\gamma^{ij}\, p_i\, p_j \notag \\
=&~ 1 + p_\varphi^2(u^{K}_c)^2 + \dfrac{p^2_r}{B^{K}}.
\end{align}
One can check that Eq.~\eqref{kerrGss} is consistent with Eq.~(3.18) 
Ref.~\cite{Barausse:2009xi} once specified to equatorial orbits.

\section{The Hamiltonian of \TEOBResumS{}}
\label{sec:TEOB}
In the following,
we will consider a binary system with masses $m_{i}$ and
spin vectors $\bm{S_i}$, with $i = 1,2$. The projections of the spins along
the direction of the orbital angular momentum are denoted by
$S_i \equiv \bm{L \cdot S_i}$. 
We denote the total mass by $M \equiv m_1 + m_2$ and the reduced mass as $\mu \equiv (m_1 m_2)/M$. 
We adopt the convention that $m_1 \geq m_2$. We hence define the mass ratio $q = m_1/m_2 \geq 1$ 
and symmetric mass ratio $\nu \equiv \mu/M= (m_1 m_2)/(m_1 + m_2)^2$. 
The mass fractions are expressed as $X_i \equiv m_i/M$. 
The dimensionless spin variables we 
use are $\chi_i \equiv S_i/m_i^2$ and $\tilde{a}_i \equiv Si/(m_i M)=X_i \chi_i$,
together with their combinations
\begin{align}
\hat{S}&\equiv \frac{S_1+S_2}{M^2}, \\ 
\label{eq:Sstar}
\hat{S}_*&\equiv \frac{1}{M^2}\left(\dfrac{m_2}{m_1}S_1 + \dfrac{m_1}{m_2}S_2\right),
\end{align}
and
\begin{align}
\tilde{a}_0 \equiv&~ \tilde{a}_1 + \tilde{a}_2 = \hat{S}+\hat{S}_*, \\
\tilde{a}_{12} \equiv&~ \tilde{a}_1 - \tilde{a}_2 = \frac{\hat{S}-\hat{S}_*}{X_{12}},
\end{align}
where $X_{12} \equiv X_1 - X_2$.
Like the case of a spinning particle on Kerr seen above, 
for spin-aligned binaries the four-dimensional phase space is described by 
$(\varphi,P_\varphi,R,P_{r_*})$ where $\varphi$ is the orbital phase, $P_\varphi$ 
the orbital angular momentum, $R$ the radial separation and $P_{R_*} \equiv \sqrt{A/B} P_R$ 
the conjugate radial momentum with respect to the tortoise radial coordinate. 
Dimensionless phase space variables are $r\equiv R/M$, $p_{r_*} \equiv P_{r_*}/\mu$ and $p_\varphi \equiv P_\varphi/ (\mu M)$,
while dimensionless time is denoted as $t\equiv T/M$.

The {\tt TEOBResumS} model~\cite{Nagar:2018zoe} stems from
the (equatorial) Hamiltonian introduced in Ref.~\cite{Damour:2014sva}.
An important element of the latter is the centrifugal radius 
that is used to incorporate, in a resummed way, spin-spin effects within
the Hamiltonian.
The EOB Hamiltonian reads
\begin{equation}
\label{eq:HTEOB}
\hat{H}_{\rm EOB} \equiv \dfrac{H_{\rm EOB}}{\mu}= \frac{1}{\nu}\sqrt{1+2 \nu \left( \hat{H}_{\rm eff} - 1 \right)}.
\end{equation}
The effective Hamiltonian $\hat{H}_{\rm eff}\equiv H_{\rm eff}/\mu$ is constructed so as to closely mimic 
the structure of the (spinning) test-particle one described in Eq.~\eqref{kerrsph} and is written as
\begin{equation}
\label{eq:HeffTEOB}
\hat{H}_{\rm eff} = \hat{H}^{\rm orb}_{\rm eff} + \left( G_S \hat{S} + G_{S_*} \hat{S}_* \right) p_\varphi,
\end{equation}
where $\hat{S}$ and $\hat{S}_*$ reduce to the spin of the primary 
object and of the particle respectively when $m_1 \gg m_2$. 

\subsection{Orbital Hamiltonian}
The orbital effective Hamiltonian in Eq.~\eqref{eq:HeffTEOB} reads
\begin{equation}
\hat{H}_{\rm orb}^{\rm eff} = \sqrt{A \left( 1+\dfrac{p_\varphi^2}{r_c^2} + 2 \nu (4-3\nu) \dfrac{p_{r_*}^4}{r_c^2} \right) + p_{r_*}^2},
\end{equation}
where $r_c$ is the EOB centrifugal radius that takes into account spin-spin interactions (see below)
and the $A$ function is written as 
\begin{equation}
A = A_{\rm orb} \left(u_c\right) \dfrac{1 + 2 u_c}{1+2 u},
\end{equation}
where
\begin{equation}
A_{\rm orb} \left(u_c\right) = P^1_5 \left[A^{\rm 5PN}_{\rm orb} \right](u_c),
\end{equation}
is the orbital potential resummed with a $(1,5)$ Pad\'e approximant. 
The PN expanded orbital potential, at 5PN formal accuracy, reads
\begin{align}
&A^{\rm PN}_{\rm orb}(u) = 1 - 2 u + 2 \nu u^3 + \left( \dfrac{94}{3} - \dfrac{41 \pi^2}{32} \right) \nu u^4 + \notag \\ &\hspace{1cm} + \left( a_5^c + a_5^{\rm log} \log{u} \right) u^5 + \nu \left( a_6^c + a_6^{\rm \log} \log{u} \right) u^6.
\label{ucorbpot}
\end{align}
The 4PN and the logarithmic 5PN term are analytically known,
\begin{align}
a_5^c =& \left( \dfrac{2275 \pi^2}{512} - \dfrac{4237}{60} + \dfrac{128}{5} \gamma_{\rm E} + \dfrac{256}{5} \log{2} \right) \nu + \notag \\
&+ \left( \dfrac{41 \pi^2}{32} - \dfrac{221}{6} \right) \nu^2, \nonumber \\
a_5^{\log} =& ~\dfrac{64}{5} \nu, \nonumber \\
a_6^{\log} =& - \dfrac{7004}{105} \nu - \dfrac{144}{5} \nu^2,
\end{align}
where $\gamma_{\rm E}=0.57721\dots$ is Euler's constant and the (effective)
5PN term $a_6^c$ is informed by NR
simulations~\cite{Nagar:2015xqa,Nagar:2017jdw,Nagar:2018zoe}
(see Sec.~\ref{sec:nr_seob} below).

All terms proportional to even powers of the spins are incorporated
in the EOB centrifugal radius $r_c$. This function is understood as
a {\it deformation} of the Kerr one, Eq.~\eqref{eq:kerr_rc}, which reads
\begin{equation}
\label{rcTEOB}
r_c^2 \equiv r^2 + \tilde{a}_0^2\left(1+\dfrac{2}{r}\right) + \dfrac{\delta a^2}{r},
\end{equation}
where the dimensionless Kerr spin is replaced by the dimensionless effective spin $\tilde{a}_0$.
The function $\delta\tilde{a}^2$ is introduced here to incorporate spin-spin terms
beyond LO. The BBH sector of \TEOBResumS{} only includes 
next-to-leading order (NLO) spin-spin terms, so that this function explicitly reads~\cite{Nagar:2018plt}
\begin{equation}
\delta a^2 \equiv -\dfrac{1}{8} \Bigg\{9\, \tilde{a}_0^2 
+ \left(1 + 4\nu\right) \tilde{a}_{12}^2 - 10 X_{12}\, \tilde{a}_0\, \tilde{a}_{12}
\Bigg\}.
\end{equation}
The other metric potential $B$ is obtained through the $D$ function, whose PN expression is 
\begin{equation}
D^{\rm PN}_{\rm orb}(u) = 1 - 6 \nu u^2 - 2\left(26-3 \nu \right)\nu u^3.
\end{equation}
Within \TEOBResumS{}, this is resummed as 
\begin{equation}
D \equiv A B = \dfrac{r^2}{r_c^2} D_{\rm orb} (u_c),
\end{equation}
with 
\begin{equation}
D_{\rm orb} \left(u_c\right) = P^0_3 \left[D^{\rm 5PN}_{\rm orb} \right](u_c)
\end{equation}
being the inverse resummation of its PN series.

\subsection{Spin-orbit Hamiltonian}
The spin-orbit contributions are encoded into the gyro-gravitomagnetic functions ($G_S,G_{S_*}$) 
of Eq.~\eqref{kerrsph}. 
In \TEOBResumS{} they are written in factorized form
\begin{align}
G_S &= G_S^0~ \hat{G}_S, \\
G_{S_{*}} &= G_{S_{*}}^0 \hat{G}_{S_{*}},
\end{align}
where 
\begin{equation}
G_S^0 = 2 u u_c^2
\end{equation}
is the Kerr spin orbit coupling structure, in which $r_c^K$ 
is replaced by the one defined in Eq.~\eqref{rcTEOB} above. 
$G_{S_{*}}^0$ is the leading
PN correction (that can be also obtained by Taylor-expanding Eq.~\eqref{kerrGss}) 
where $u$ is replaced by $u_c$ and reads 
\begin{equation}
G_{S_{*}}^0 = \frac{3}{2} u_c^3. 
\end{equation}

In this respect, one should be reminded that Ref.~\cite{Damour:2014sva} 
chose, for simplicity, to only use {\it part of} the analytical information encoded into
the Hamiltonian of a spinning particle, Eq.~\eqref{kerrGss}, i.e. restricting it to the case of
a Schwarzschild background and expanding it up to (next-to)$^3$-leading order (N$^3$LO). 
We stress this was a choice prompted both by the desire of constructing a rather simple
model using the Damour-Jaranowski-Sch\"afer (DJS) gauge~\cite{Damour:2008qf}, where all dependence on
the angular momentum $p_\varphi$ is removed from $(G_S,G_{S_*})$, and by the idea that,
in the physically relevant case of BBH systems, the $G_{S_*}$-type coupling is always
secondary with respect to the $G_S$-type one because it contains an extra factor
$\nu$ (with $\nu \leq 1/4$ in all cases). 
In the DJS gauge~\footnote{
	The DJS gauge has the disadvantage
of introducing formal (and fictitious \cite{Akcay:2012ea}) singularities at the light ring,
but it has many other useful properties:
(i) it minimizes the effect of non-circularities during the late inspiral and the premerger phase;
(ii) it allows, in principle, a clean separation between spin-orbit (odd in spin) and spin-spin (even in spin) effects.
},
the Hamiltonian of a spinning particle (either Schwarzschild or Kerr) becomes singular
at light ring. So, the only way of incorporating some of this analytical information is by
PN-expanding the corresponding $G_{S_*}$, that is then eventually resummed after 
in a different way. Note however that the {\it full} spinning-particle information can
be incorporated also in a special flavor of \TEOBResumS{}, notably in factorized form.
To do so, however, a different spin gauge should be chosen. We shall briefly comment
on this at the end of Sec.~\ref{sec:conclusions}. 

Finally, $\hat{G}_S$ and $\hat{G}_{S_{*}}$ are PN correcting factors that in \TEOBResumS{} are inverse-resummed as
\begin{align}
\label{eq:GS_T}
\hat{G}_S =&~ \big( 1+c_{10} u_c +c_{20} u_c^2 +c_{30} u_c^3 + c_{02} p_{r^*}^2
 + \notag \\ &+c_{12} u_c p_{r^*}^2 +c_{04} p_{r^*}^4 \big)^{-1} ,\\
\label{eq:GSs_T}
\hat{G}_{S_{*}} =&~ \big( 1+c_{10}^{*} u_c +c_{20}^{*} u_c^2 +c_{30}^{*} u_c^3 +c_{40}^*u_c^4+ c_{02}^{*} p_{r^*}^2 + \notag \\ & +c_{12}^{*} u_c p_{r^*}^2 +c_{04}^* p_{r^*}^4 \big)^{-1}.
\end{align}
All coefficients are fully known analytically, with their complete
$\nu$ dependence, except for $(c_{30}^*,c^*_{40})$, which are those
corresponding to the PN expansion of the spin-orbit sector of the
Hamiltonian of a spinning particle on a Schwarzschild
background~\cite{Damour:2014sva}. In addition, the $\nu$-dependence of
$c_{30}$ and $c_{30}^*$ is informed by NR simulations.
More precisely, we use $c_{30}\equiv \nu c_3$ and $c_{30}^*=135/32+\nu c_3$,
where $135/32$ is the spinning-particle value and $c_3$ is an NR-tuned
effective N$^3$LO parameter. The numerical values of the other
coefficients in the DJS gauge are listed in Appendix~\ref{app:GScoeff}.

\subsection{Numerical-relativity informed functions}
\label{sec:nr_seob}
The dynamics of \TEOBResumS{} depends on two free functions
(or flexibility parameters), $a_6^c$ and $c_3$, that are determined
by comparison with NR simulations. The orbital Hamiltonian is NR-informed
through $a_6^c$, that explicitly reads~\cite{Nagar:2015xqa,Nagar:2017jdw,Nagar:2018zoe}
\begin{equation}
a_6^c =3097.3 ~\nu^2-1330.6 ~\nu+81.38 \, .
\end{equation}
The spin-orbit sector is instead calibrated using 
\begin{align}
c_3 =&~ p_0 \frac{1 + n_1 \tilde{a}_0 + n_2 \tilde{a}_0^2}{1 + d_1 \tilde{a}_0} + \notag \\
&+\big(p_1 \nu + p_2 \nu^2 + p_3 \nu^3\big)\tilde{a}_0X_{12} + p_4 \tilde{a}_{12} \nu^2,
\end{align}
with
\footnote{In the equal-mass case, since the last term is not symmetric under the exchange
	of $\chi_1$ and $\chi_2$, $c_3$ is computed adopting the convention $|\chi_1| > |\chi_2|$.}
\begin{align}
p_0 &= 43.371638, \hspace{1.85cm} p_1 = 929.579, \notag \\
n_1 &= -1.174839, \hspace{1.75cm} p_2 = -9178.87, \notag \\
n_2 &= 0.354064, \hspace{2cm} p_3 = 23632.3, \notag \\
d_1 &= -0.151961, \hspace{1.75cm} p_4 = -104.891.
\end{align}

Similar to what will be seen to occur also for \SEOBNRvq{}, the spin-dependence of the NR-informed
parameters violates the clear distinction between spin-orbit and spin-spin
Hamiltonians. In this case, $c_3$ introduces even-in-spin
terms in $G_{S_*} \hat{S}_*$.

\section{The Hamiltonian of \SEOBNRvq{}}
\label{sec:seobNRv4}

The Hamiltonian used in the \SEOBNRvq{}~\cite{Bohe:2016gbl} model was structurally
introduced in Ref.~\cite{Barausse:2009xi} for the case of generally oriented
spins. In order to compare it to the \TEOBResumS{} one, here we only focus
on the spin-aligned case (the generic scenario is discussed in
Appendix~\ref{App:seob_general}).

The \SEOBNRvq{} Hamiltonian is obtained as the result of a certain deformation of
the Hamiltonian of a spinning particle on a Kerr background.
First, the dimensionless BH spin $\hat{a}$ is replaced by the effective spin $\hat{S}$ (instead of $\tilde{a}_0=\hat{S} + \hat{S}_*$ used in \TEOBResumS{}).
Second, the functions $(\Delta^K, \Sigma, \Lambda)$ entering the Kerr metric, Eq.~\eqref{kerr_metric_inv},
are deformed by adding $\nu$-dependent PN information.
These functions are resummed so
as to obtain a robust behavior in the strong-field regime. 
Finally, one adds to the latter
Hamiltonian additional terms that are obtained by similarly deforming the spin-orbit coupling 
function of a spinning particle on a Kerr BH. 

We now denote the EOB Hamiltonian as
\begin{equation}
\hat{H}_{\rm EOB} \equiv \frac{1}{\nu}\sqrt{1+2 \nu \left( \hat{H}_{\rm eff}^{\rm SEOB} - 1 \right)},
\end{equation}
with $\hat{H}_{\rm eff}^{\rm SEOB}$ replacing the generic $\hat{H}_{\rm eff}$ of Eq.~\eqref{eq:HTEOB}
and where, following Refs.~\cite{Barausse:2009xi,Barausse:2011ys}, we define the effective EOB Hamiltonian as
\begin{equation}
\label{eq:HBB}
\hat{H}_{\rm eff}^{\rm SEOB}=\hat{H}_{\rm NS} + \hat{H}_{\rm SO}+\hat{H}_{\rm SS}^{\rm eff}.
\end{equation}
Here, $\hat{H}_{\rm NS}$ denotes the (deformed)
Hamiltonian of a nonspinning particle; $\hat{H}_{\rm SO}$ indicates the $\nu$-deformed
spin-orbit coupling of the spinning particle and $\hat{H}_{\rm SS}^{\rm eff}$
refers to an additional spin-spin contribution. 
In this respect one has to
be aware that {\it part of} the spin-orbit and spin-spin interaction is also
incorporated in $\hat{H}_{\rm NS}$, as it is inherited by the structure
of the Hamiltonian of a test-particle moving in a Kerr metric.

The aim of this section is to illustrate that it is possible to recast the
spin-aligned Hamiltonian of \SEOBNRvq{} in a way that is formally close
to the one of \TEOBResumS{} as defined in Eq.~\eqref{eq:HeffTEOB},
modulo the additional spin-spin contribution. The final result will
be an expression of the form
\begin{equation}
\label{eq:Hseob_recast}
\hat{H}_{\rm eff}^{\rm SEOB} =
\hat{{\mathbb H}}_{\rm orb}^{\rm eff}+\left(\bar{G}_S\hat{S}+\bar{G}_{S_*}\hat{S}_*\right)p_\varphi + \hat{H}_{\rm SS}^{\rm eff} \ ,
\end{equation}
where: (i) the orbital Hamiltonian $\hat{{\mathbb H}}_{\rm orb}^{\rm eff}$ is
formally analogous to $\hat{H}_{\rm orb}^{\rm eff}$, although
the metric functions and the centrifugal radius are replaced by different 
analytical expressions; (ii) similarly, the spin-orbit sector (i.e., odd-in-spins) 
will resemble the \TEOBResumS{} one, with the gyro-gravitomagnetic
functions $(\bar{G}_S,\bar{G}_{S_*})$ replacing $(G_S,G_{S_*})$ being different both 
in the gauge choice and the resummation approach. 
By contrast, the even-in-spin terms, that in \TEOBResumS{} are
entirely contained in $\hat{H}_{\rm orb}^{\rm eff}$, are partly
incorporated within ${\mathbb H}_{\rm orb}^{\rm eff}$ and partly
in $\hat{H}_{\rm SS}^{\rm eff}$, as detailed below.

\subsection{Rewriting of $\hat{H}_{\rm NS}$: the centrifugal radius $\bar{r}_c$}
\label{sec:HNS}
Following Ref.~\cite{Barausse:2009xi}, $\hat{H}_{\rm NS}$ is written following the structure of Eq.~\eqref{kerrhnsp} and reads
\begin{equation}
\label{eq:Hns1}
\hat{H}_{\rm NS} = \alpha \sqrt{1+\gamma^{ij}\, p_i\, p_j + Q_4(p)} + \beta^i \, p_i,
\end{equation}
where $Q_4(p)$ is a PN term quartic in the momenta that will be defined below and vanishes in the Kerr limit.
The functions $(\alpha,\beta_i,\gamma^{ij})$ have the same structure of Eqs.~\eqref{eq:alpha}--\eqref{eq:gamma},
but different explicit form, since the components of the $\nu$-deformed metric introduced
in Ref.~\cite{Barausse:2011ys}, for equatorial orbits, are
\begin{align}
\label{eq:metric1}
g^{tt} &= -\dfrac{\Lambda_t}{\Delta_t \Sigma}, \\
g^{rr} &= \dfrac{\Delta_r}{\Sigma}, \\
g^{\theta \theta} &= \dfrac{1}{\Sigma}, \\
g^{\varphi \varphi}& = \dfrac{1}{\Lambda_t} \left( -\dfrac{\widetilde{\omega}^2_{fd}}{\Delta_t \Sigma} + \Sigma \right), \\
g^{t \varphi} &= -\dfrac{\widetilde{\omega}_{fd}}{\Delta_t \Sigma},
\label{eq:metric2}
\end{align}
where
\begin{align}
\Delta_t &= r^2\, \Delta_u, \\
\Delta_r &= \Delta_t\, \DD^{-1}, \\
\Lambda_t &= \left(r^2+\hat{S}^2 \right)^2-\hat{S}^2 \Delta_t, \\
\Sigma &= r^2, \\
\widetilde{\omega}_{fd} &= 2\, \hat{S}\, r,
\end{align}
which mimic the Kerr functions\footnote{
In general, $\widetilde{\omega}_{fd}$ reads
\begin{equation}
\label{eq:w_fd}
\widetilde{\omega}_{ fd} = 2\,\hat{S}\, r + \nu\, \omega^0_{fd}\,\hat{S} + \nu\, \omega^1_{fd} \,\hat{S}\, r.
\end{equation}
With respect to Eq.~(36) of Ref.~\cite{Barausse:2011ys}, we already
gauge-fixed the two frame-dragging parameters to zero, i.e. $\omega^0_{fd} = \omega^1_{fd} = 0$. 
} and the Kerr BH spin $\hat{a}$ is 
replaced by the effective spin $\hat{S}$.
Note that the function $\Delta^K$ appears in both the $g^{tt}$ and
the $g^{rr}$ components of the Kerr metric. 
This implies that, in Kerr, $\Delta^K$ is also part of the $B$ function. 
In EOB models the connection between the metric potentials is more complicated because of the presence of the $D$ function. 
Hence, $\Delta^K$ was replaced by $\Delta_t$ in the $g^{tt}$ metric component
and by $\Delta_r$ that appears in $g^{rr}$. 
The $\nu$-deformation is implemented as follows.
At 4PN accuracy, we define the function
\begin{equation}
\label{eq:Delta_t}
\Delta_t^{\rm 4PN} \equiv r^2 A^{\rm 4PN}_{\rm orb}(u) + \hat{S}^2,
\end{equation}
where the terms $1-2 u$ appearing in the Kerr function $\Delta^K$
is replaced by the PN-expanded EOB orbital potential at 4PN accuracy,
as obtained from Eq.~\eqref{ucorbpot} dropping the 5PN,
effective, correction. In \SEOBNRvq{}, the resummation procedure
is implemented on the $\Delta_u$ function, that at 4PN reads 
\begin{equation}
\label{eq:Du_pn}
\Delta^{\rm 4PN}_u \equiv u^2\Delta^{\rm 4PN}_t =A^{\rm 4PN}_{\rm orb}(u)+u^2 \hat{S}^2.
\end{equation}
Two Kerr-like horizons $u_{\pm}$ are imposed and the residual function is then resummed using a global logarithmic function as
\begin{align}
\label{Du_resum}
\Delta_u =&~ \hat{S}^2 \left( u - u_{+} \right) \left( u - u_{-} \right) \times \notag \\
&\times \left[ 1 + \nu \Delta_0 + \log{ \left( 1+\sum_{i=1}^{5} \Delta_i u^i \right) } \right].
\end{align}
Here $(\Delta_0, \Delta_i)$ are $\nu$-dependent coefficients that are obtained imposing
that the PN-expansion of Eq.~\eqref{Du_resum} coincides with the one of Eq.~\eqref{eq:Du_pn},
see Ref.~\cite{Taracchini:2013rva}. The two horizons are placed at
\begin{equation}
r_{\pm} \equiv \frac{1}{u_\pm} = \left( 1 \pm \sqrt{1 - \hat{S}^2}\right) \left(1 - K \nu\right),
\end{equation}
where $K$ is a free parameter in the model that is calibrated to NR simulations~\cite{Bohe:2016gbl}. 
Note that also that the various functions $(\Delta_0,\Delta_i)$ depend on this parameter 
and can be found in Appendix~A of Ref.~\cite{Steinhoff:2016rfi}. We also list them for
completeness in our Appendix~\ref{app:Alog}.

Finally, the $D$ function is also resummed using a global overall logarithm instead
of the Pad\'e approximant used in \TEOBResumS{}. It reads
\begin{equation}
\DD = \left[1 + \log \left(1+6 \nu u^2 + 2 \left(26-3 \nu \right)\nu u^3 \right)\right]^{-1}.
\end{equation}

We can then rewrite Eq.~\eqref{eq:Hns1} in the following form
\begin{equation}
\label{eq:Hns}
\hat{H}_{\rm NS} = \hat{\mathbb H}^{\rm eff}_{\rm orb} + \bar{G}^0_S \, \hat{S}\, p_\varphi,
\end{equation}
in which we defined the Kerr-like gyro-gravitomagnetic function $\bar{G}^0_S$ as
\begin{equation}
\label{GsB}
\bar{G}^0_S \equiv \dfrac{\tilde{\omega}_{fd}}{\Lambda_t \hat{S}} = 2\, u\, \bar{u}_c^2.
\end{equation}
Moreover, the effective {\it orbital} Hamiltonian $\hat{\mathbb H}^{\rm eff}_{\rm orb}$ reads
\begin{equation}
\label{eq:HseobEff}
\hat{\mathbb H}^{\rm eff}_{\rm orb}= \sqrt{\AA \bigg( 1 + p_\varphi^2 \bar{u}_c^2 + 2 \nu (4- 3\nu) \: u^2 \bar{p}_{r_*}^4 \bigg) + \bar{p}_{r_*}^2},
\end{equation}
where we expanded $Q_4(p) \equiv 2 \nu (4-3\nu) u^2 \bar{p}_{r_*}^4$ and we now define the momentum conjugate to the tortoise radial coordinate as $\bar{p}_{r_*} \equiv \sqrt{\AA / \BB}~ p_r$. 
The functions $(\AA,\BB,\QQ)$
are expressed in terms of the $\nu$-deformed metric functions as
\begin{align}
\label{Abb}
\AA &\equiv \dfrac{\Delta_t \Sigma}{\Lambda_t}= \dfrac{\bar{u}_c^2}{u^2}\Delta_u,\\
\label{Bbb}
\BB &\equiv \dfrac{\Sigma}{\Delta_r} = \dfrac{\DD}{\Delta_u},\\
\label{Qbb}
\QQ &\equiv
 1+\gamma^{ij}\, p_i\, p_j = 1 + p_\varphi^2 \bar{u}_c^2 + \dfrac{\bar{p}_{r_*}^2}{\AA}.
\end{align}
The functions $(\AA,\BB,\DD,\QQ)$ are analogous to $(A,B,D,Q)$ used within \TEOBResumS{} 
and, although different, they reduce to the same corresponding Kerr functions in 
the $\nu\to 0$ limit. 
In Eq.~\eqref{eq:HseobEff} we also introduced $\bar{u}_c \equiv 1/\bar{r}_c$,
where $\bar{r}_c$ is a {\it new} centrifugal radius. 
This function is a $\nu$-deformation of the Kerr $r_c^K$, but differs from the \TEOBResumS{} one, $r_c$, and explicitly reads
\begin{equation} 
\label{raggiocentrifugobb}
\bar{r}_c^2 \equiv \dfrac{\Lambda_t}{\Sigma} =\dfrac{(r^2+\hat{S}^2)^2}{r^2}-\hat{S}^2 \Delta_u.
\end{equation}

Writing an \textit{orbital} Hamiltonian for \SEOBNRvq{}, $\hat{\mathbb H}^{\rm eff}_{\rm orb}$,
that mimics $\hat{H}^{\rm eff}_{\rm orb}$ makes it clearer where the differences between
the models arise, though the two expressions look formally the same. The functions $(A,r_c)$
and $({\mathbb A},\bar{r}_c)$ are different from one another, even if they correctly
reproduce the corresponding Kerr functions when $\nu\to 0$. To understand how this is possible,
let us go back to the definitions of the centrifugal
radius $r_c^K$ and of the potential $A^K$ for the Kerr metric, Eqs.~\eqref{eq:kerr_rc} and~\eqref{eq:kerr_A} respectively.
These can be written in two, analytically equivalent, forms, namely
\begin{subequations}
\begin{align}
\label{rc:seob}
\left[r_c^K\right]^2 &\equiv \dfrac{(r^2+\hat{a}^2)^2}{r^2}-\hat{a}^2 \Delta^K \\
\label{rc:teob}
&= r^2+\hat{a}^2\left(1+\dfrac{2}{r}\right),
\end{align}
\end{subequations}
and
\begin{subequations}
\begin{align}
\label{A:seob}
A^K &\equiv \dfrac{\left(u_c^K\right)^2}{u^2}\Delta^K \\
\label{A:teob}
&= \left( 1 - 2 u^K_c \right) \frac{1+2 u^K_c}{1+2 u}.
\end{align}
\end{subequations}
In \SEOBNRvq{} one obtains $\bar{r}_c$ and $\AA$ using Eqs.~\eqref{rc:seob} and \eqref{A:seob},
without expanding the expression of $\Delta^K$, and then substituting $\hat{a}\to \hat{S}$ and $\Delta^K\to \Delta_u$.
On the other hand, in \TEOBResumS{} $r_c$ and $A$ are obtained through Eqs.~\eqref{rc:teob} and~\eqref{A:teob},
where the expressions have been simplified and bear no memory of the original function $\Delta^K$
that appears in the Kerr metric. Then, one substitutes $\hat{a}\to \tilde{a}_0$ and $(1 - 2u_c^K) \to A_{\rm orb}$.
In conclusion, as well as different spin variable and resummation choices, $\bar{r}_c$ differs
from $r_c$ because it contains includes additional $\nu$-dependent corrections that come from $\Delta_u$. 
Hence, although the two function share the same $\nu=0$ limit, the spin-square contributions
that they incorporate differ already at linear order in $\nu$.

\subsection{$\hat{H}_{\rm SO}$ and the spin-orbit sector}
\label{spin_orbit_seob}
Let us turn now to rewriting $\hat{H}_{\rm SO}$ using a different notation
consistent with the orbital part. Our starting point is $\hat{H}_{\rm SO}$
as given by Eq.~(4.18) of Ref.~\cite{Barausse:2009xi} (and re-written in Appendix~\ref{App:seob_general}). Once restricted to
spin-aligned systems, this gives 
\footnote{Note that our notation differs from Ref.~\cite{Barausse:2011ys}. We define their $\nu$ as $\tilde{\nu}$, not to confuse it with the symmetric mass ratio.
Also, we use explicitly $\widetilde{B}_r = \widetilde{B}\,' - \widetilde{B}/\widetilde{J}$ and $\mu_r = \mu\,' - 1/\widetilde{J}$.}
\begin{equation}
\label{eq:HSO1}
\hat{H}_{\rm SO} =\dfrac{e^{2\tilde{\nu}-\tilde{\mu}}}{\widetilde{B}^2 \sqrt{\QQ{}}} \bigg\{ e^{\tilde{\mu}+\tilde{\nu}}- \widetilde{J} \widetilde{B}\,' + \frac{1 +2\sqrt{\QQ{}}}{1 + \sqrt{\QQ{}}}\widetilde{J} \widetilde{B}\, \tilde{\nu}\,' \bigg\} p_\varphi \hat{\mathbb S}_*,
\end{equation}
where the functions of Ref.~\cite{Barausse:2011ys} are connected to the metric ones as
\begin{align}
\label{eq:emu}
e^{2\tilde{\mu}} =&~ \hspace{0.25cm} \Sigma \hspace{0.25cm} = r^2, \hspace{1.3cm} e^{2\tilde{\nu}} = \frac{\Delta_t \Sigma}{\Lambda_t} = \AA, \\
\label{eq:BJ}
\widetilde{B} =&~ \sqrt{\Delta_t} = \sqrt{\AA}\, \bar{r}_c, \hspace{0.95cm} \widetilde{J} = \sqrt{\Delta_r} = \frac{r}{\sqrt{\BB}},
\end{align}
and the prime indicates derivative with respect to $r$. In addition, the $\nu$-dependent PN results for
the spin-orbit coupling functions are included in Eq.~\eqref{eq:HSO1} 
through a (gauge-dependent) mapping~\cite{Barausse:2011ys} between the spin variables that naturally enter the 
PN-expanded effective Hamiltonian, $(\hat{S},\hat{S}_*)$, that are used in \TEOBResumS{}, and the effective spin 
variables $(\hat{\mathbb S},\hat{\mathbb{S}}_*)$ that appear in \SEOBNRvq{}. These spin quantities are intended
to be the spin of an effective particle, $\hat{\mathbb S}_*$, moving around an effective
Kerr BH whose spin is $\hat{{\mathbb S}}$.
Following Ref.~\cite{Barausse:2011ys}, such spin mapping is defined as
\begin{align}
\label{eq:S_BB}
\hat{{\mathbb S}} &= \hat{S} + \dfrac{1}{c^2} \Delta_{\sigma}^{(1)}+ \dfrac{1}{c^4} \Delta_{\sigma}^{(2)}, \\
\label{eq:Sstar_BB}
\hat{{\mathbb S}}_* &= \hat{S}_* + \dfrac{1}{c^2} \Delta_{\sigma^*}^{(1)}+ \dfrac{1}{c^4} \Delta_{\sigma^*}^{(2)} + \dfrac{1}{c^6} \Delta_{\sigma^*}^{(3)},
\end{align}
where the functions $(\Delta_\sigma^{(i)},\Delta_{\sigma^*}^{(i)})$ are gauge-dependent function that are chosen
so to incorporate the high-order $\nu$-dependent PN information. 
Ref.~\cite{Barausse:2011ys} fixes the gauge imposing that
$\Delta_\sigma^{(1)}=\Delta_\sigma^{(2)}=0$, so that
\be
\hat{{\mathbb S}}\equiv \hat{S}.
\ee
On the other hand, the functions $\Delta_{\sigma^*}^{(1)}$ and $\Delta_{\sigma^*}^{(2)}$
are fixed in such a way that, once the SEOB Hamiltonian is 
PN-expanded, the spin-orbit PN contributions up to NNLO are
correctly recovered. Moreover, the spin-orbit sector is NR-informed
by an additional N$^3$LO effective correction of the form
\begin{equation}
\Delta_{\sigma^*}^{(3)} = \frac{d_{\rm SO}\, \nu}{r^3} \hat{S},
\end{equation}
whose explicit expression can be found below.

Using the definitions of Eqs.~\eqref{eq:emu} and \eqref{eq:BJ}, $\hat{H}_{\rm SO}$ can be rewritten as
\begin{equation}
\label{eq:HSO_new}
\hat{H}_{\rm SO} = {\mathbb G}_{{\mathbb S}_{*}}\, p_\varphi\, \hat{{\mathbb S}}_{*},
\end{equation}
where we defined
\begin{equation}
\label{GsStarB}
{\mathbb G}_{{\mathbb S}_{*}} \equiv \frac{1}{\left(\bar{r}_c\right)^2} \Bigg\{ \dfrac{\sqrt{\AA}}{\sqrt{\QQ}} \left[ 1 - \frac{\left(\bar{r}_c\right)'}{\sqrt{\BB}} \right] +
\dfrac{\bar{r}_c}{2\left(1+\sqrt{\QQ}\right)}\dfrac{\AA'}{\sqrt{\AA \BB}} \Bigg\},
\end{equation}
that formally coincides with 
Eq.~\eqref{kerrGss}, having replaced the Kerr functions $(A^K,B^K,Q^K,r_c^K)$ with $(\AA,\BB,\QQ,\bar{r}_c)$. 

We found it convenient to write the complete spin-orbit content of \SEOBNRvq{} in a form that is close to the
one of \TEOBResumS{}, so to similarly define two gyro-gravitomagnetic functions. To do so, we define 
the complete spin-orbit sector of $\hat{H}_{\rm eff}^{\rm SEOB}$ as 
\begin{equation}
\label{eq:Heff_SEOB}
\hat{{\mathbb H}}_{\rm SO}=\left(\bar{G}^0_S\hat{S}+ {\mathbb G}_{{\mathbb S}_{*}}\hat{{\mathbb S}}_{*}\right)p_\varphi.
\end{equation}
Since $\hat{{\mathbb S}}_{*}$ is a linear combination of $(\hat{S},\hat{S}_*)$, one sees that the above function
can be written precisely as the corresponding function in \TEOBResumS{}, though the gyro-gravitomagnetic functions
will eventually be different. 
We see that the $\Delta_{\sigma^*}^{(i)}$ that appear in Eq.~\eqref{eq:Sstar_BB} are functions of $(r, p_{r_*}, p_\varphi)$, 
with some additional gauge-freedom that can be fixed at will (see below). These latter formally read
\begin{align}
\label{eq:delta_s1}
\Delta_{\sigma^{*}}^{(1)} =&~ c_u u + c_{\QQ} \left( \QQ -1 \right) + c_{p_r^2} \frac{p_r^2}{\BB}, \\
\label{eq:delta_s2}
\Delta_{\sigma^{*}}^{(2)} =&~ c_{u^2} \, u^2 + c_{\QQ^2} \left(\QQ-1 \right)^2 + c_{u \QQ} \, u \left(\QQ-1 \right) + \nonumber\\
&+ c_{p_r^4} \, \frac{p_r^4}{\BB^2} + c_{up_r^2} \, u \frac{p_r^2}{\BB} + c_{p_r^2\QQ} \frac{p_r^2}{\BB}(\QQ-1).
\end{align}
The explicit expression of the $c_{X}$ coefficients can be obtained comparing Eqs.~\eqref{eq:delta_s1} and~\eqref{eq:delta_s2} to Eqs.~(51) and~(52) of
Ref.~\cite{Barausse:2011ys} and are recalled in Appendix \ref{app:delta_*}.
All these coefficients are linear functions of $(\hat{S},\hat{S}_*)$.
Thus, we can write 
$\Delta_{\sigma^{*}}^{(1)} = c^{(1)}_S \hat{S} + c^{(1)}_{S_*} \hat{S}_{*}$
and
$\Delta_{\sigma^{*}}^{(2)} = c^{(2)}_S \hat{S} + c^{(2)}_{S_*} \hat{S}_{*}$,
and, substituting them into Eq.~\eqref{eq:Heff_SEOB}, we obtain
\begin{equation}
\hat{{\mathbb H}}_{\rm SO} \equiv \left(\bar{G}_S\hat{S}+\bar{G}_{S_*}\hat{S}_*\right)p_\varphi ,
\end{equation}
where we defined two {\it new} gyro-gravitomagnetic functions 
\begin{align}
\bar{G}_S &\equiv \bar{G}_S^0 + \left(c^{(1)}_{S}+c^{(2)}_{S} \right) {\mathbb G}_{{\mathbb S}_{*}}, \\
\bar{G}_{S_*} &\equiv \left( 1 + c^{(1)}_{S_{*}}+c^{(2)}_{S_{*}}\right) {\mathbb G}_{{\mathbb S}_{*}}.
\end{align}
The explicit forms of $\big[c_S^{(i)},c_{S*}^{(i)}\big]$ are also reported in Appendix \ref{app:delta_*}.
In inspecting those expressions, one should be aware that the two models adopt
two different gauges in the spin-orbit sector. On the one hand, \TEOBResumS{} is written in
the DJS gauge~\cite{Damour:2008qf,Nagar:2011fx} that is designed to cancel all the dependence
on $\bm{p}^2$ in the gyro-gravitomagnetic functions. On the other hand, within \SEOBNRvqT{}
one makes the minimal gauge choice and sets all gauge parameters to be zero. More details can
be found in Appendix~\ref{app:gauge}.

\subsection{$\hat{H}_{\rm SS}^{\rm eff}$ and the spin-spin sector}
\label{sec:seob_SS}
Moving finally to the spin-spin sector, we define $\hat{H}_{\rm SS}^{\rm eff}$ as
\begin{equation}
\label{eq:HeffSS}
\hat{H}_{\rm SS}^{\rm eff} = \hat{H}_{\rm SS} - \dfrac{1}{2} u^3 \big(\hat{{\mathbb S}}_{*}\big)^2 + \frac{d_{\rm SS}\, \nu}{r^4}\left(X_1^4\chi_1^2+X_2^4\chi_2^2\right).
\end{equation} 
The first term in the r.h.s. of the above equation, for equatorial orbits
(see Eq.~(4.19) of Ref.~\cite{Barausse:2009xi} or Appendix~\ref{App:seob_general} for generic ones) explicitly reads
\begin{align}
\label{eq:HSSnew}
&\hat{H}_{\rm SS} = \omega\, \hat{\mathbb S}_{*} + \frac{e^{-3 \tilde{\mu}-\tilde{\nu}} \widetilde{J}}{2 \widetilde{B} \sqrt{\QQ} \left(1 + \sqrt{\QQ}\right)} \times \notag \\
&\times \left\{ e^{2\tilde{\mu}+2\tilde{\nu}}p_\varphi^2 + e^{2\tilde{\mu}}\sqrt{\QQ}\left(1+\sqrt{\QQ}\right)\widetilde{B}^2 - \widetilde{J}^2 p_r^2 \widetilde{B}^2 \right\} \omega'\, \hat{\mathbb S}_{*},
\end{align}
where 
\begin{equation}
\label{eq:omega}
\omega \equiv \dfrac{\tilde{\omega}_{fd}}{\Lambda_t} = \bar{G}^0_S\, \hat{S}.
\end{equation}
Using Eqs.~\eqref{eq:emu},~\eqref{eq:BJ} and~\eqref{eq:omega},
$\hat{H}_{\rm SS}$ can be rewritten as
\begin{align}
\label{eq:Hss_seob}
\hat{H}_{\rm SS} =&~ \Bigg\{ \bar{G}^0_S + \frac{\bar{r}_c}{2 \sqrt{\BB}} \Bigg[1 - \frac{1}{\sqrt{\QQ}\left(1 + \sqrt{\QQ}\right)} \times \notag \\
& \times \left(p_\varphi^2 \bar{u}_c^2 - \frac{p_{r_*}^2}{\AA}\right)\Bigg] \left(\bar{G}^0_S\right)' \Bigg\}
\,\hat{S}~ \hat{\mathbb S}_{*}. 
\end{align}
The second term in the r.h.s. of Eq.~\eqref{eq:HeffSS} was introduced in Ref.~\cite{Barausse:2009xi}
[see Eqs.~(5.59), (5.60) and (5.70) therein and related discussion] to correctly account
for the LO spin-spin coupling. One easily checks that PN-expanding the
whole $\hat{H}_{\rm SS}^{\rm eff}$ together with $\hat{\mathbb H}^{\rm eff}_{\rm orb}$
is necessary to correctly recover the LO spin-spin contribution in the full Hamiltonian, 
$\hat{H}^{\rm LO}_{\rm SS}=-u^3(\hat{S}+\hat{S}_*)^2/2 = -\tilde{a}_0^2/2$. 
This constitutes the main structural difference between \TEOBResumS{} and \SEOBNRvq{}
in the spin-spin sector. In fact, in the former even-in-spin terms are fully resummed
through $r_c$, while in the latter these terms are partially resummed within $\bar{r}_c$
and partly added to the Hamiltonian as they are.

We also note in passing that, by expanding ${\mathbb H}^{\rm eff}_{\rm orb}$,
one also finds the Kerr-like quartic-in-spin term $\hat{S}^4/2$. 
This term takes into account only a fraction of the analytically known
LO quartic-in-spin Hamiltonian. 
By contrast, it was shown in Ref.~\cite{Nagar:2018plt} that this
is completely incorporated in the \TEOBResumS{} Hamiltonian because
of the use of effective spin $\tilde{a}_0$ within $r_c$.

Finally, Eq.~\eqref{eq:HeffSS} also features the presence of an effective
NLO spin-spin correction, with the adjustable parameter $d_{\rm SS}$
that will be discussed below.

\subsection{Numerical relativity calibrated functions}
As briefly mentioned above, the \SEOBNRvq{} analytic structure is completed
by 3 functions that are calibrated to NR simulations.
These functions are: (i) $K$, that enters $\Delta_u$; (ii) $d_{\rm SO}$, that is found
in the definition of the effective spin variable $\, \hat{{\mathbb S}}_*$; and (iii)
$d_{\rm SS}$ that affects the spin-spin coupling.
The NR-calibrated expression of $K$ was obtained in
Ref.~\cite{Bohe:2016gbl} and reads
\begin{equation}
K=K_{|\chi=0}+K_{|\chi\neq 0},
\end{equation}
where we introduced the functions
\begin{align}
K\vert_{\chi=0}=&~ 267.788247\nu^3 - 126.686734\nu^2 + 10.257281\nu + \nonumber\\
&+1.733598,\\
K\vert_{\chi\neq 0}=&~ - 59.165806\,\chi^3\nu^3
- 0.426958\,\chi^3\nu
+ 1.436589\,\chi^3 + \nonumber \\
& + 31.17459\,\chi^2\nu^3
+ 6.164663\,\chi^2\nu^2
- 1.380863\,\chi^2 + \nonumber \\
& - 27.520106\,\chi \nu^3
+ 17.373601\,\chi\nu^2
+ 2.268313\,\chi\nu + \nonumber \\
& - 1.62045\,\chi,
\label{eq:K:spin}
\end{align}
where
\begin{equation}
\chi\equiv \chi_S+X_{12}\dfrac{\chi_A}{1-2\nu}=\dfrac{\hat{S}}{X_1^2 + X_2^2},
\end{equation}
with $\chi_S=(\chi_1+\chi_2)/2$ and $\chi_A=(\chi_1-\chi_2)/2$.
The spin-orbit sector presents an additional N$^3$LO effective
correction that reads
\begin{eqnarray}
d_{\textrm{SO}} =&~ 147.481449\,\chi^3\nu^2 - 568.651115\chi^3\nu\nonumber \\
& + 66.198703\,\chi^3 - 343.313058\,\chi^2\nu \nonumber \\
& + 2495.293427\,\chi\nu^2 - 44.532373\,.
\end{eqnarray}
Finally, the NLO effective spin-spin correction that enters $\hat{H}^{\rm eff}_{\rm SS}$
is NR-calibrated through the parameter
\begin{align}
d_{\textrm{SS}} =&~ 528.511252\,\chi^3\nu^2 - 41.000256\,\chi^3\nu \nonumber \\
& + 1161.780126\,\chi^2\nu^3 - 326.324859\,\chi^2\nu^2 \nonumber \\
& + 37.196389\,\chi\nu + 706.958312\,\nu^3 \nonumber \\
& - 36.027203\,\nu + 6.068071.
\end{align}
As all these coefficients depend on multiple powers of the individual spins,
a clear distinction between the spin-orbit and spin-orbit sectors is impossible.

\section{Select comparison between \TEOBResumS{} and \SEOBNRvq{}}
\label{sec:comparisons}
We have seen that the \TEOBResumS{} and \SEOBNRvq{} Hamiltonians are constructed
rather differently. They differ in the amount of analytical information
that is included, the spin-gauge, the resummation procedures and the
way they are informed (or calibrated) to NR simulations. 
Still, both models deliver waveforms that are {\it faithful}
with state-of-the-art NR simulations at $1\%$ level or better~\cite{Bohe:2016gbl,Nagar:2018zoe}. This is possible because,
on top of the tunable functions that enter the dynamics of the two models, $(a_6^c,c_3) $ and $(K,d_{\rm SO},d_{\rm SS})$
the waveforms are also NR-completed through merger and ringdown in some way.
The aim of this section is to attempt to quantify the differences entailed
by the two NR-informed Hamiltonians. To do so, we focus on the gauge-invariant
relation between energy and angular momentum (or orbital frequency) and we calculate
them both in the adiabatic approximation as well as non-adiabatically, switching on
some analytical radiation reaction to account for the angular momentum losses. 

\subsection{Adiabatic dynamics}
\begin{figure*}[t]
\center
\includegraphics[width=0.43\textwidth]{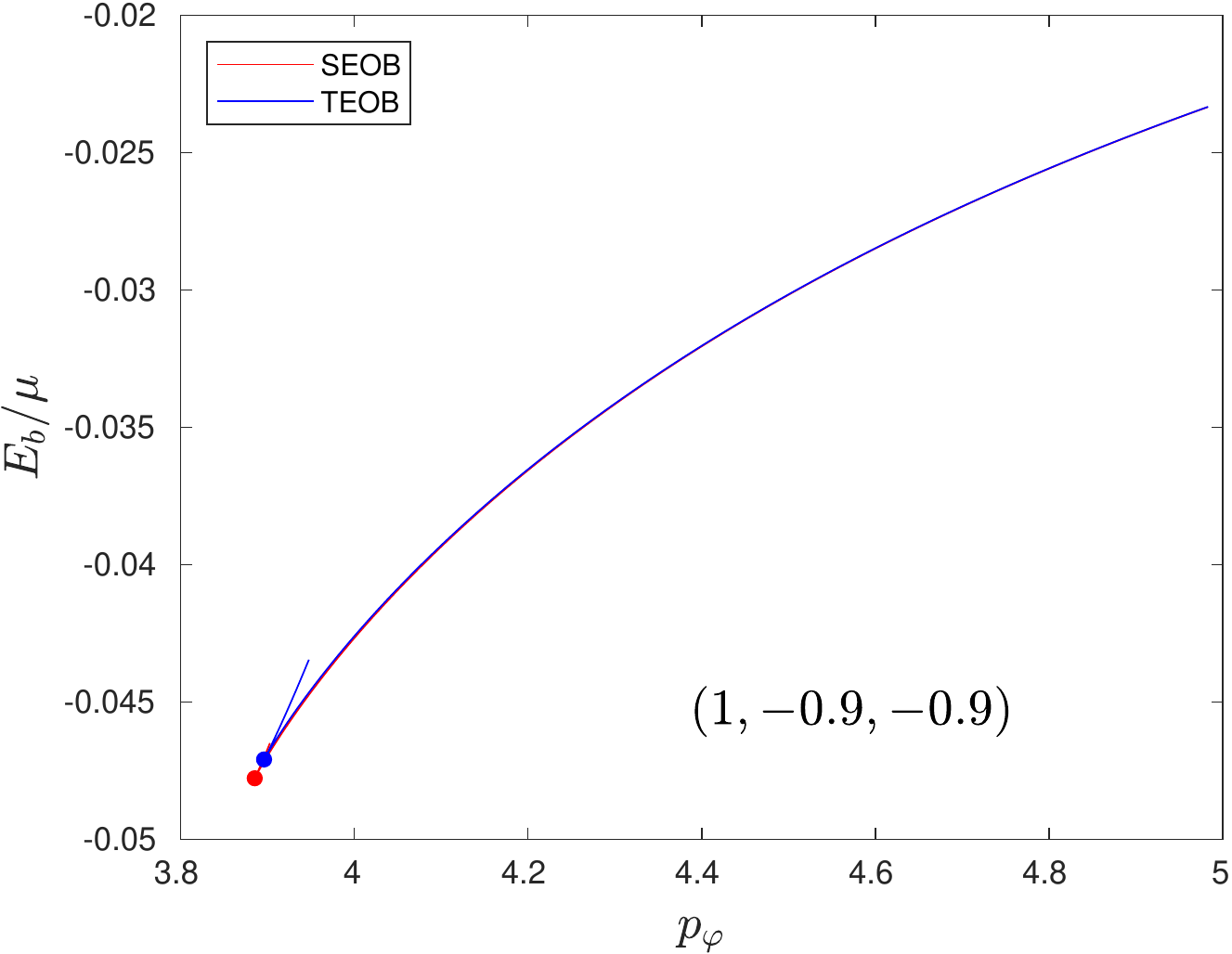}
\hspace{1.5cm}
\includegraphics[width=0.43\textwidth]{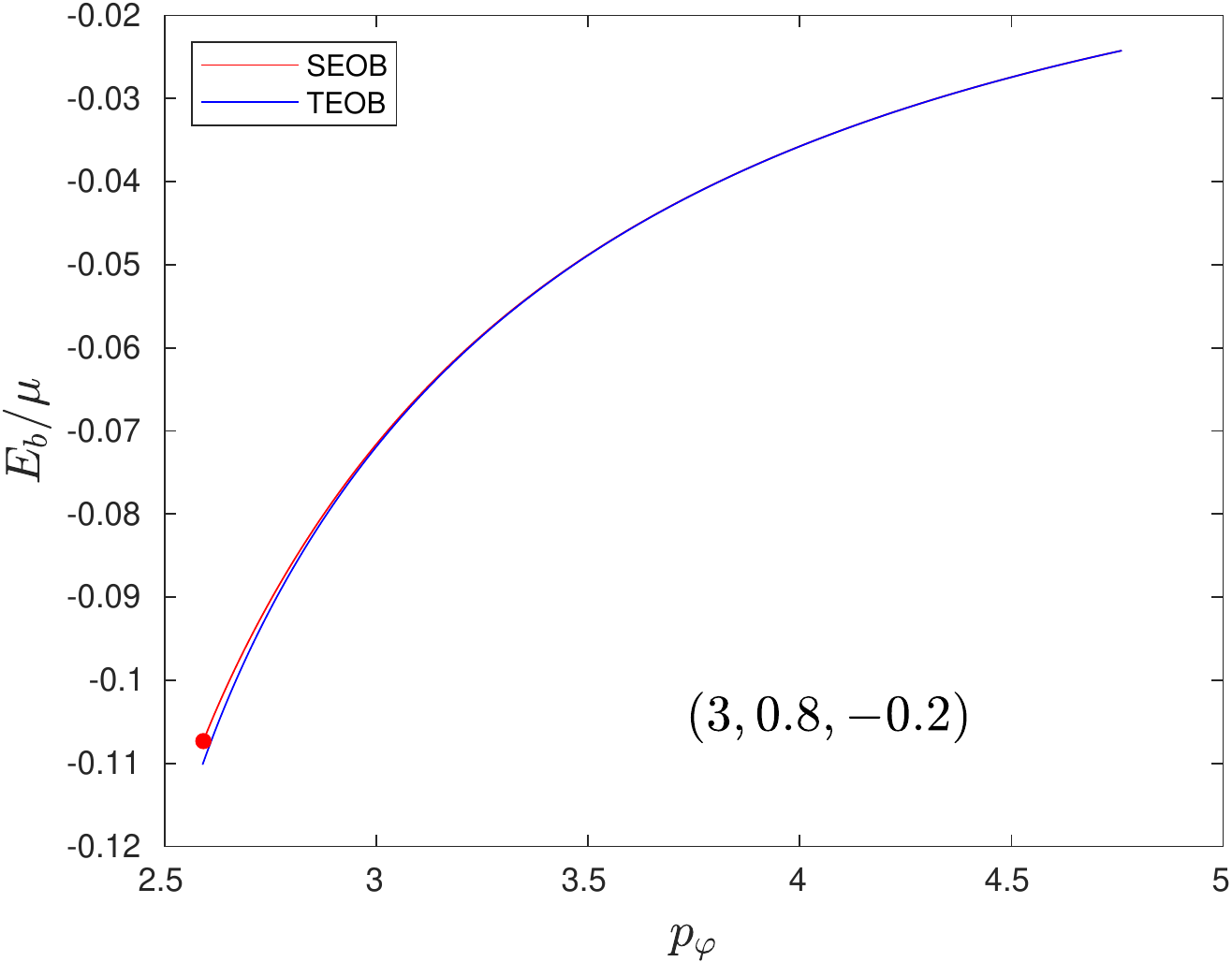}\\	
\vspace{0.3cm}	\hspace{0.2cm}\includegraphics[width=0.43\textwidth]{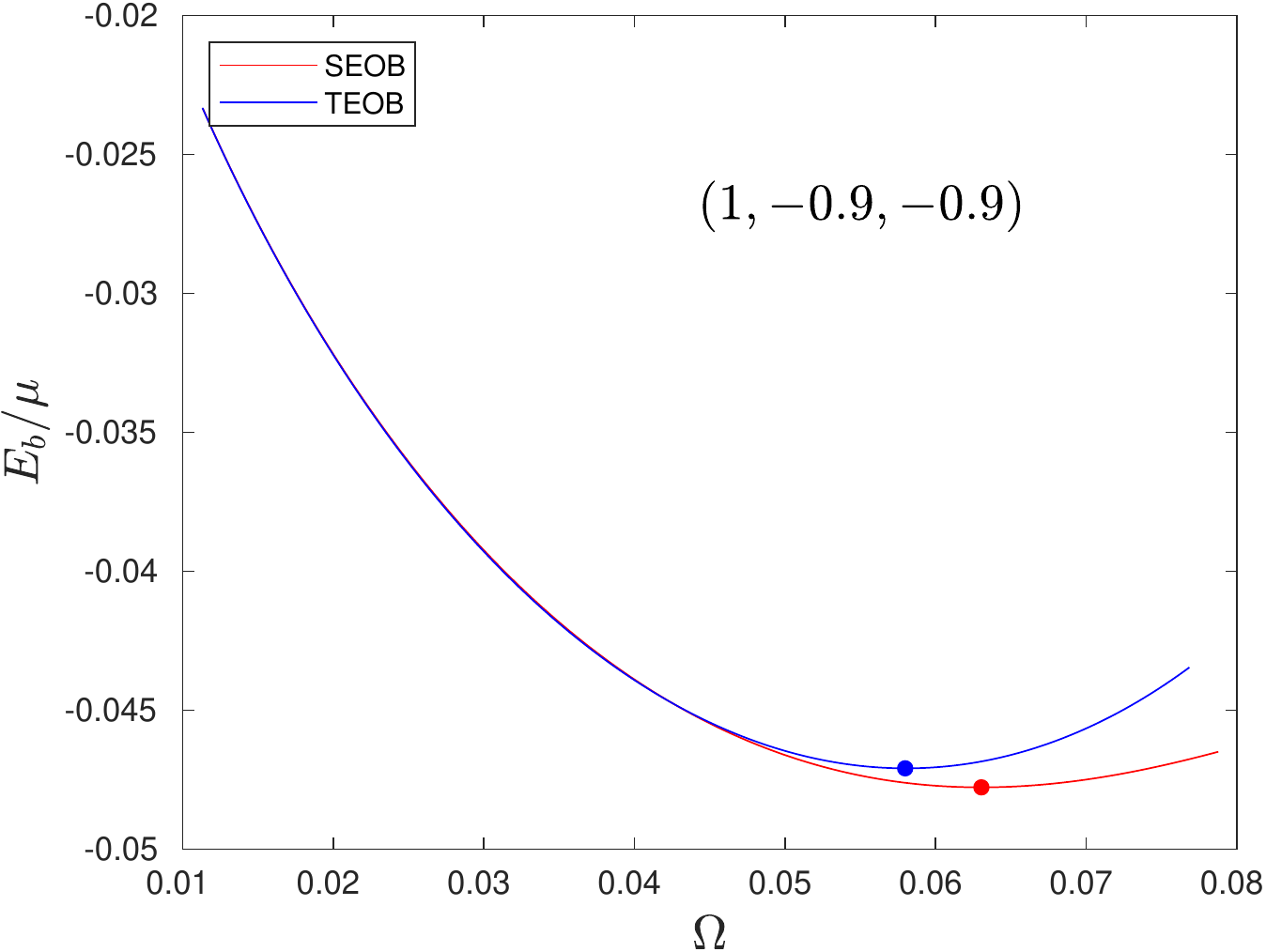}
\hspace{1.5cm}
\includegraphics[width=0.43\textwidth]{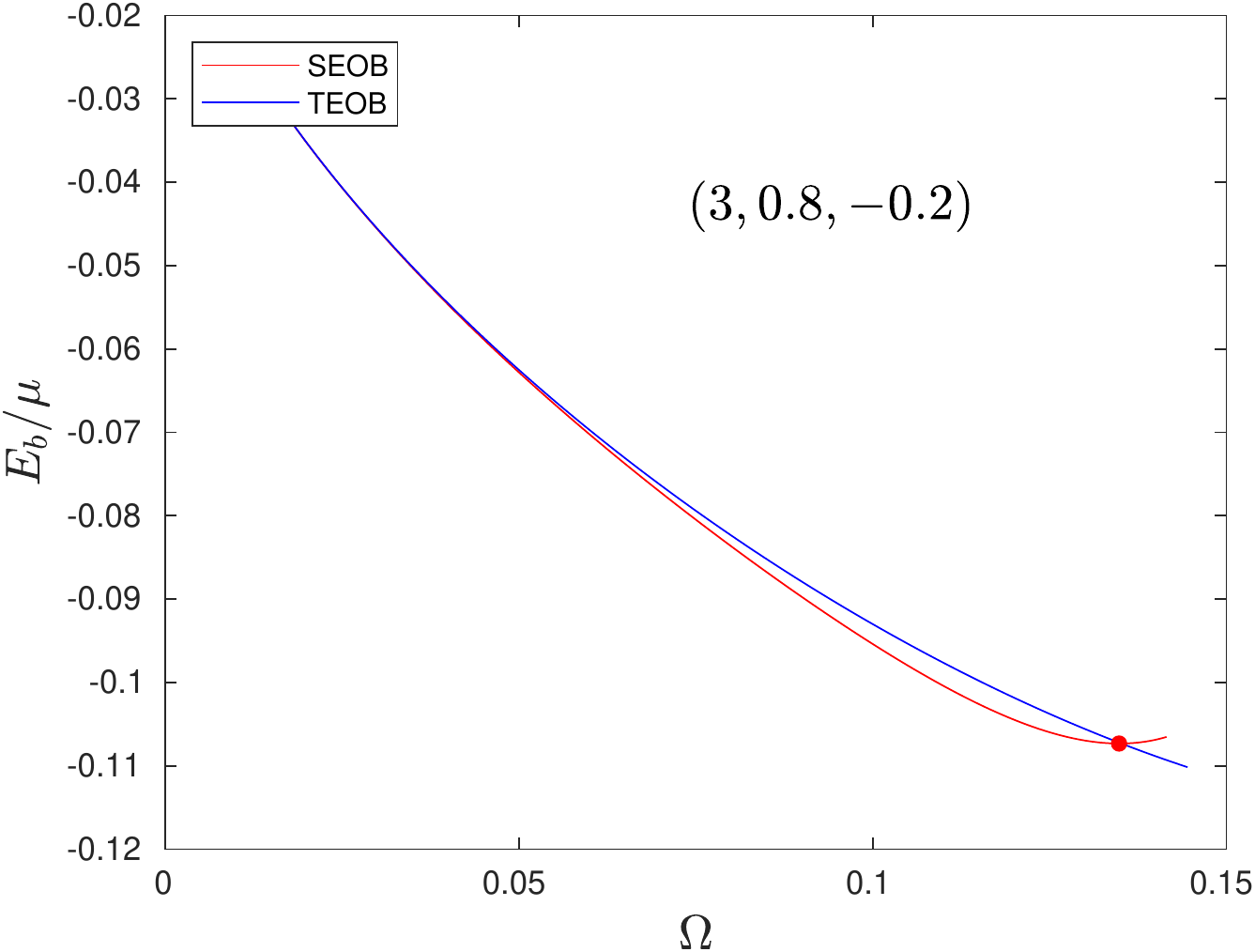}
\caption{
	\label{fig:Eb_adiab}
	Gauge-invariant relation between $E_b$ and $p_\varphi$ and $\Omega$ in the adiabatic case.
	The markers correspond to the LSO position (not present in \TEOBResumS{} for large aligned spins).
}
\center
\end{figure*}

\begin{figure*}[t]
\center
\includegraphics[width=0.335\textwidth]{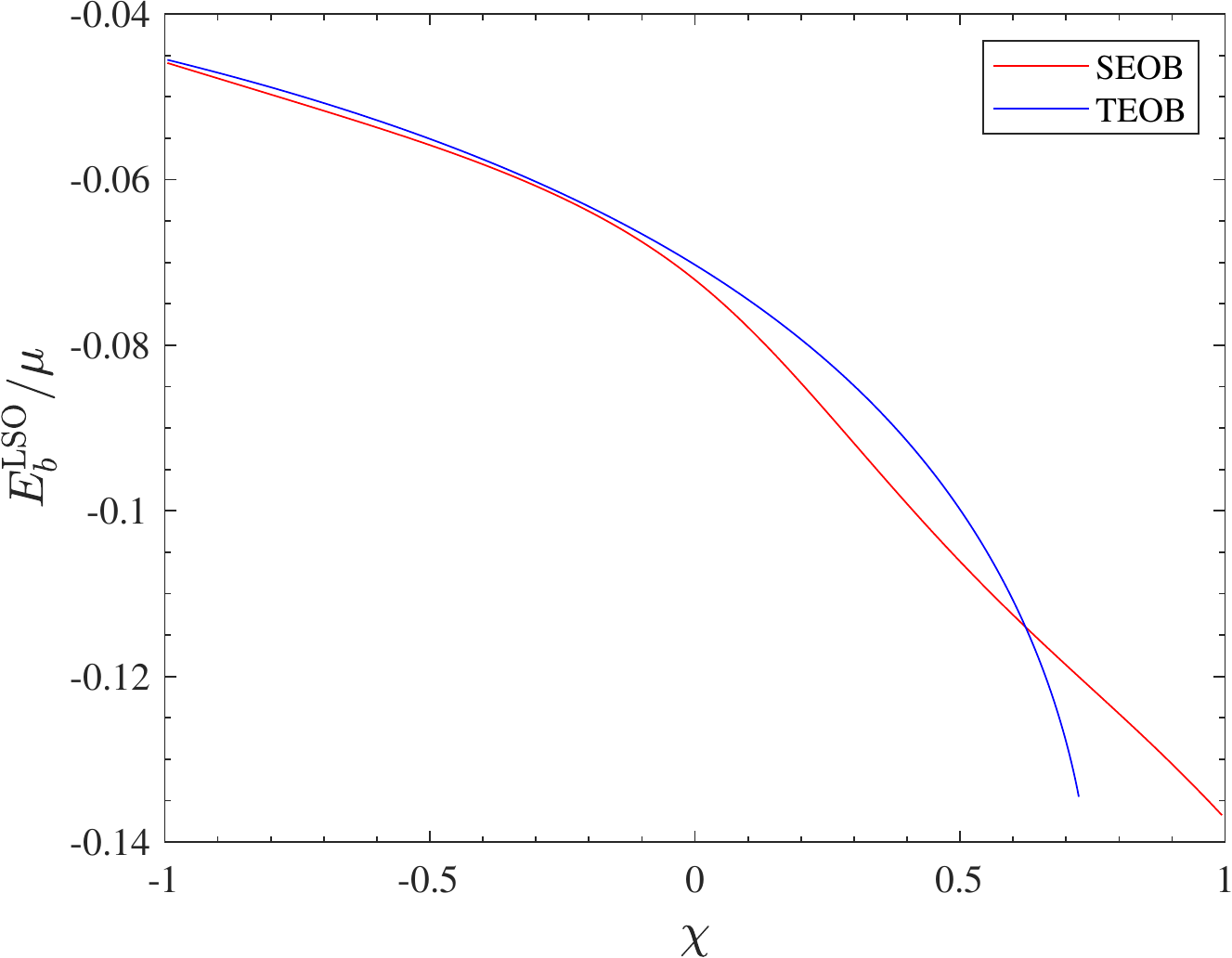}
\includegraphics[width=0.325\textwidth]{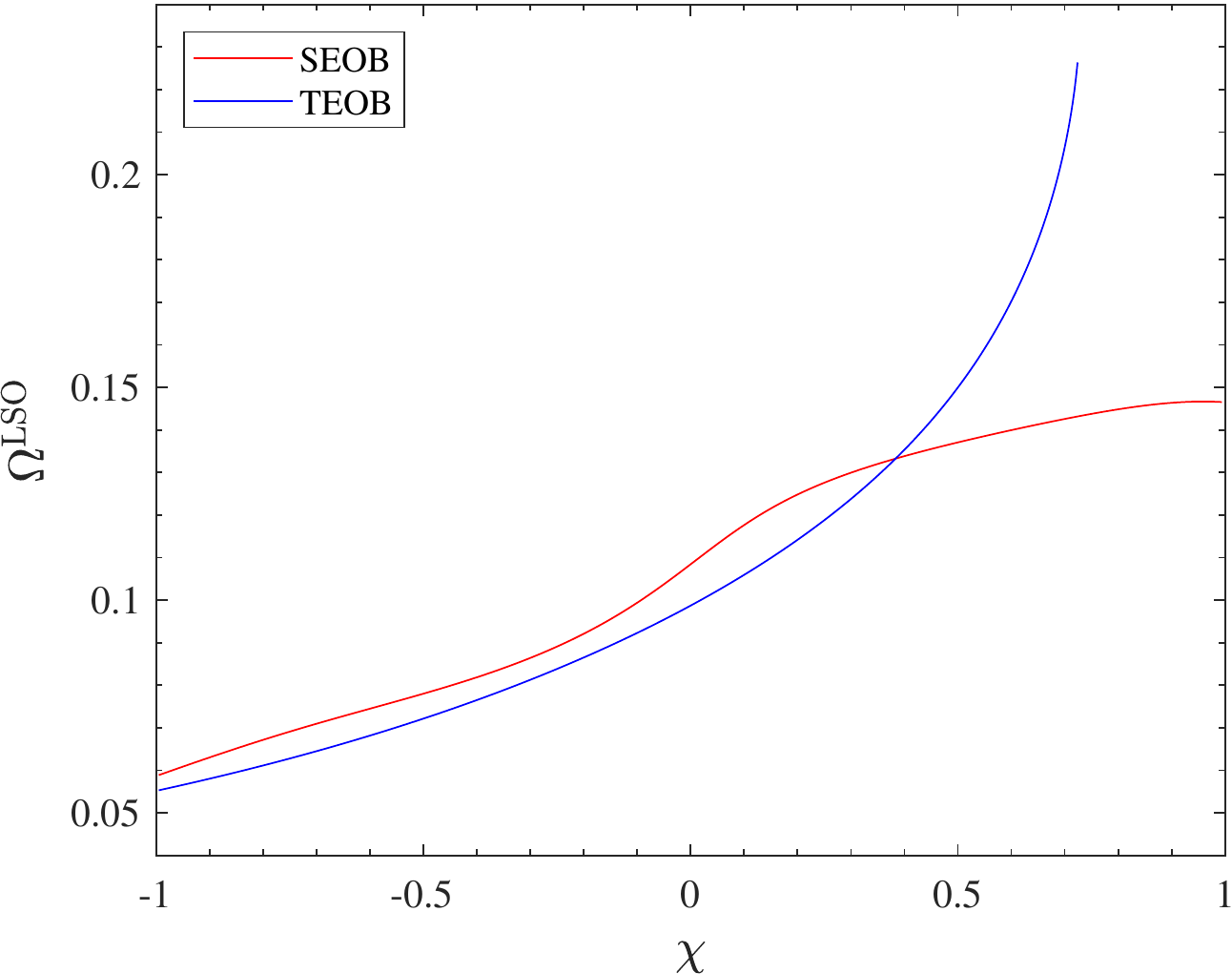}
\includegraphics[width=0.325\textwidth]{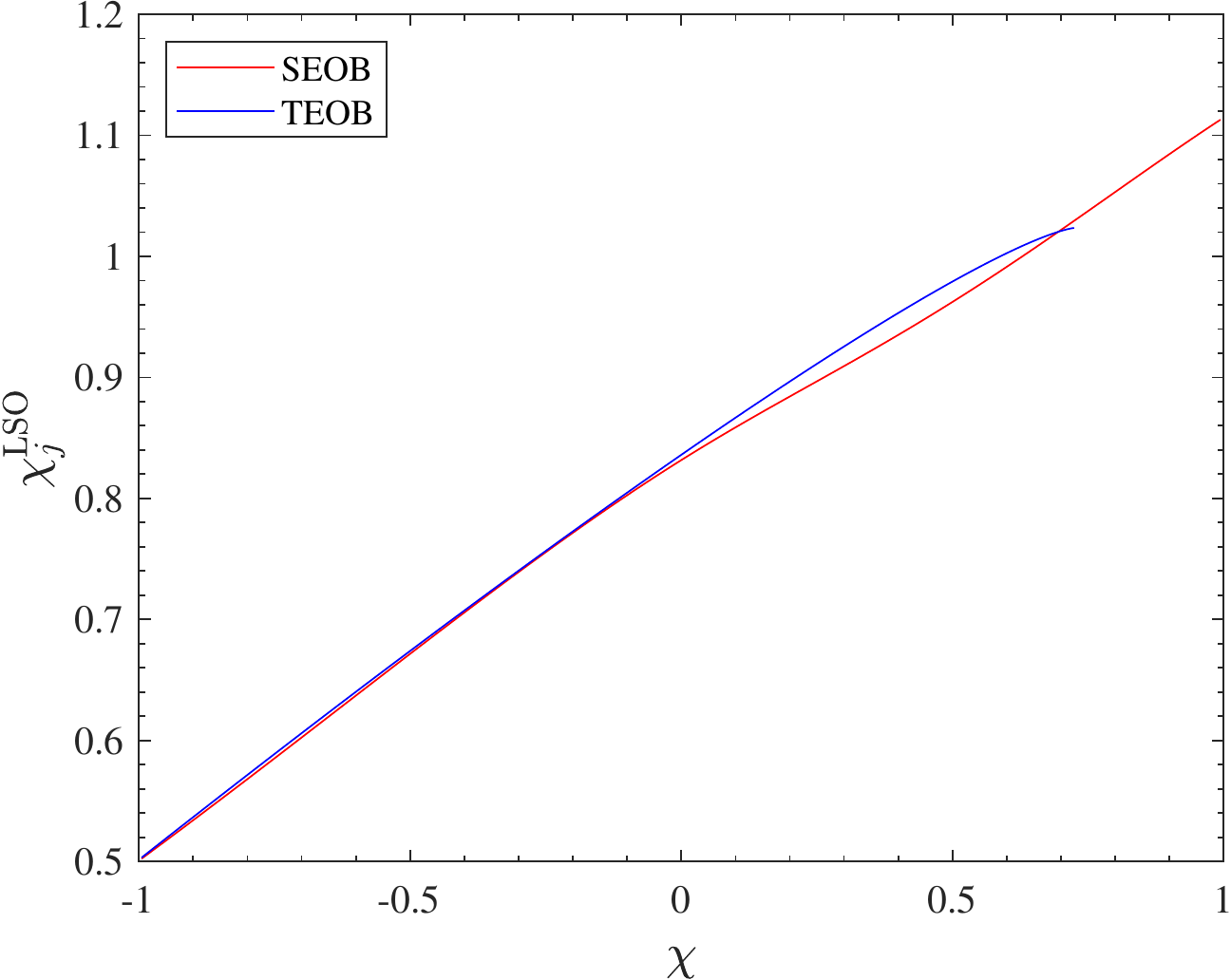}
\caption{
	\label{fig:LSO}
	Gauge-invariant quantities computed at the LSO for $q=1$. 
	Note that the \TEOBResumS{} Hamiltonian does not have an
	LSO after $\chi\approx 0.7$ and thus the corresponding curves terminate there.
}
\center
\end{figure*}

\begin{figure*}[t]
\center
\includegraphics[width=0.45\textwidth]{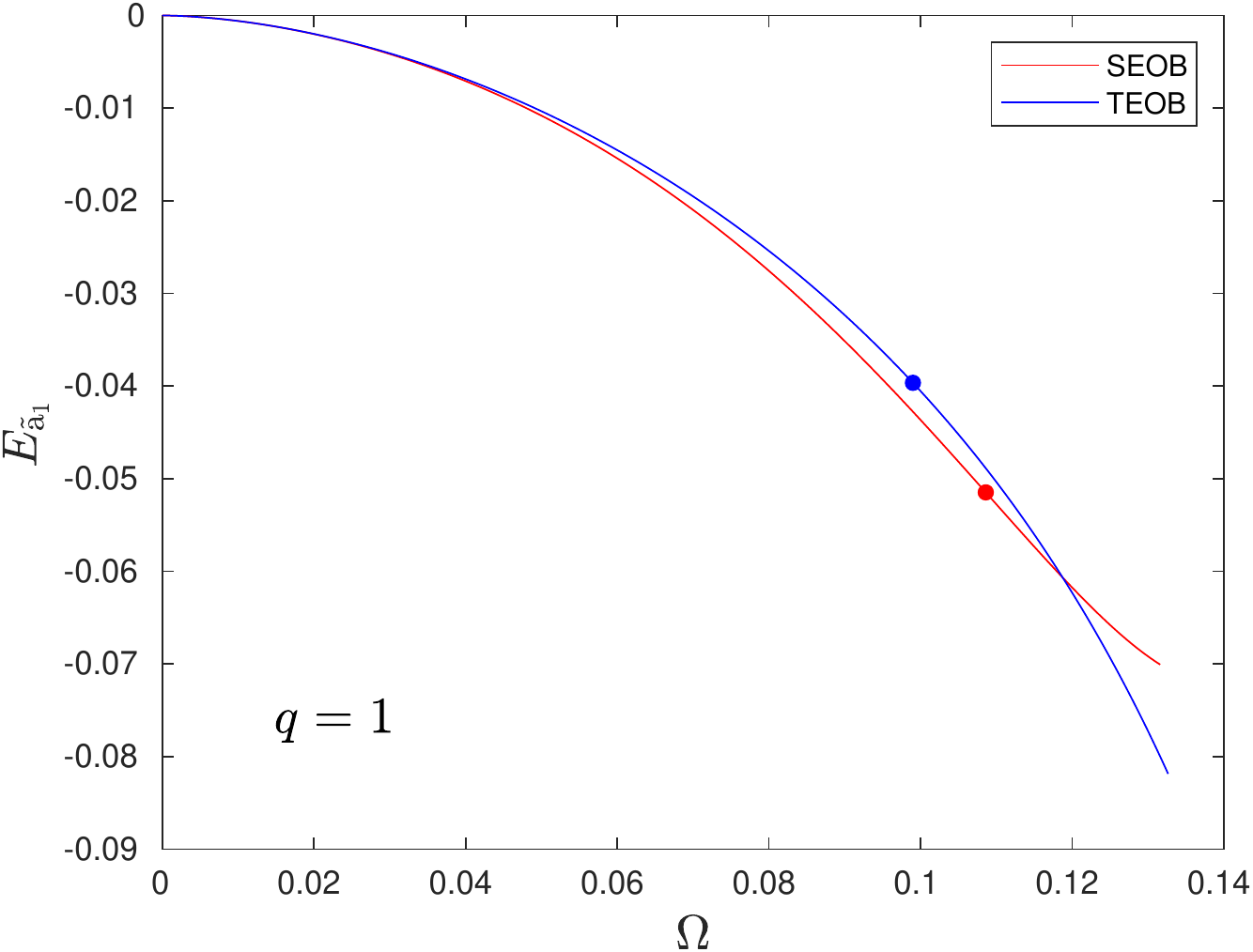}
\hspace{0.5cm}
\includegraphics[width=0.44\textwidth]{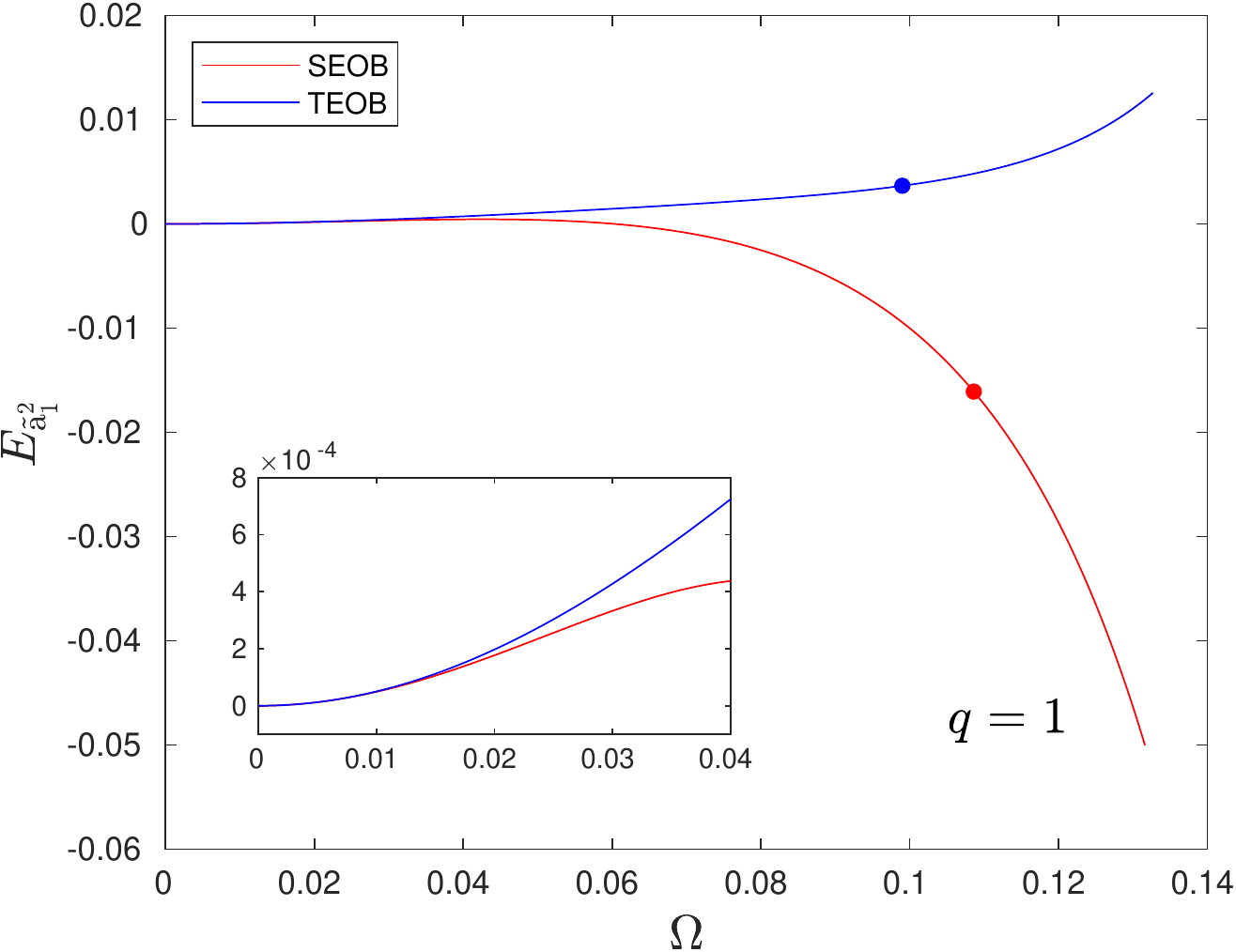}
\caption{
	Linear and quadratic-in-spin Hamiltonian contributions in the equal-mass case.
	$E_{\rm \tilde{a}_1}$ qualitatively agrees between the two models.
	The quadratic-in-spin behaviour is instead completely different, although it is similar in the PN regime.
	We remind the reader that for these systems $E_{\tilde{a}_1} = E_{\tilde{a}_2}$ and $E_{\tilde{a}_1^2} = E_{\tilde{a}_2^2}$, while $E_{\tilde{a}_1 \tilde{a}_2}$, though not shown, displays a similar behavior to $E_{\tilde{a}_1^2}$.
	The markers highlight the nonspinning LSO position.
}
\label{fig:Ha1_q1}
\center
\end{figure*}

\begin{figure*}[t]
\center
\includegraphics[width=0.45\textwidth]{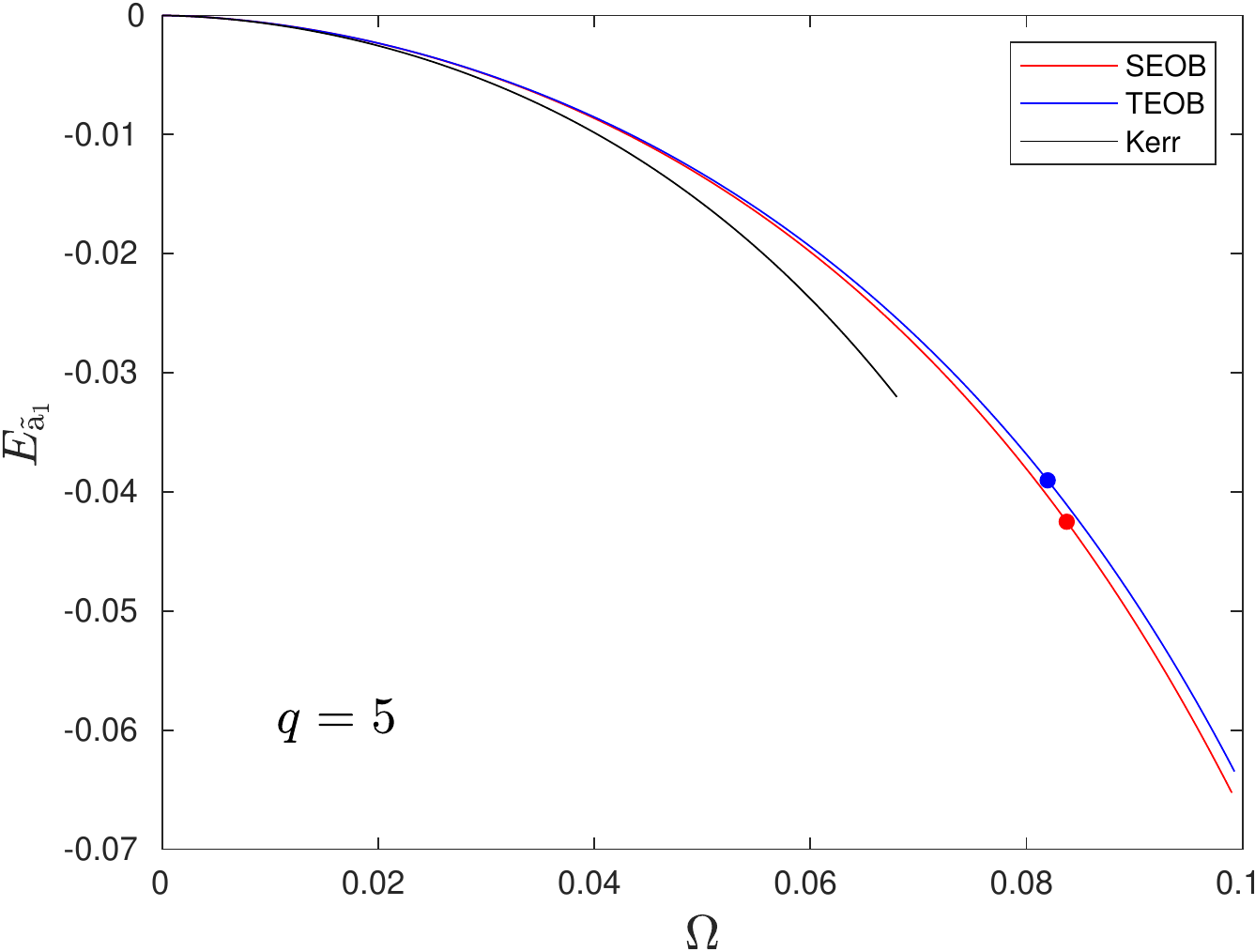}
\hspace{0.5cm}
\includegraphics[width=0.44\textwidth]{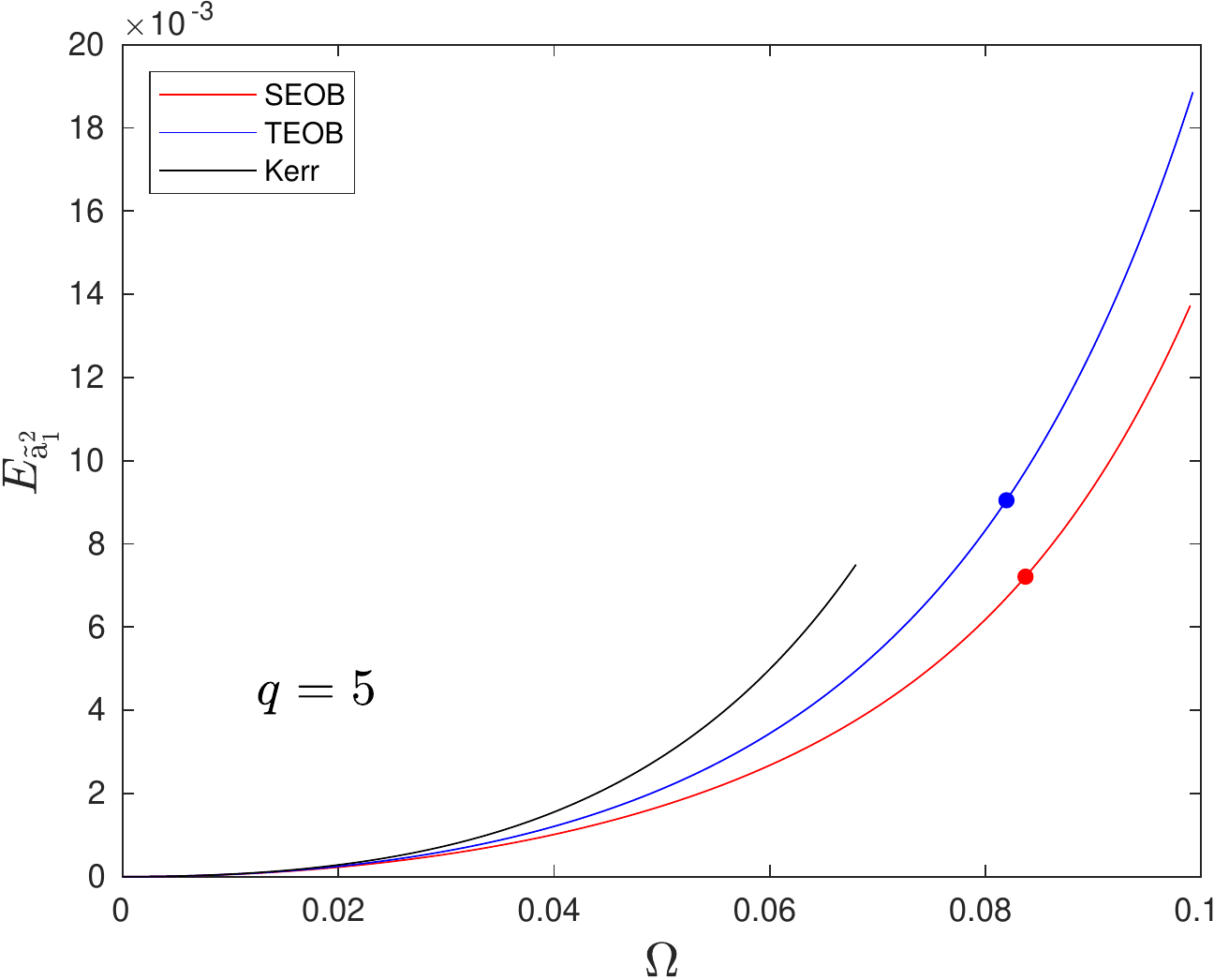}
\caption{
Same as Fig.~\ref{fig:Ha1_q1} for systems of $q = 5$.
We also added the Kerr corresponding functions up to the Schwarzschild LSO.
In this case, all three curves agree qualitatively.
}
\label{fig:Ha1_q5}
\center
\end{figure*}

\begin{figure*}[t]
\center
\includegraphics[width=0.325\textwidth]{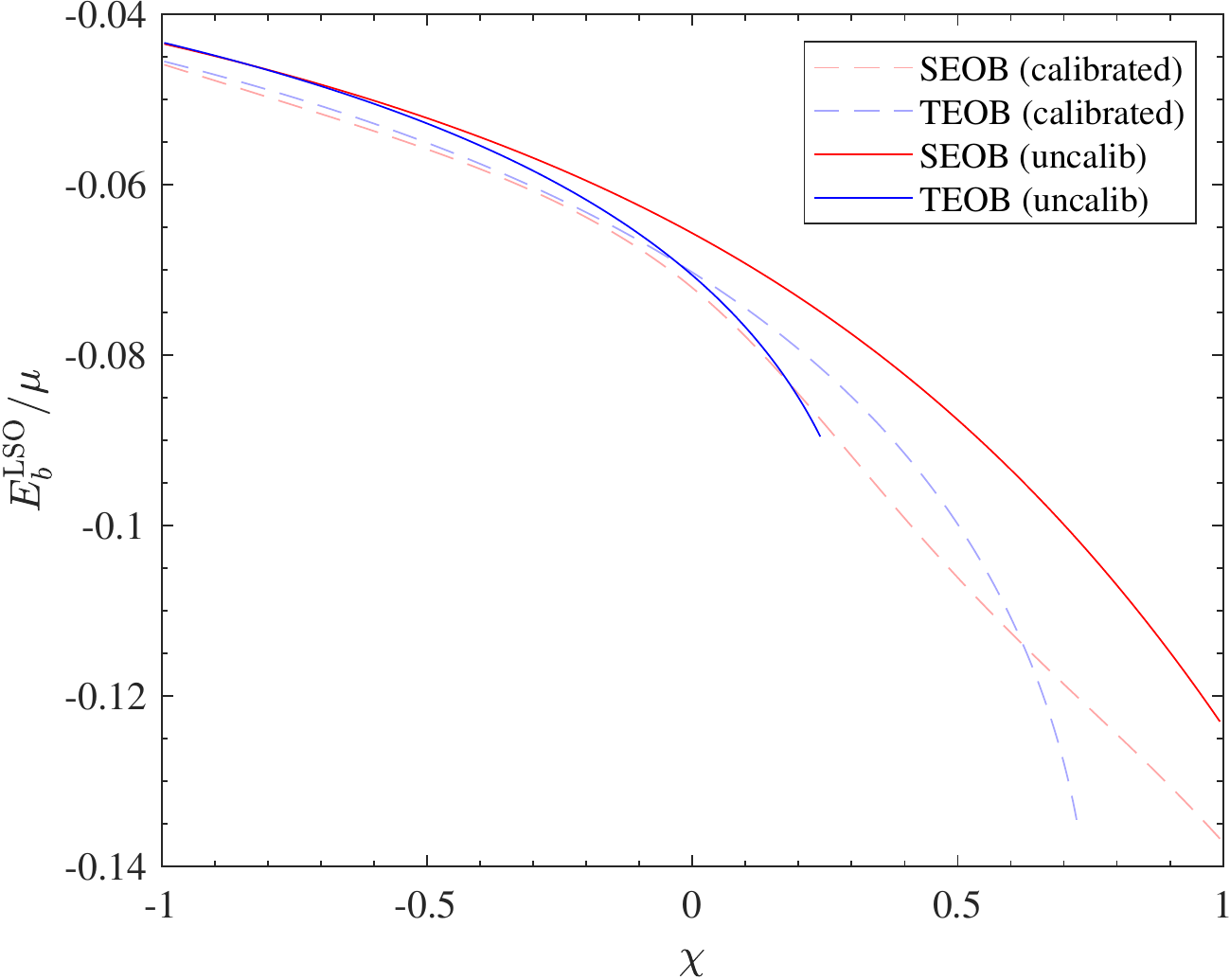}
\includegraphics[width=0.33\textwidth]{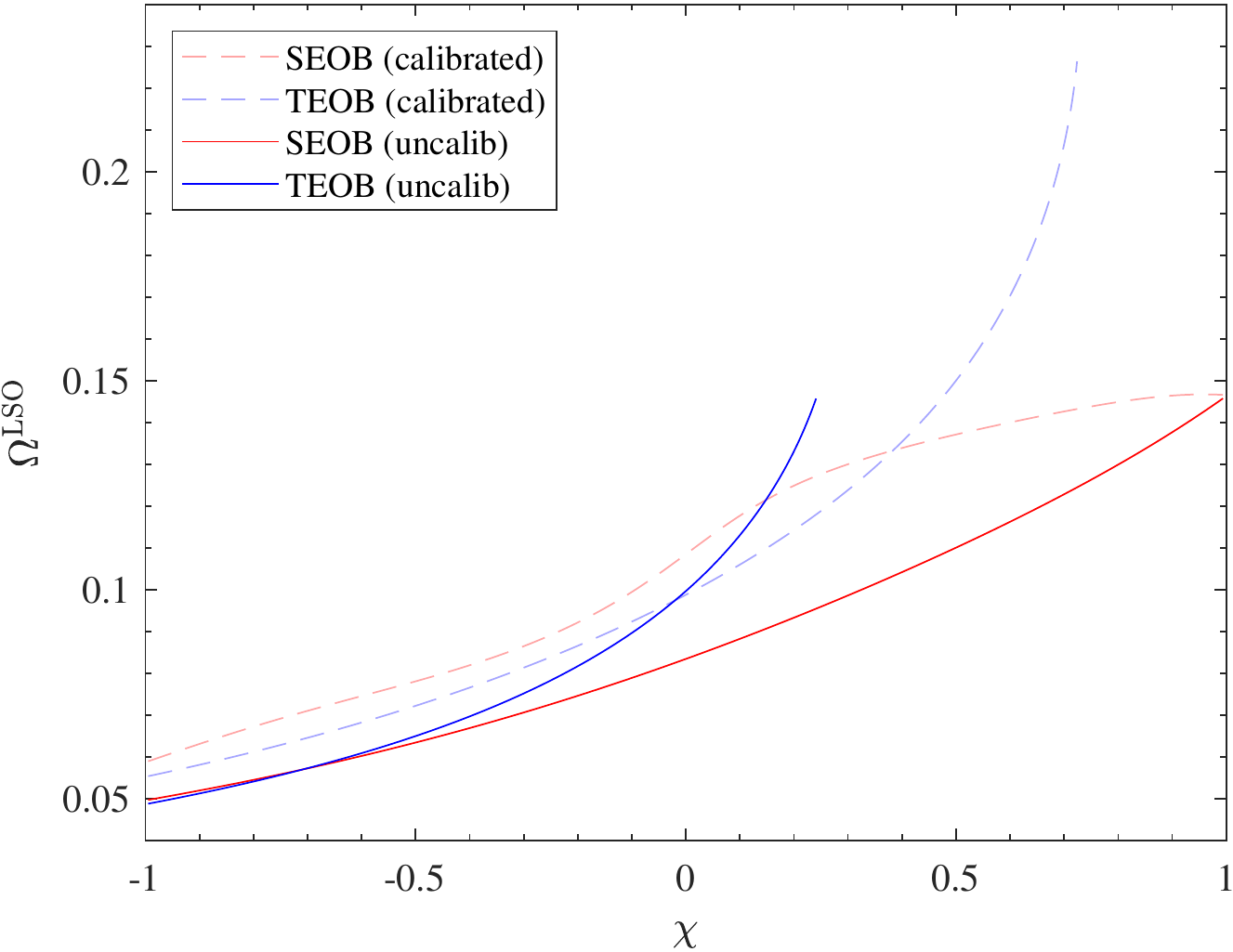}
\includegraphics[width=0.325\textwidth]{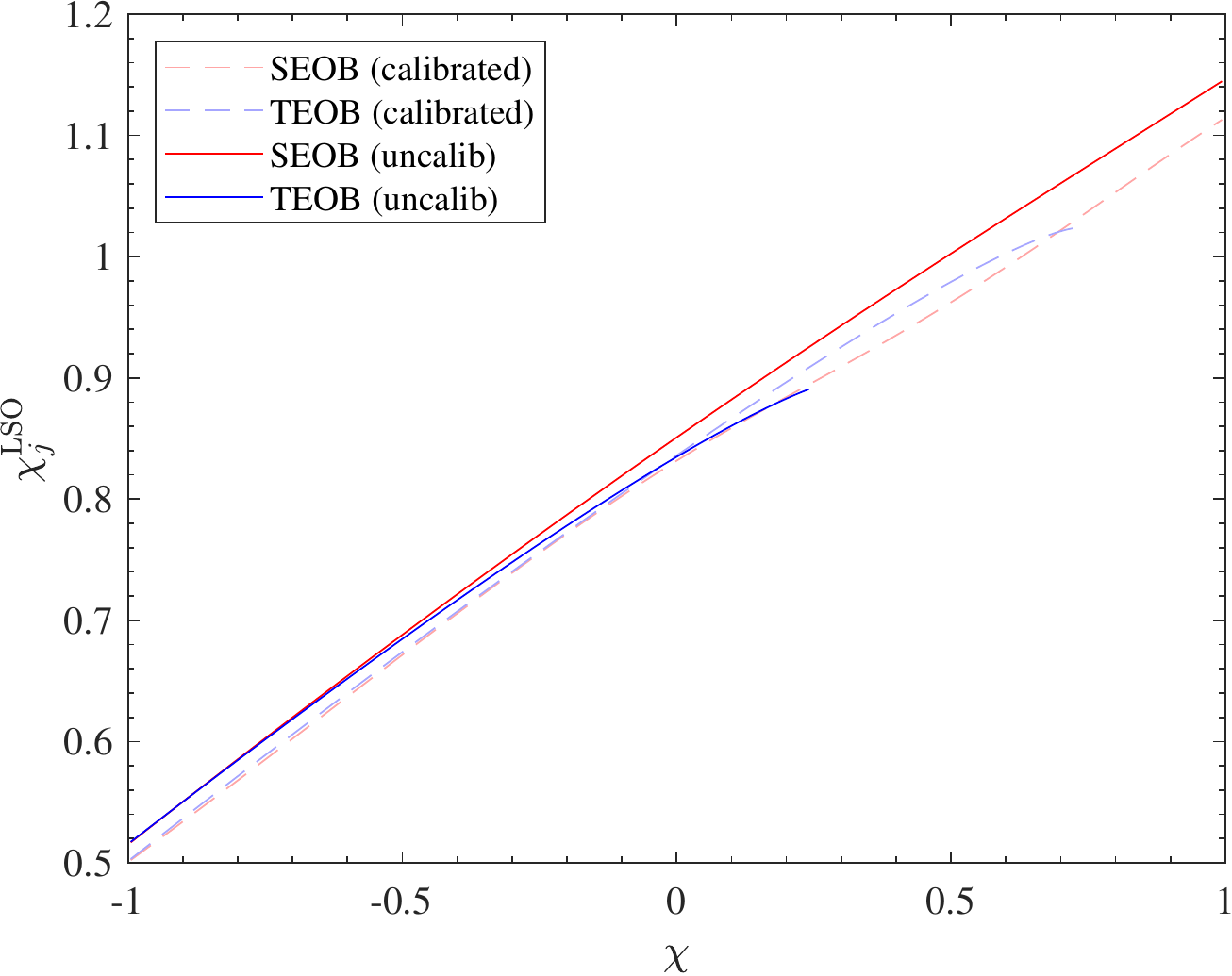}
\caption{
	\label{fig:LSO_nocalib}
	Same as Fig.~\ref{fig:LSO} using models without NR information, i.e. setting all calibration coefficients to zero.
	The NR-informed \TEOBResumS{} and \SEOBNRvq{} are indicated with dashed lines.
	Without calibration, \TEOBResumS{} does not have an LSO after $\chi\approx 0.3$.
}
\center
\end{figure*}

\begin{figure*}[t]
\center
\includegraphics[width=0.45\textwidth]{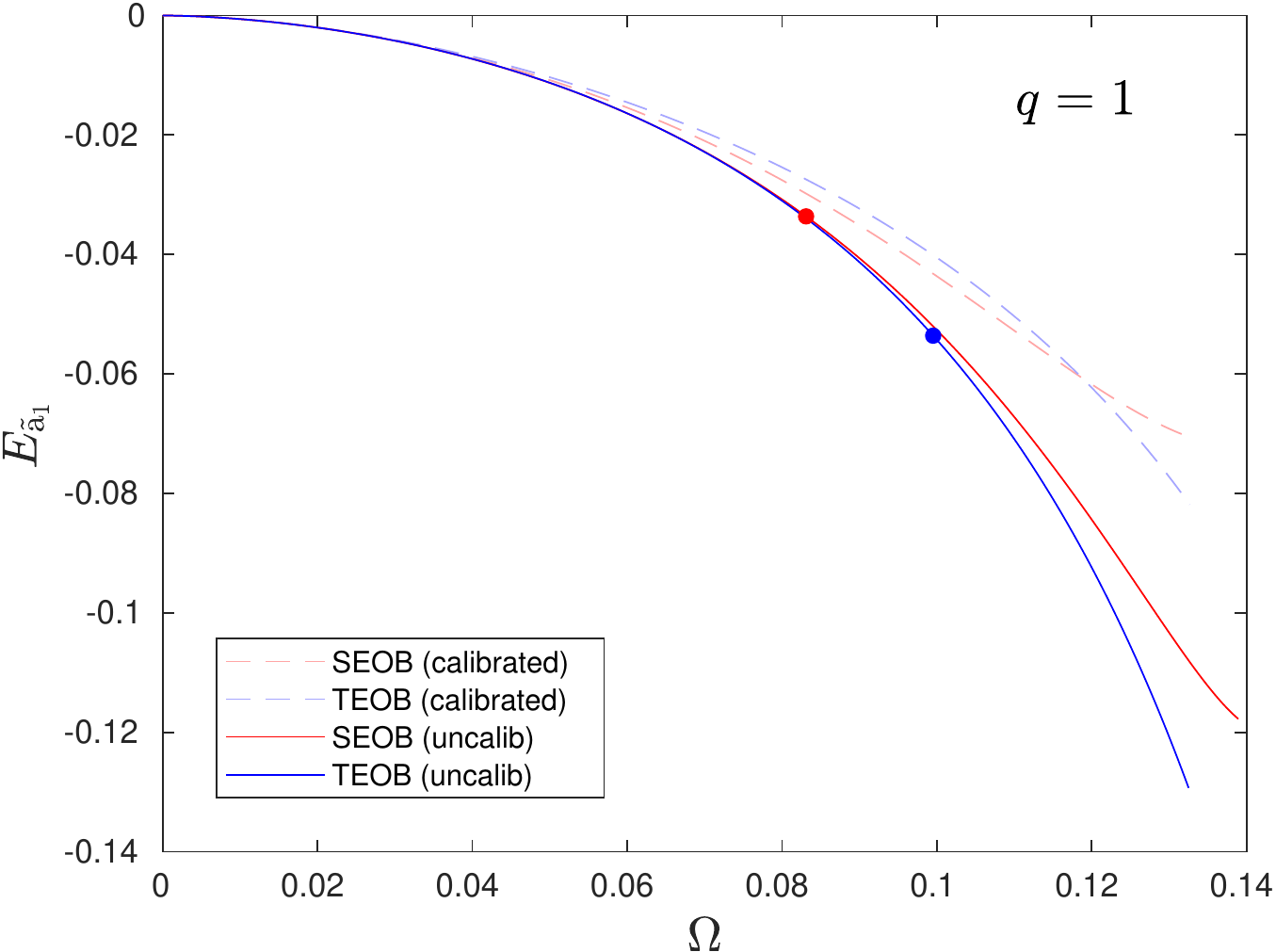}
\hspace{0.5cm}
\includegraphics[width=0.44\textwidth]{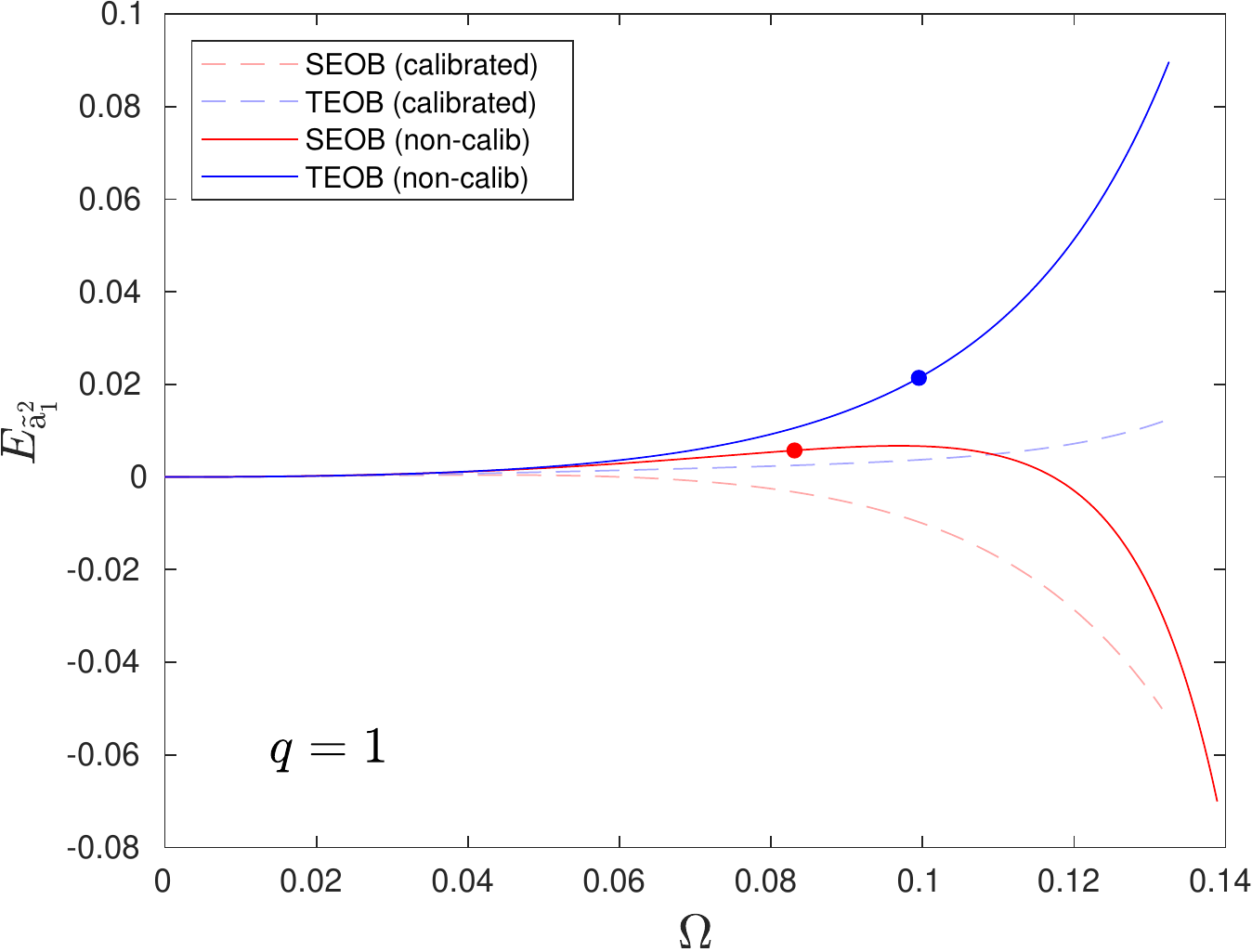}
\caption{
	Comparison of $E_{\rm \tilde{a}_1}$ and $E_{\rm \tilde{a}_1^2}$ for \TEOBResumS{} and \SEOBNRvq{} without NR calibration.
	The dashed lines correspond to the curves of Fig.~\ref{fig:Ha1_q1}.
	It is possible to notice how the non-calibrated curves are closer and have a more similar behavior up the LSO.
	Moreover, we can see that both the spin-orbit and spin-spin interactions are tempered by the use of NR information.
}
\label{fig:Ha1_nocalib}
\center
\end{figure*}

Our interest is to make some comparative statements between the dynamics of the two models. 
Since the models are calibrated to NR, and moreover are expressed in different gauges, 
direct comparisons between the analytical expressions discussed above are not informative.
On the contrary, comparisons between gauge-invariant quantities are meaningful and
we start by considering the {\it adiabatic} approximation to the dynamics, i.e.
a sequence of circular orbits.
We hence set $p_{r_*} = 0$ and compute, at each given radius, the circular angular
momentum $p_\varphi^{\rm circ}$ solving $\de \hat{H}_{\rm eff}(r,p_\varphi^{\rm circ})/\de r=0$.
We can then compare the rescaled binding energy of a system $\hat{E}_b\equiv (E-M)/\mu$ of the two models,
when plotted as a function of the angular momentum $p_\varphi$ or of
the dimensionless orbital frequency $\Omega \equiv M \Omega_{\rm phys} = \de \hat{H}_{\rm EOB}/\de p_\varphi$.

The results of these comparisons are shown in Fig.~\ref{fig:Eb_adiab}.
From simplicity, in the following we will often denote \TEOBResumS{} as TEOB and \SEOBNRvq{} as SEOB.
The markers highlight the location of the last stable orbit (LSO),
which corresponds to the inflection point of the Hamiltonian and is
thus found imposing $\de \hat{H}_{\rm eff}/\de r=\de^2 \hat{H}_{\rm eff}/\de r^2=0$.
As expected, the binding energies are similar but not exactly overlapping.
It is difficult to quantify the effects of this difference, but it
is probably tapered in the full models, when taking into account
the respective radiation reactions. In the next section, we will
compare binding energy in the non-adiabatic scenario,
adding the same radiation reaction to both models.

The general characteristics of the dynamics can be also summarized by 
inspecting various gauge-invariant quantities at the LSO, i.e. binding energy,
orbital frequency and the dimensionless Kerr parameter 
\begin{equation}
\chi_J \equiv \frac{1}{\nu} \frac{j_{\rm tot}}{\hat{H}_{\rm EOB}^2},
\end{equation}
where $j_{\rm tot}$ is total angular momentum and reads
\begin{equation}
j_{\rm tot} = p_\varphi + \frac{X_1}{X_2}\chi_1 + \frac{X_2}{X_1}\chi_2 \, .
\end{equation} 
This is done in Fig.~\ref{fig:LSO}, that refers to the equal-mass, equal-spin case. 
On the $x$-axis we put $\tilde{a}_0 = \chi_1 = \chi_2$. 
Note that the curve for \TEOBResumS{} stops at 
$\tilde{a}\approx 0.7$ because the LSO does not exist for higher spins. 
We will comment more on this aspect in the conclusions. 
It is interesting to note that for large, positive spins \TEOBResumS{}
predicts values of the LSO frequency larger than the \SEOBNRvq{} ones.

The last piece of information that can be extracted from the two Hamiltonians
in the adiabatic case concerns the spin-orbit and spin-spin contributions.
In fact, if we consider small spins, $\tilde{a}_i \ll 1$, we can expand
the Hamiltonian as
\begin{align}
\label{eq:expH}
\hat{H}_{\rm EOB}\left(\nu,\tilde{a}_1,\tilde{a}_2\right) \sim&~ 
E_0(\nu) + E_{\tilde{a}_1}(\nu)\, \tilde{a}_1 + E_{ \tilde{a}_2}(\nu)\, \tilde{a}_2 + \notag \\
&+ E_{\tilde{a}_1^2}(\nu)\, \tilde{a}_1^2 + E_{\tilde{a}_1 \tilde{a}_2}(\nu)\, \tilde{a}_1 \tilde{a}_2 + \notag \\
&+ E_{\tilde{a}_2^2}(\nu)\, \tilde{a}_2^2 + \O[\tilde{a}_i^3].
\end{align}
In this situation, the $E_X$ functions are well defined and depend on the mass ratio
and dynamical variables but not on the spin values. These functions hence
encode the way the linear and quadratic-in-spin terms are described in the two models. 
We can obtain each contribution analytically differentiating
$\hat{H}_{\rm EOB}$, e.g. $E_{ \tilde{a}_1} = (\de \hat{H}_{\rm EOB} / \de \tilde{a}_1)|_{\tilde{a}_i = 0}$.
For simplicity, we instead compute them numerically, considering
very small (positive or negative) spins and suitably summing/subtracting
the corresponding energies so to obtain the coefficients. For example,
$E_{\rm \tilde{a}_1} = \big[\hat{H}_{\rm EOB}|_{(\tilde{a}_1 = a,\, \tilde{a}_2 = 0)} - \hat{H}_{\rm EOB}|_{(\tilde{a}_1 = -a,\, \tilde{a}_2 = 0)}\big]/(2a)$, with $a \sim 10^{-4}$.

Note that in the adiabatic case, using $p_\varphi$ as a variable is problematic,
as it presents a cusp at the LSO, when the stable and unstable orbits branches meet.
Moreover, the spin-squared contributions are singular at the same point when plotted
versus the angular momentum. Conversely, the orbital frequency is continuous and well-behaved
near the LSO, making it more useful for comparisons. The results for equal-mass
systems are exhibited in Fig.~\ref{fig:Ha1_q1}.

The figure illustrates that $E_{\rm \tilde{a}_1}$ is reasonably consistent between the
two models, although it has a slightly different behavior after the nonspinning LSO.
$E_{\rm \tilde{a}_1^2}$, instead, is completely different. The two curves behave similarly
in the PN regime (for small values of $\Omega$) but quickly start to disagree and
even change sign well before the LSO. Some difference was to be expected due to the
different included PN information and way to include spin-spin terms within the EOB framework.

Since these functions are universal, we can extract the linear-in-spin contribution
for any value of the spins as $E_{\tilde{a}_1} \tilde{a}_1 + E_{ \tilde{a}_2} \tilde{a}_2$,
even if for large spins the expansion of Eq.~\eqref{eq:expH} is no longer valid and
higher order contributions become non-negligible. Thus, we expect that these differences
will be more pronounced for large aligned spins, when the LSO occurs at higher frequencies.

As a consistency test, we show in Fig.~\ref{fig:Ha1_q5} the same comparison
for $q = 5$, together with the Kerr corresponding curves.
As expected, since the two models share the same $\nu \to 0$ and PN limits,
in this case the curves have a similar behavior and are close to the Kerr functions. 

As we briefly mentioned in the previous sections, the NR-informed parameters
introduce a complicated spin-dependence in both models. In order to remove these
effects, we compare in Fig.~\ref{fig:LSO_nocalib} the LSO quantities for \TEOBResumS{} and \SEOBNRvq{},
after eliminating the NR-calibration, i.e. we impose $a_6^c = c_3 = 0$ and
$K = d_{\rm SS} = d_{\rm SO} = 0$ respectively. Two features become evident:
(i) \TEOBResumS{} does not display an LSO for $\chi \geq 0.3$; (ii) \SEOBNRvq{}
has a behavior that is Kerr-like and does not display a change of concavity.

We conclude this section by showing in Fig.~\ref{fig:Ha1_nocalib} the comparisons between
$E_{\tilde{a}_1}$ and $E_{\tilde{a}_1^2}$. We can see that that the main effect of
using NR information is a decrease in the importance of the spin terms. However,
NR-calibrated terms also change the behavior of the spin interaction. Without these,
the \TEOBResumS{} and \SEOBNRvq{} curves are closer and $E_{\tilde{a}_1^2}$
is positive for both models up to the LSO.

\subsection{Non adiabatic dynamics}
\label{sec:nonadiab}
\begin{figure*}[t]
\center
\includegraphics[width=0.43\textwidth]{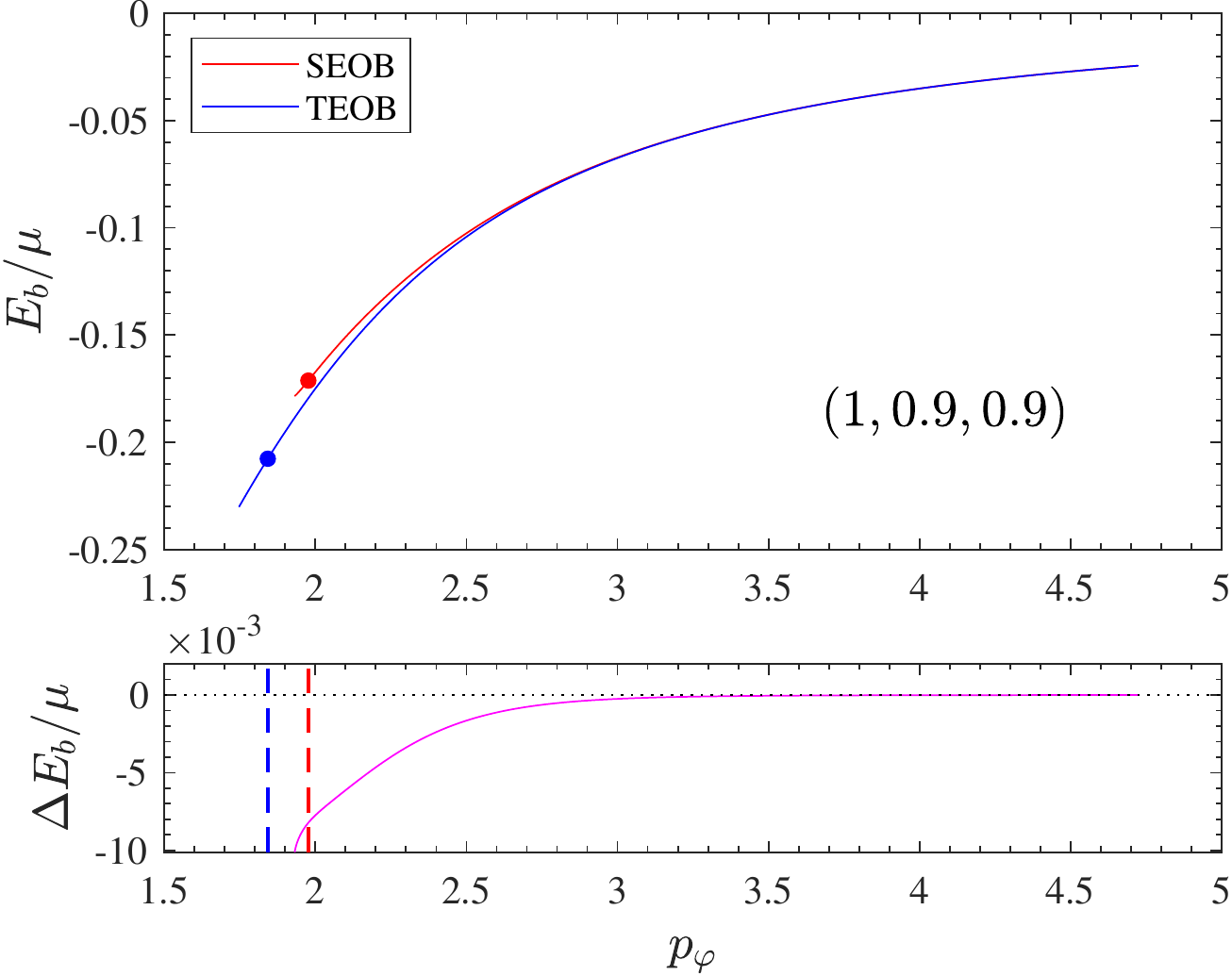}
\hspace{1cm}
\includegraphics[width=0.43\textwidth]{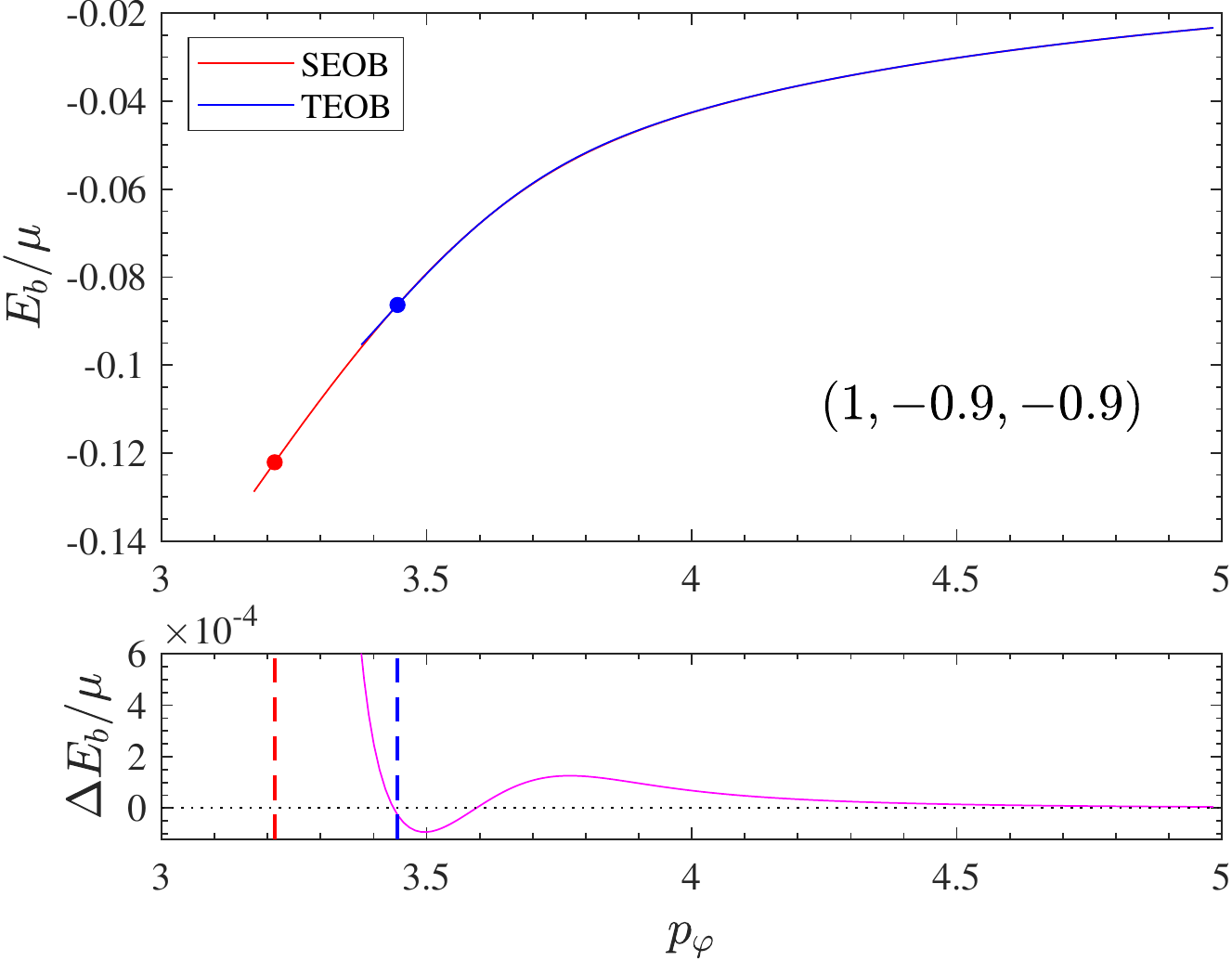}\\	
\vspace{0.3cm}
\includegraphics[width=0.43\textwidth]{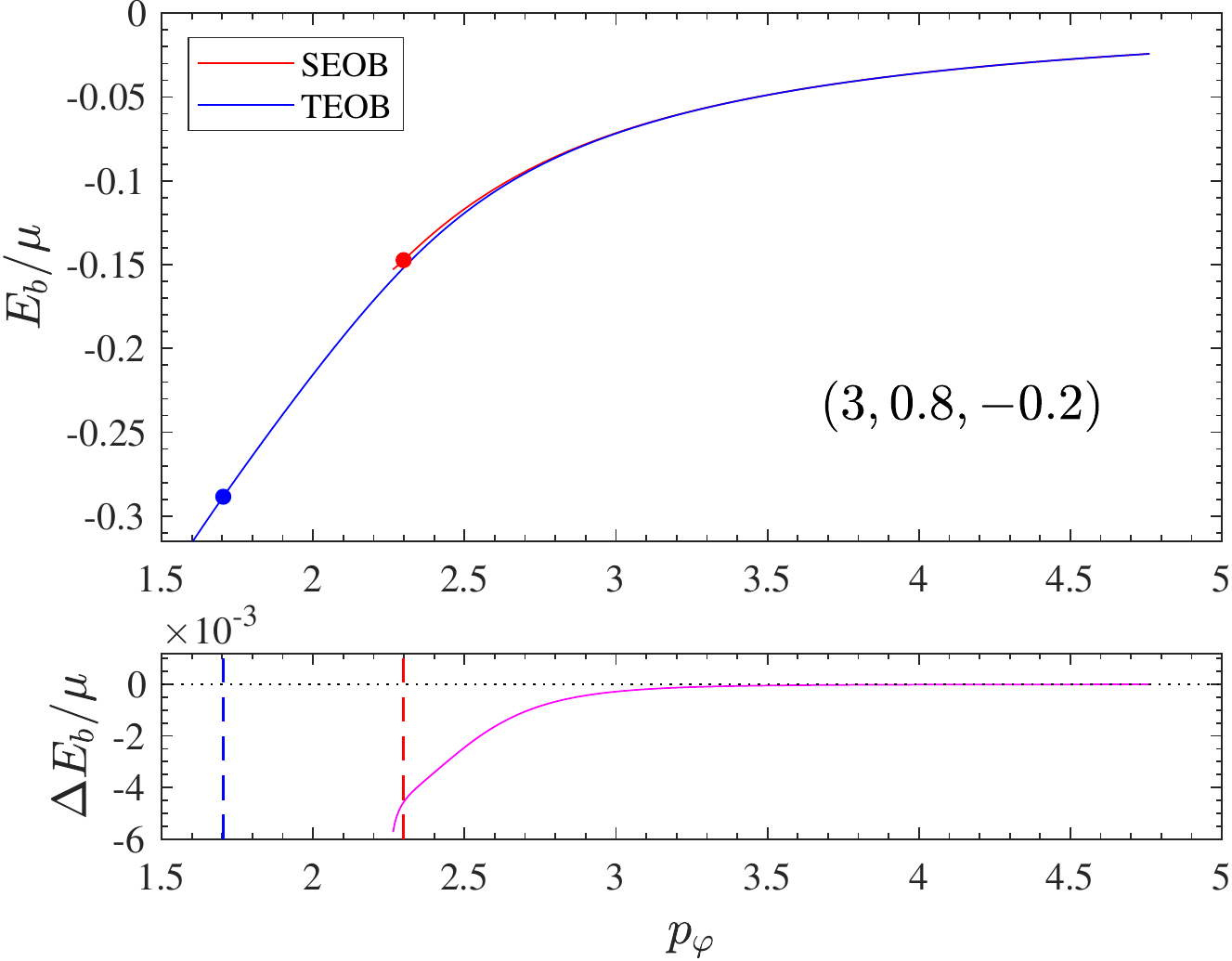}	
\hspace{1cm}
\includegraphics[width=0.43\textwidth]{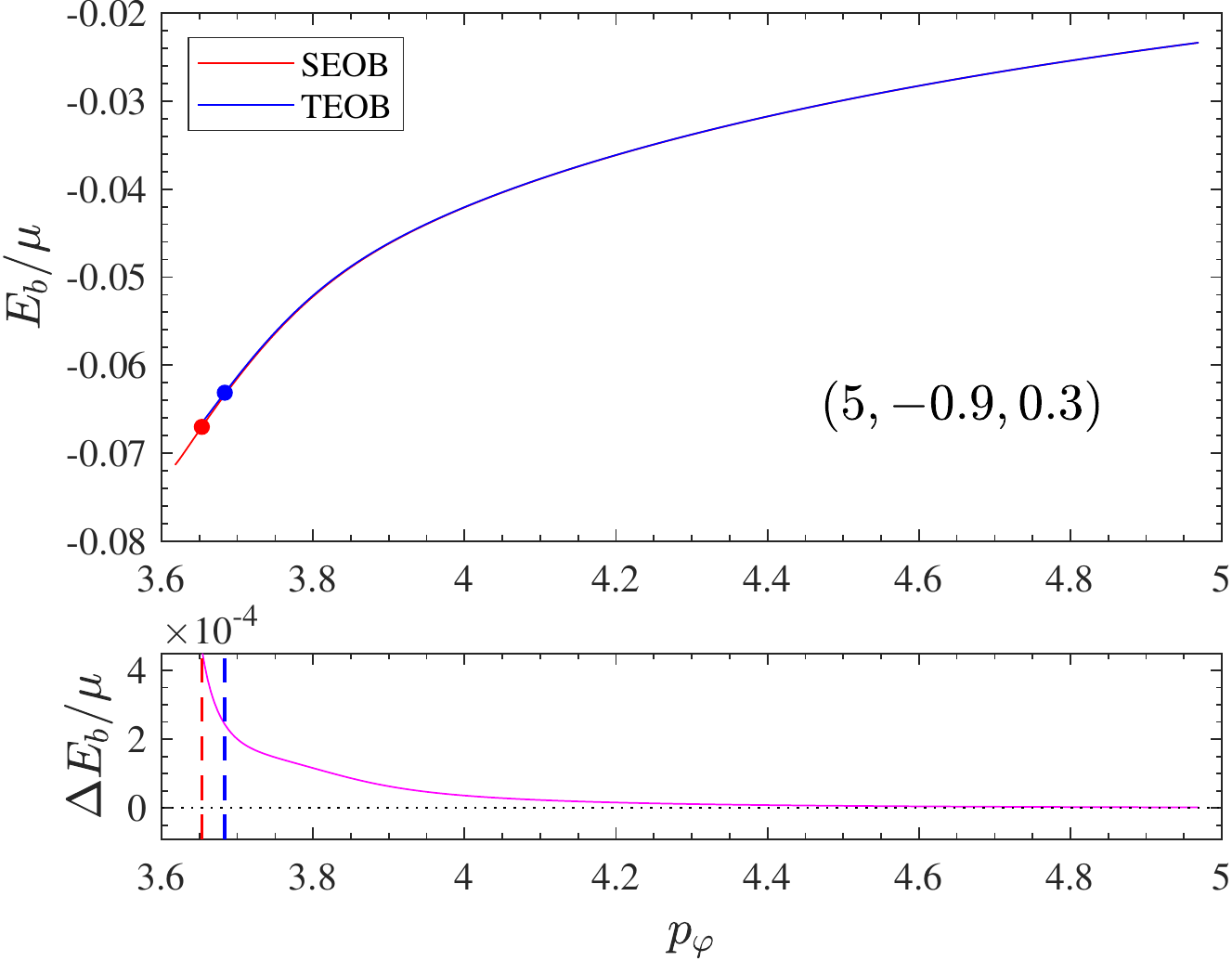}
\caption{
	Non-adiabatic evolution: gauge-invariant relation between binding energy $E_b/\mu$ and orbital angular momentum
	$p_\varphi$ obtained using the two different \TEOBResumS{} and \SEOBNRvq{} Hamiltonians, 
	but the same radiation reaction.
	The markers highlight the position of the peak of the $(2,2)$ mode.
	The lower panel shows the difference $\Delta E_b/\mu = (E_b^{\rm TEOB} - E_b^{\rm SEOB})/\mu$. 
}
\label{fig:Ebfull}
\center
\end{figure*}
Let us now complement the above section with similar comparisons based on {\it non-adiabatic}
evolutions, so to get up to merger. 
To do so, for both Hamiltonians we write Hamilton's equations with the
{\it same} radiation reaction ${\cal F}_\varphi$. 
For consistency between \TEOBResumS{} and \SEOBNRvqT{}, we use the formal expression
of ${\cal F}_\varphi$ discussed in Ref.~\cite{Nagar:2018zoe}, where however the argument
$x$ is taken to be $x = \Omega^2 (\Omega|_{p_r = 0})^{-4/3}$.
We stress that this choices does not correspond to neither the \TEOBResumS{} nor the \SEOBNRvq{} one.
The purpose of this section is to purely explore the
structure of the Hamiltonians in the strong field, and compare them. 
It is intended that the full energetics obtained from this dynamics is not expected
to be fully compatible with the corresponding NR one, like
it is for the NR-completed model~\cite{Nagar:2015xqa}.
Similarly, we don't improve the inspiral EOB analytical waveform
with a NR-improved description of the merger (i.e., next-to-quasi-circular corrections)
nor ringdown, but we adopt it as is. However, since its amplitude has a peak that
is known to be close (bot in location and amplitude) to the actual merger
point obtained by NR simulations, we use it as an {\it approximate} merger point (note that this is the 
choice usually adopted in the analytical description of coalescing and merging BNS).
Such approximate merger location will be useful below.
Fig.~\ref{fig:Ebfull} compares the relation $E_b(p_\varphi)$ of the two models
for a few configurations. The approximate merger point (as defined above) for each model
is shown as a colored marker. One sees that, on randomly chosen configurations,
the global differences are non negligible. In particular, they are larger than the expected
uncertainty on the corresponding NR curves ($\sim 10^{-4}$). Moreover, the position of
the $(2,2)$ peak is often very different, with \TEOBResumS{} merging later for large
aligned spins and sooner for anti-aligned ones.

\subsection{Modifying the resummation of the $A$-potential of \TEOBResumS{}: EOB/NR phasing and unfaithfulness}
\label{sec:exploringA}
\begin{figure*}[t]
\center
\includegraphics[width=0.45\textwidth]{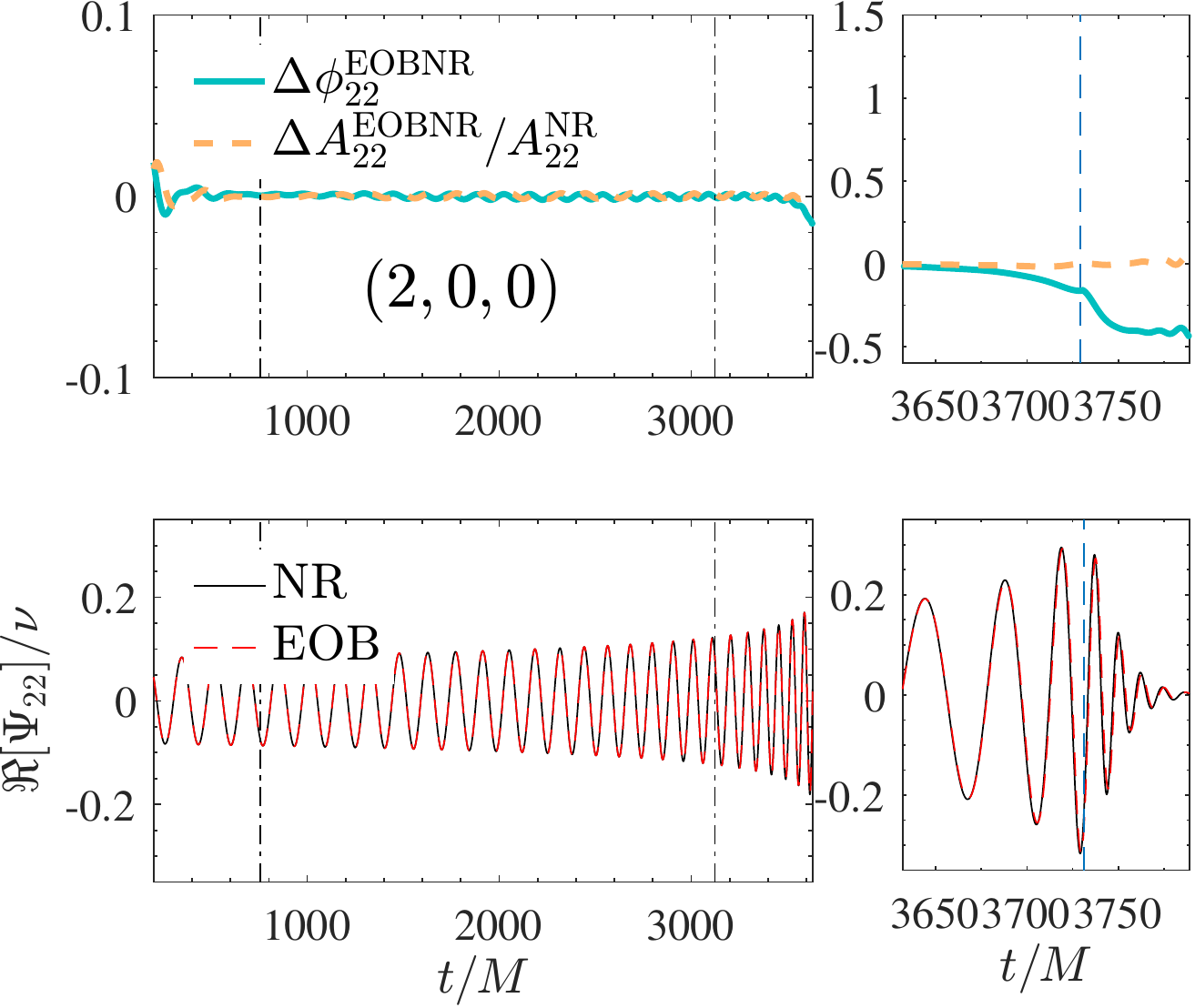}
\includegraphics[width=0.45\textwidth]{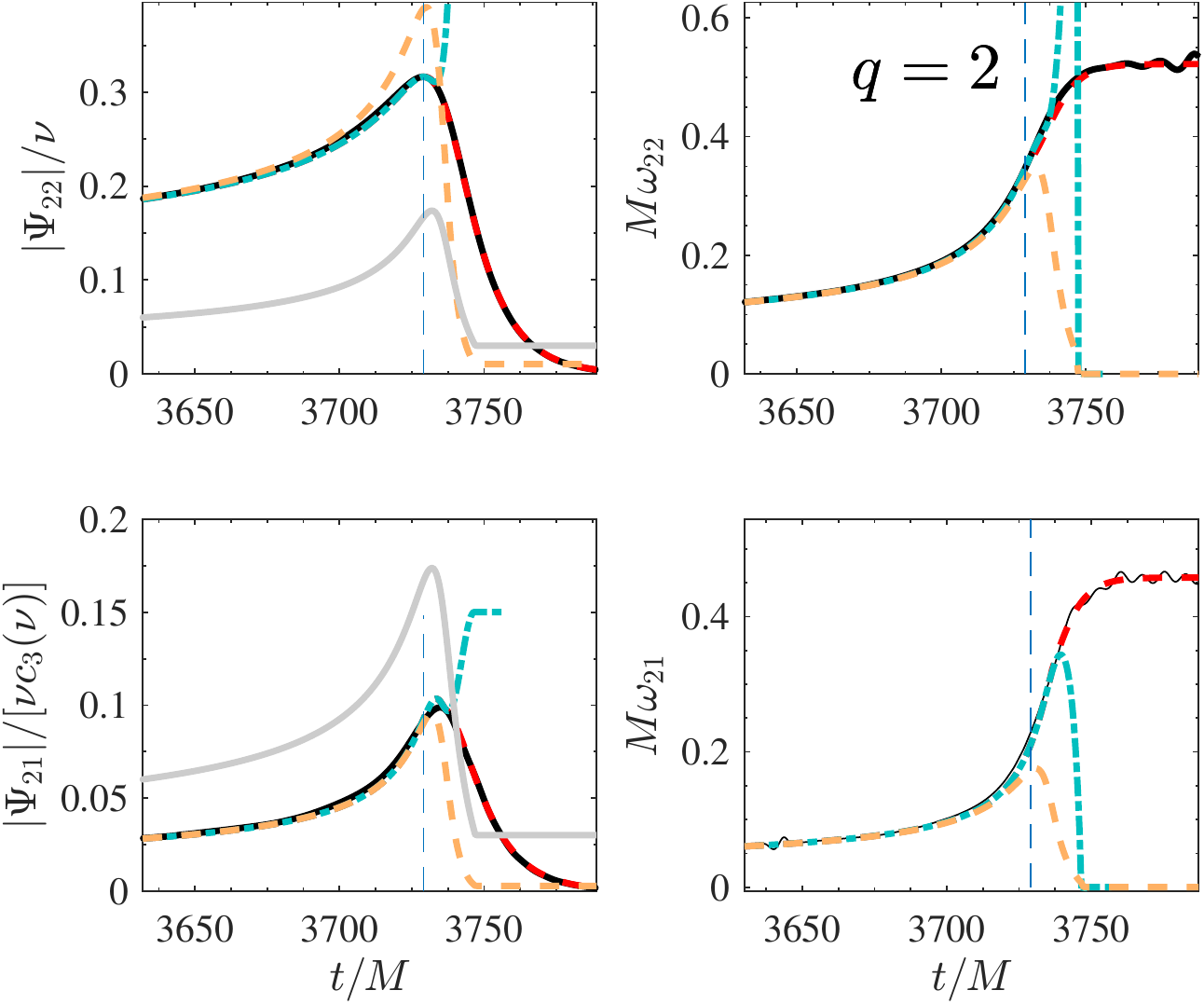}	
\caption{EOB/NR phasing comparison with {\tt TEOBiResumMultipoles}~\cite{Nagar:2019wds},
with the Pad\'e resummed $A$ potential.
Left panel: the $(2,2)$ phasing. Right panel, $(2,2)$ and $(2,1)$ amplitude and frequency. The plots also
show: (i) the orbital frequency (gray lines): (ii) the bare EOB inspiral waveform (orange); (ii) the NQC improved waveform
(blue); (iii) the NQC-ringdown completed waveform (red). The EOB/NR consistency for the $(2,1)$ mode is excellent. 	
\label{l2_teob}}
\center
\end{figure*}

\begin{figure*}[t]
\center
\includegraphics[width=0.45\textwidth]{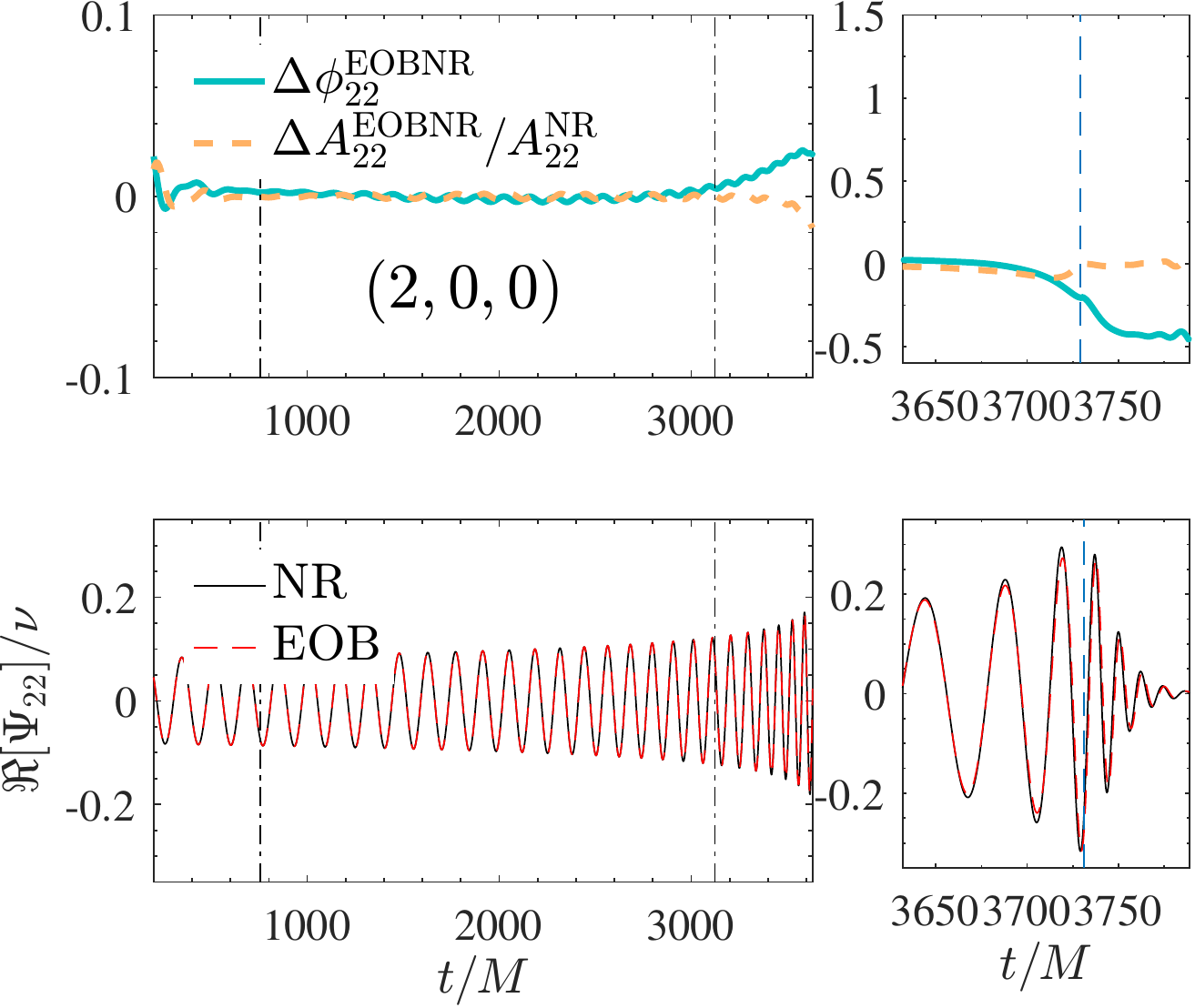}
\includegraphics[width=0.45\textwidth]{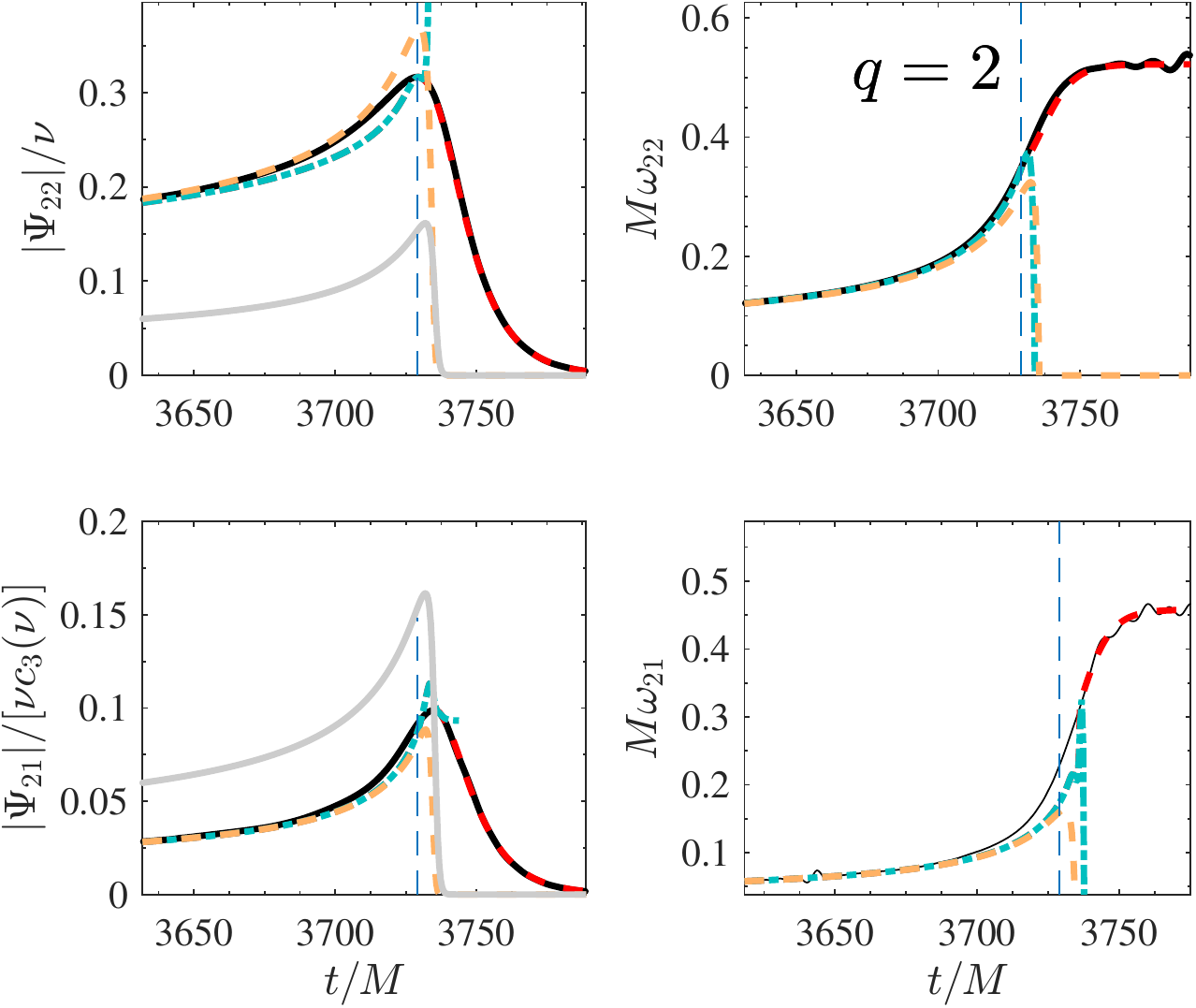}	
	\caption{Same as Fig~\ref{l2_teob} where the Pad\'e resummed potential was replaced
	by the log-resummed potential with $K$ given by Eq.~\eqref{eq:newK}. Though the agreement
	between the $(2,2)$ mode is rather acceptable. the NQC correction to the EOB instantaneous
	frequency is unable to provide a smooth matching with the postpeak part. This feature is related
	to the drop of the orbital frequency (gray line), that is much faster than the \TEOBResumS{} one
        in Fig.~\ref{l2_teob}; such behavior stems from the shape of the log-resummed $A$ potential. 
	See text for additional discussion.}
	\label{l2_seob}
	\center
	\end{figure*}
\begin{figure*}[t]
\center
\includegraphics[width=0.45\textwidth]{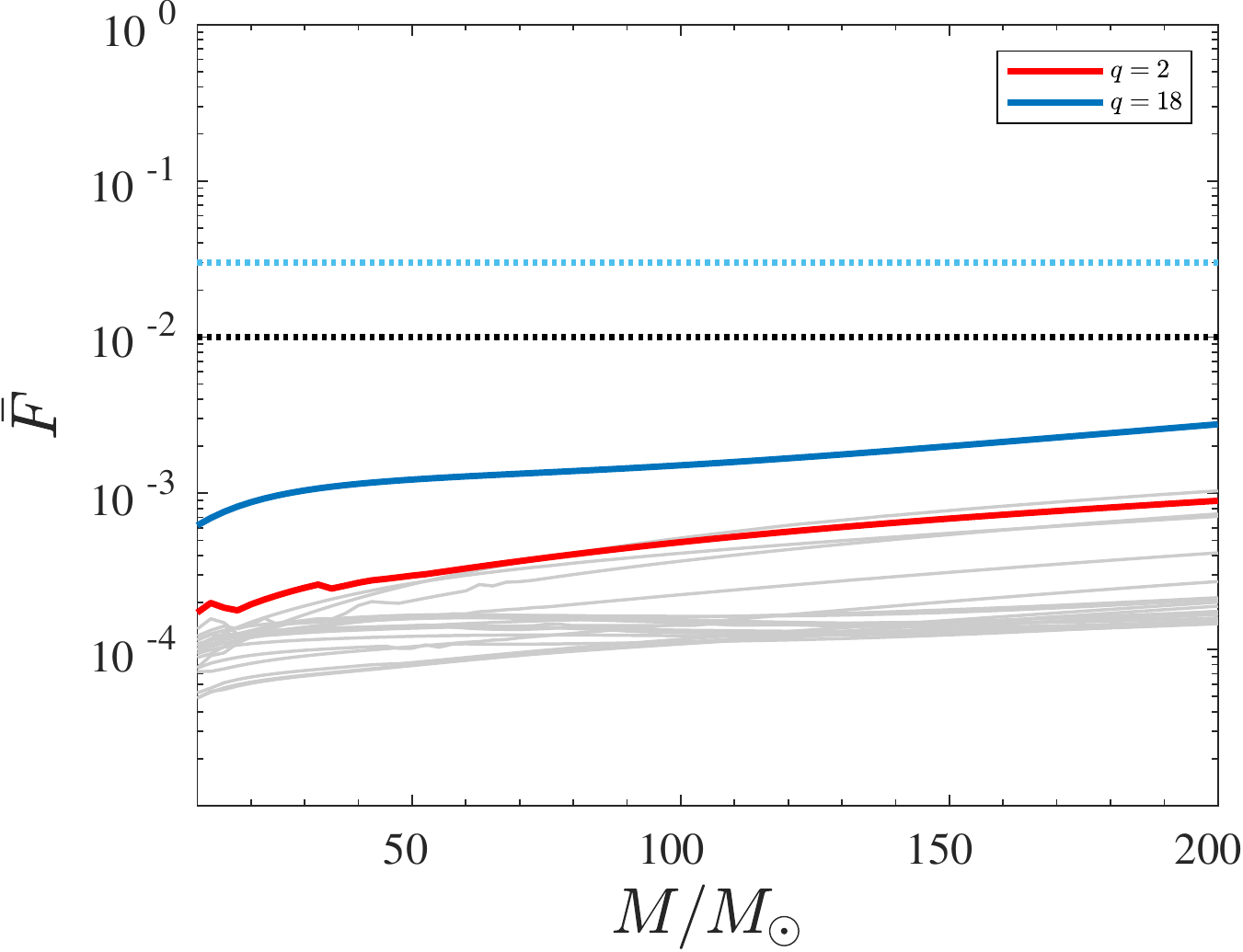}
\includegraphics[width=0.45\textwidth]{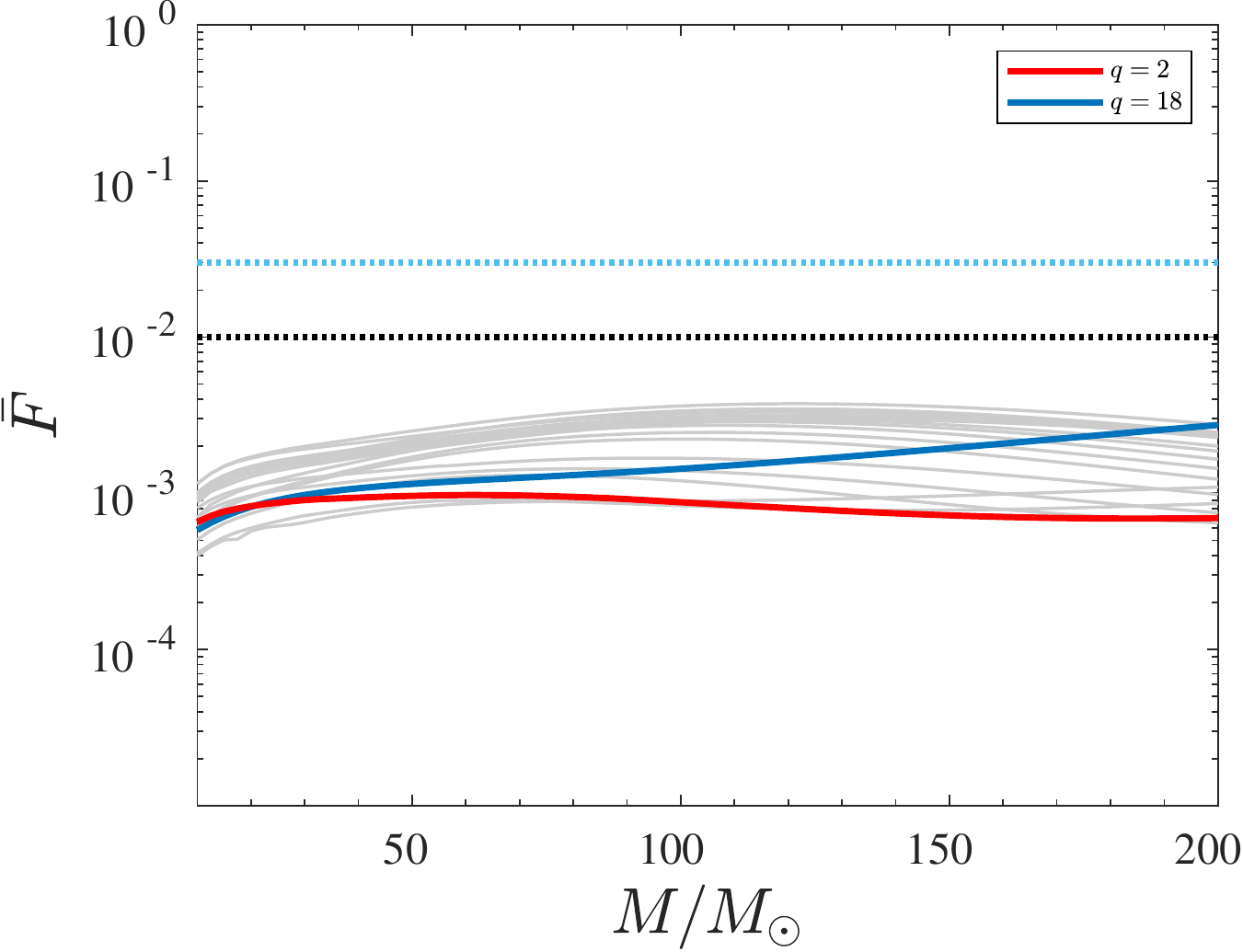}	
	\caption{EOB/NR unfaithfulness $\bar{F}$ computation for the $(2,2)$ mode with the
	Pad\'e resummed potential (left panel) and the log-resummed potential (right-panel).
	We are collecting here all SXS NR simulations considered in Ref.~\cite{Nagar:2019wds},
	that span the mass ratio range $1\leq q \leq 10$, plus a BAM dataset with $q=18$.
	Although $\max(\bar{F})$ is well below the $1\%$ threshold, the behavior of the function 
	$\bar{F}$ between the two potentials is actually very different. The red lines highlight
	the $q=2$ data of Figs.~\ref{l2_teob}-\ref{l2_seob}. See text for discussion.}
	\label{fig:Fbar}
\center
\end{figure*}
We have seen above that, due to the several structural differences between
the two models, it is difficult to understand clearly what special physical
element is responsible for some specific dynamical behavior. Generally
speaking one sees that the two models implement a fundamentally different 
description of the spin-spin interaction and this eventually reflects on
all diagnostics that we have analyzed.

To shed more light on the impact of the various analytical
structure, we focus here on a specific analytical element, the $A$
potential, and explore the consequences of its resummation. We do so
considering only the {\it nonspinning} sector of \TEOBResumS{} 
and \SEOBNRvq{} by performing the following exercise: we replace 
the Pad\'e resummed $A$ potential of \TEOBResumS{} by 
the log-resummed one of \SEOBNRvq{}, with some undetermined value of the 
parameter $K$. We ask then the following question: is it possible to tune the value of $K$ within such potential to obtain a new, NR-faithful, 
nonspinning EOB model {\it without} changing anything else? 
As a reference waveform model with a Pad\'e resummed potential we
use here the improved nonspinning version of \TEOBResumS{},
called of {\tt TEOBiResumMultipoles}, introduced in Ref.~\cite{Nagar:2019wds}.
Analytically, the model features an improved description of the multipolar
waveform amplitude (and thus radiation reaction) that in general incorporate
up to 6PN test-mass information in Pad\'e resummed form. The only
change we adopt here with respect to that model is that the residual waveform amplitude
$\rho_{22}^{\rm orb}$ is kept in its Taylor-expanded form at $3^{+2}$PN accuracy.
In addition, the model is based on an improved determination of $a_6^c$,
partly due to the different analytical framework and partly due to the
comparison with NR simulations with smaller numerical uncertainties.
Moreover, {\tt TEOBiResumMultipoles} also incorporates all modes up
to $\ell=m=5$ (included) completed through merger and ringdown.
To determine the new values of $K$ we follow the procedure discussed
in Ref.~\cite{Nagar:2019wds}: we align the EOB and NR waveforms in the
early inspiral and then determine $K$ so that the EOB/NR phase difference
at merger is of the order of the numerical uncertainty. In doing so, we
also keep attention that the corresponding behavior of the EOB frequencies
during the plunge is consistent with the NR one. We do so on seven mass
ratios $q=\{1,2, 2.5, 3, 6,8,18\}$ and determine some good values of $K$.
These values are then fitted with the following functional form
 \be
\label{eq:newK}
K=k_0\dfrac{1+n_1\nu + n_2\nu^2}{1+d_1\nu + d_2\nu^2}
\ee
where the fitted parameters are
\begin{align}
k_0&=0.13707,\nonumber\\
n_1&=191.5614,\nonumber\\
n_2&=-221.8992,\\
d_1&=-12.2328,\nonumber\\
d_2&=187.6895.\nonumber
\end{align}
Just as a illustrative comparison, the left panel of Figs.~\ref{l2_teob} and \ref{l2_seob}
exhibit the EOB/NR $\ell=m=2$ phasing comparison obtained with {\tt TEOBiResumMultipoles}
and with its $K$-avatar. The right panels of the same figures also show the corresponding 
$\ell=m=2$ and $\ell=2$, $m=1$ amplitude and frequency. Following the approach of 
Ref.~\cite{Nagar:2019wds}, to which we refer the reader for additional details, we show
together the (i) bare EOB waveform (orange, dashed); (ii) the NQC completed 
one (blue, dash-dotted) and (iii) the complete waveform with the postmerger (postpeak)
part (red-dashed). The most interesting result concerns the $(2,1)$ mode.
The figure 
pinpoints the fact that, while for TEOB the NQC basis is efficient in correcting the $\omega_{21}$
frequency so to allow a smooth connection to the ringdown part, it is unable to do so
also when the log-resummed $A$ potential is used. The reason for this behavior can be
traced back on the structure of the NQC basis, that depends on the inverse of the orbital
frequency. For the Pad\'e-resummed $A$ function, $\Omega$ decreases in a relatively mild
way after its peak; on the contrary, for the log-resummed potential the decrease of
$\Omega$ is very sharp, until it crosses zero very close to the $(2,1)$ waveform peak.
At a practical level, the fact that $\Omega\simeq 0$ when the relative separation $r$
is small, though finite, implies that the frequency-related NQC functions
$n^{21}_3\equiv p_{r_*}/(r\Omega)$ and $n^{21}_4\equiv p_{r_*}/(r\Omega)\Omega^{2/3}$
(see~\cite{Nagar:2019wds}) become very large and prevent the related
NQC correction to the phase to act efficiently so to correctly modify
the circular EOB waveform.

To have an idea of the reliability of our new $K$-resummed EOB model, we computed
EOB/NR faithfulness curves versus total mass, and also compared it with the {\tt TEOBiResumSMultipoles},
see Fig.~\ref{fig:Fbar}. We follow precisely the procedure and notation of Ref.~\cite{Nagar:2019wds},
to which we refer the reader for details. It is interesting to note that $\bar{F}$ has worsened with respect
to the Pad\'e resummed case, although it is still well below the usually accepted threshold of $1\%$.
Also, one should note that the global shape of the curves is different, with the \TEOBResumS{} ones
being essentially monotonic. This illustrates the worsening of the model (or sometimes also
of the NR waveform) during the ringdown, to which one is sensitive for large values of the masses.
Such clear trend is not evident for the log-resummed potential (except for the $q=18$ case): the curve
have a maximum and it is not possible an immediate, intuitive, explanation of what is seen.

\section{Post adiabatic dynamics}
\label{pa}
\begin{figure*}[t]
	\center
	\includegraphics[width=0.33\textwidth]{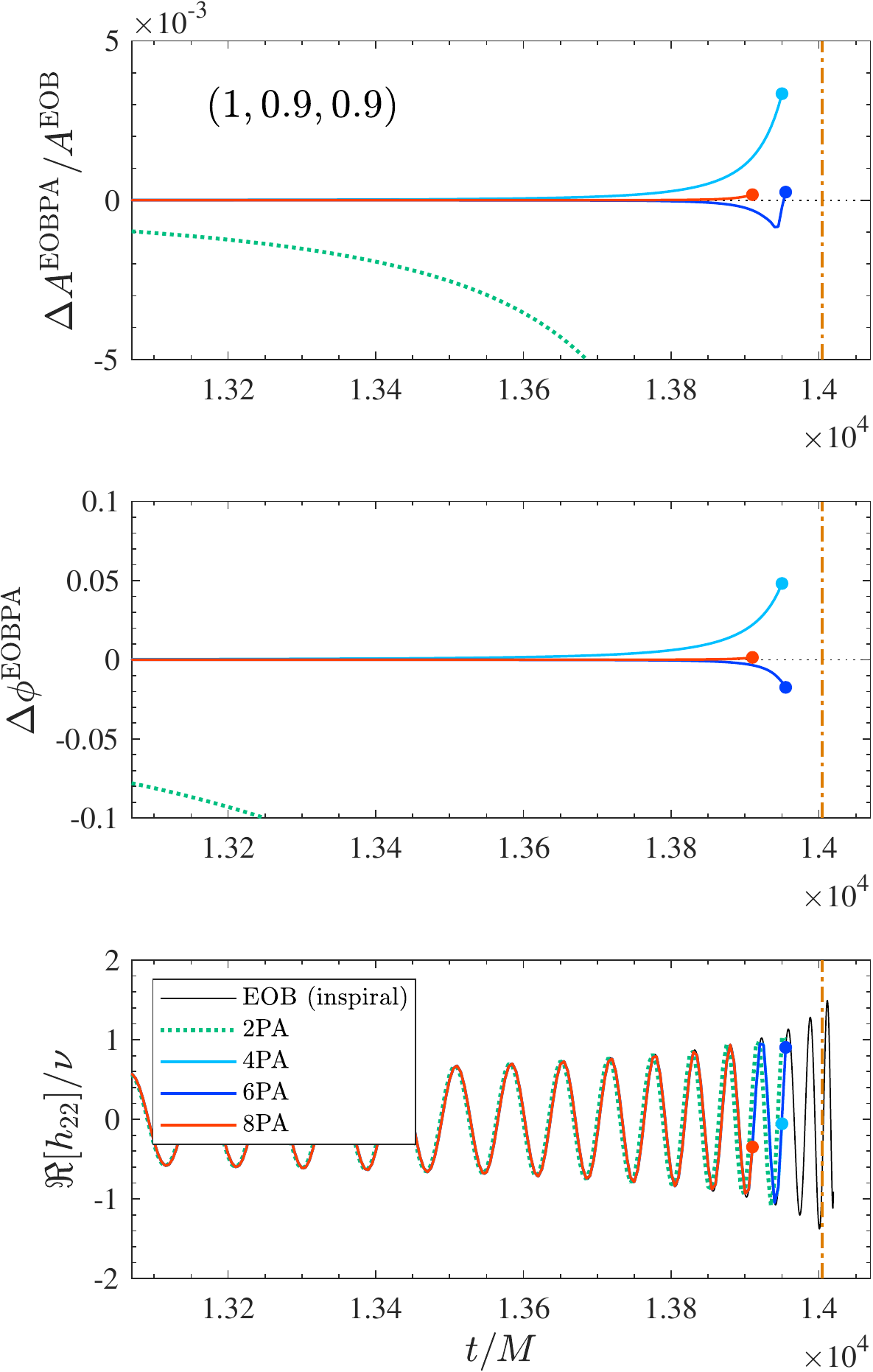}
	\includegraphics[width=0.32\textwidth]{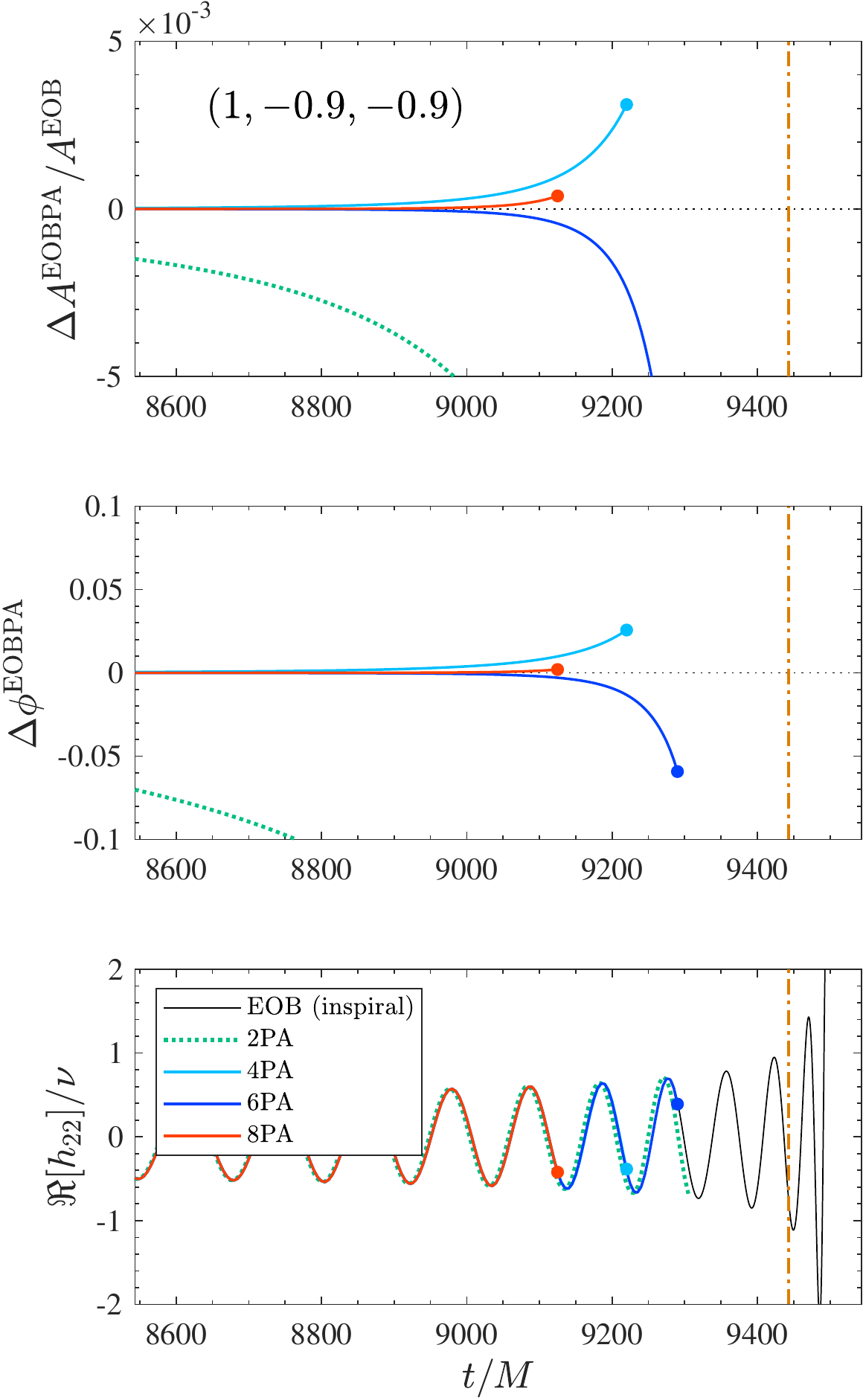}
	\includegraphics[width=0.33\textwidth]{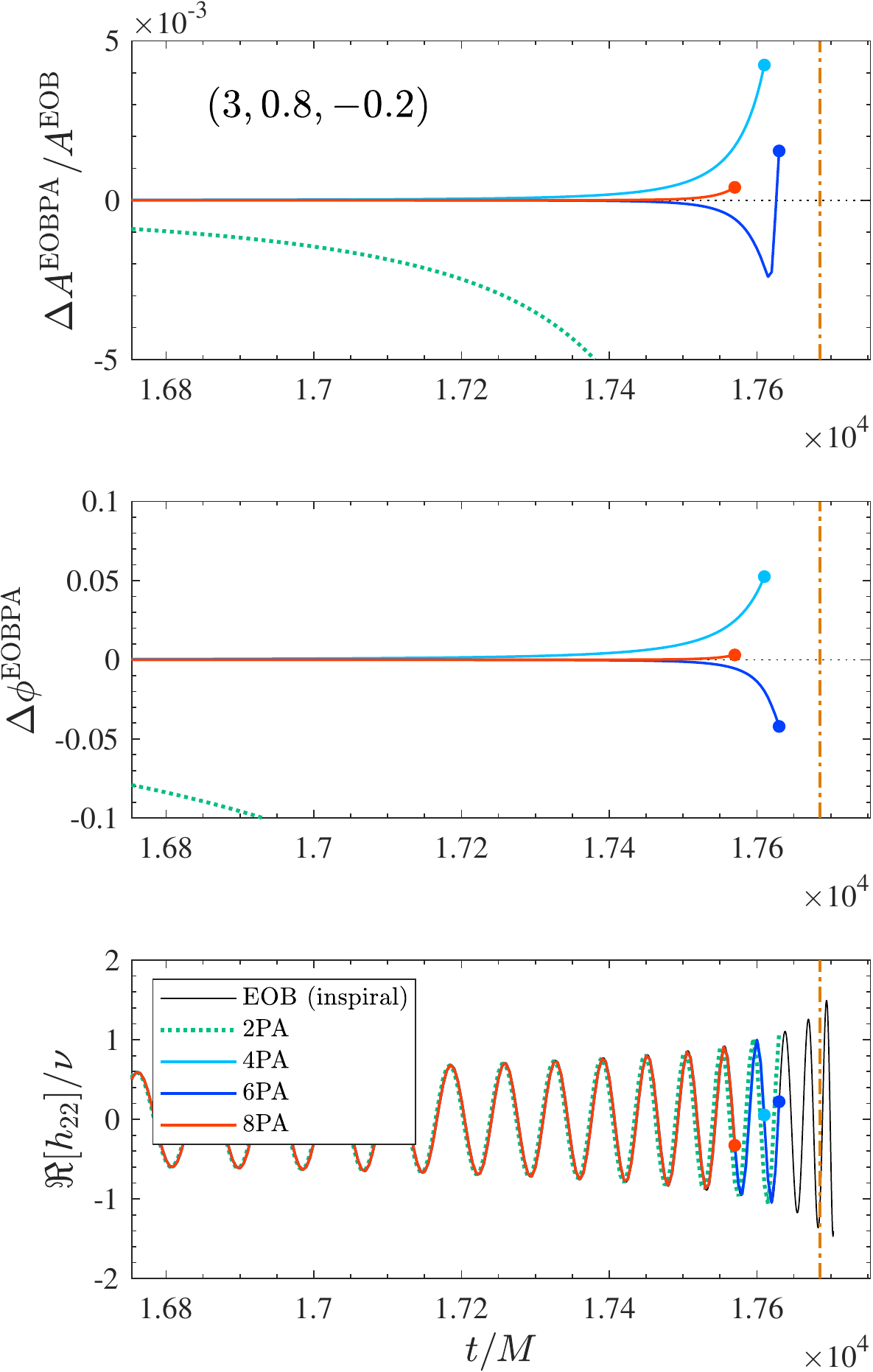}
	\caption{
		\label{fig:wave_bbh} Waveform comparison, $\ell=m=2$ strain mode:
		EOB$_{\rm PA}$ inspiral (colours) versus EOB inspiral obtained solving the ODEs (black).
		Note that the waveform is the purely analytical EOB one and no merger and ringdown
		modelization is included. The orange vertical line marks the LSO crossing
		location on the time axis. The filled markers highlight the end of the PA inspirals.}
\end{figure*}
\begin{figure}[t]
	\center
	\includegraphics[width=0.48\textwidth]{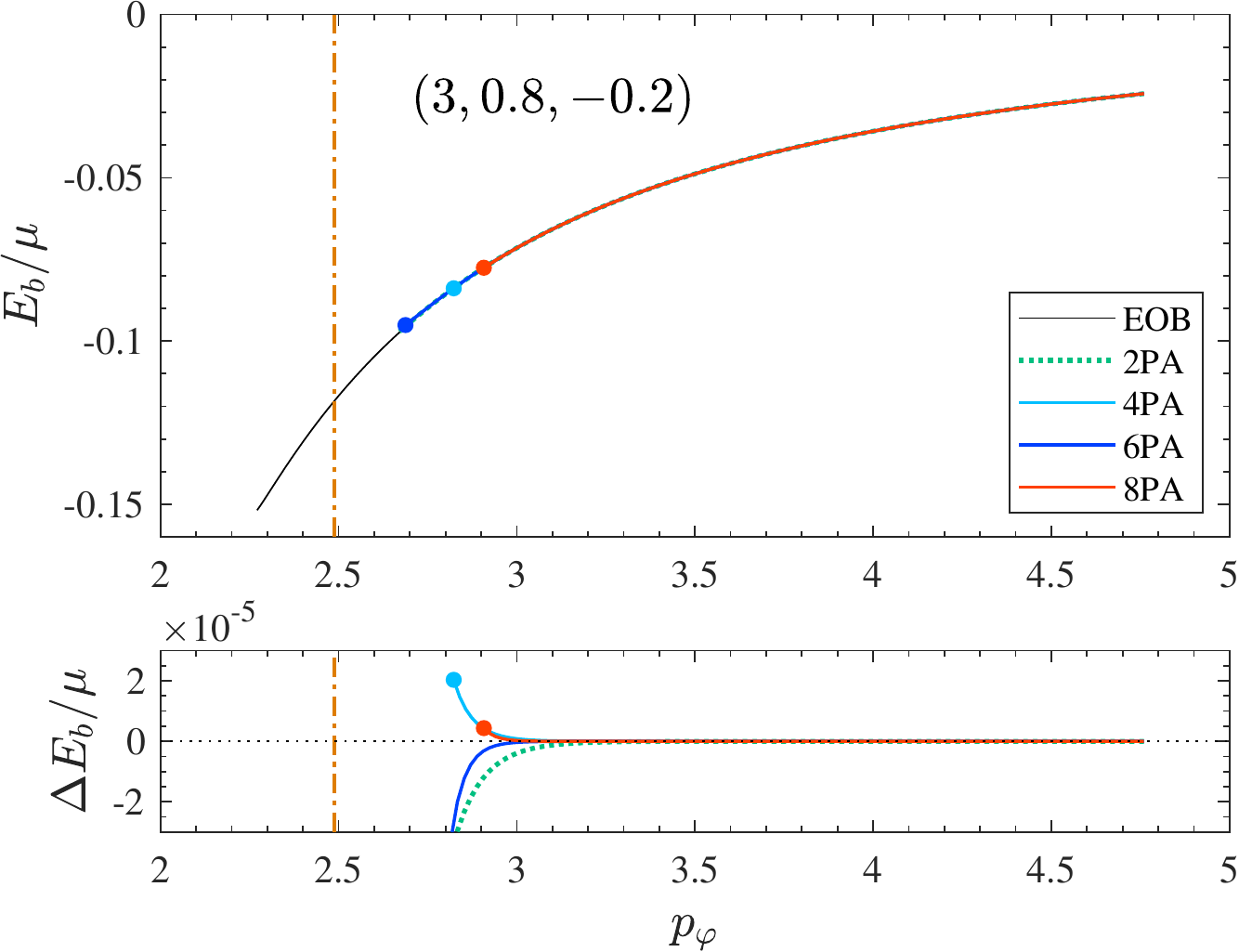}
	\caption{\label{fig:E_j} Illustrative comparison between energies
	  versus orbital angular momentum curves. The orange vertical line
          marks the EOB-LSO crossing.
	}
\end{figure}
Despite being semi-analytical, EOB models improved by NR information are, generally,
too slow to perform a parameter estimation run in a reasonable time. In order to speed
them up, reduced-order modeling~\cite{Field:2013cfa,Purrer:2014fza,Purrer:2015tud,Galley:2016mvy,Lackey:2016krb,Bohe:2016gbl}
versions of both \SEOBNRvqT{}~\cite{Bohe:2016gbl} 
and \TEOBResumS{}~\cite{Lackey:2016krb} were built.

An alternative route to improve the computational efficiency of long-inspiral
waveforms was proposed in Ref.~\cite{Nagar:2018gnk}, taking advantage of
the post-adiabatic (PA) approximation to obtain the inspiral dynamics.
The most expensive part of an EOB-waveform evaluation is in fact represented 
by the solution of the ODE system of the four Hamilton's equation, that are
solved with standard Runge-Kutta routines. 
The PA approximation can be used
to analytically (though approximately) solve two of the four Hamilton's
equations on a sparse radial grid and obtain the system momenta. 
Two quadratures allow then one to obtain the time, the orbital phase and thus the waveform.
The PA approximation is valid as long as the GW flux emitted by the binary is small.
When this is no longer the case (typically a few orbits before merger),
the PA approximation is used to generate the ODE initial conditions in
order to compute the dynamics through plunge and merger. 

Once we re-wrote the \SEOBNRvq{} Hamiltonian, we were able to repeat the procedure described 
in Ref.~\cite{Nagar:2018gnk} for \TEOBResumS{}. 
The model we tested is an hybrid, in that it uses the \SEOBNRvq{} conservative dynamics
and the \TEOBResumS{}-like radiation reaction and waveform described in the previous section.
This is clearly meant to be only a theoretical exercise to show the flexibility of 
the PA approach and it is not supposed to be, as is, faithful when compared with NR simulation data.
The construction of the PA dynamics for \SEOBNRvq{} follows closely Ref.~\cite{Nagar:2018gnk}.
We alternately solve the two equations
$dp_{r_*}/dt = (dp_{r_*}/dr)(dr/dt)$ and $dp_\varphi/dt = (dp_\varphi/dr)(dr/dt)$, in which we substitute Hamilton's equations, obtaining
\begin{align}
\label{eq:pphi}
&\left(\frac{\AA}{\bar{r}_c^2}\right)'p_\varphi^2 + 2\, \hat{\HH}_{\rm eff}^{\rm orb} \left(\frac{\p \tilde{\GG}}{\p r} + \frac{\p \tilde{\GG}}{\p \bar{p}_{r_*}} \frac{d \bar{p}_{r_*}}{d r}\right) p_{\varphi} + \nonumber \\
&~+ \AA' + z_3\left(\frac{\AA}{\bar{r}_c^2}\right)' \bar{p}_{r_*}^4 + 2\left(1+2 z_3 \frac{\AA}{\bar{r}_c^2} \bar{p}_{r_*}^2\right)\bar{p}_{r_*}\frac{d \bar{p}_{r_*}}{dr} + \notag \\
&~+ 2\, \hat{\HH}_{\rm eff}^{\rm orb} \left(\frac{\p \hat{H}^{\rm eff}_{\rm SS}}{\p r} + \frac{\p \hat{H}^{\rm eff}_{\rm SS}}{\p \bar{p}_{r_*}} \frac{d \bar{p}_{r_*}}{d r}\right) = 0, \\
\label{eq:prs}
&\bar{p}_{r_*} = \hat{\F}_\varphi\left(\frac{d p_\varphi}{d r}\right)^{-1} \left(\frac{\AA}{\BB}\right)^{-1/2} \nu \hat{H}_{\rm EOB} \hat{\HH}_{\rm eff}^{\rm orb} \ \times \nonumber \\
&~ \times \left\{1+2 z_3 \frac{\AA}{\bar{r}_c^2} \bar{p}_{r_*}^2 + 2\, \hat{\HH}_{\rm eff}^{\rm orb} \left[p_\varphi \frac{\p \tilde{\GG}}{\p \left(\bar{p}_{r_*}^2\right)}+ \frac{\p \hat{H}^{\rm eff}_{\rm SS}}{\p \left(\bar{p}_{r_*}^2\right)}\right] \right\}^{-1},
\end{align}
where we defined
\begin{equation}
\tilde{\GG} \equiv \bar{G}^0_S\hat{S}+ {\mathbb G}_{{\mathbb S}_{*}}\hat{{\mathbb S}}_{*} = \bar{G}_S\hat{S}+\bar{G}_{S_*}\hat{S}_*.
\end{equation}
Equations~\eqref{eq:pphi} and~\eqref{eq:prs} above translate Eqs.~(5) and~(6) of Ref.~\cite{Nagar:2018gnk}
into the \SEOBNRvq{} dictionary, adding a term depending on $\hat{H}^{\rm eff}_{\rm SS}$, that is not present
in \TEOBResumS{}. We denote as $n$PA order, the $n$-th iteration of this procedure,
i.e. the $n$-th correction to the adiabatic momenta $(p_\varphi, p_{r_*}) = (p_\varphi^{\rm circ}, 0)$.
The accuracy of the PA approximation versus the standard ODE approach is illustrated
in Figs.~\ref{fig:wave_bbh}-\ref{fig:E_j}. While Fig.~\ref{fig:wave_bbh} compares the
waveform phasing in different regions of the parameter space, Fig.~\ref{fig:E_j} shows
the gauge-invariant relation between binding energy and angular momentum. If, as expected,
the 2PA order is far from being consistent with the ODE solution, the successive iterations
rapidly converge to obtain a more than satisfying agreement. The 8PA order is consistent with
the ODE up to the last three orbits before merger, as found in Ref.~\cite{Nagar:2018gnk}. 
In the case of \TEOBResumS{}, this approach allowed for a drastic improvement of the waveform
generation time (see Refs.~\cite{Akcay:2018yyh,Nagar:2018plt}). We have checked that this is
the case also for the \SEOBNRvq{} Hamiltonian {\it already} within a standard {\tt Matlab}
implementation analogous to the one of Ref.~\cite{Nagar:2018gnk}, with comparable results.
Currently, efforts are in progress so to implement the PA dynamics within the complete
version of \SEOBNRvq{} used by the LIGO-Virgo collaboration. Detailed results and timing
comparisons will be presented in a forthcoming work.

\section{Conclusions} 
\label{sec:conclusions}

\begin{figure}[t]
\center
\includegraphics[width=0.48\textwidth]{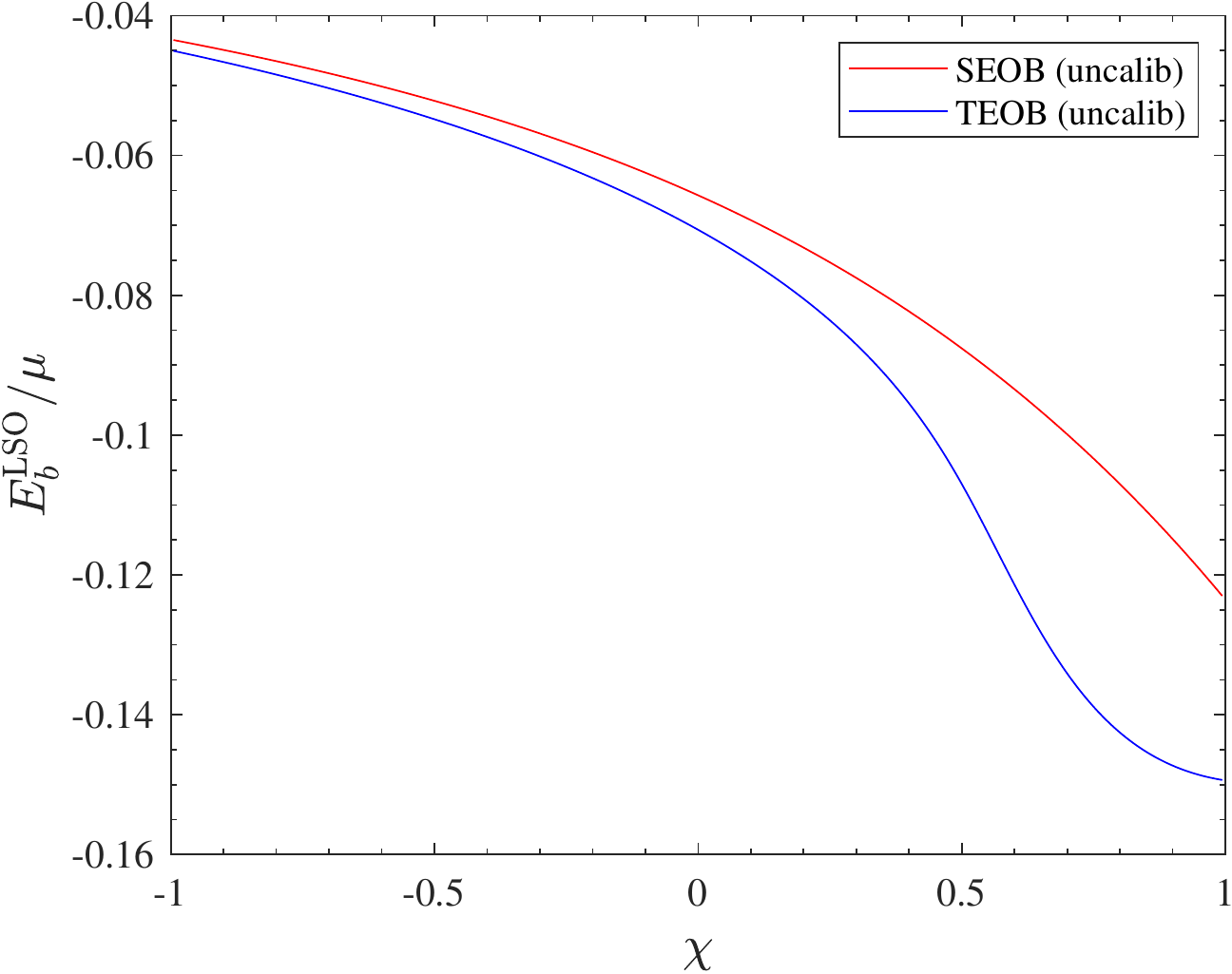}
\caption{\label{fig:EbLSO_part}
  Rescaled binding energy $\hat{E}_b \equiv E_b / \mu$ at the LSO for $q=1$.
  This is obtained with the non-calibrated Hamiltonians of \SEOBNRvq{} and \TEOBResumS{}, where the latter includes a new $G_{S_*}$
  with the complete spinning particle information.
  Note that this new flavor of \TEOBResumS{} presents an LSO for all values
  of the spin $\chi = \chi_1 = \chi_2$.
}
\end{figure}

In this paper we have performed the first comprehensive analytic comparison
between the Hamiltonians of the two state-of-the-art EOB waveform models
for coalescing BBHs, \TEOBResumS{} and \SEOBNRvq{}. In particular, we have
illustrated that the \SEOBNRvq{} Hamiltonian can be {\it formally} written
similarly to the \TEOBResumS{} one, though with different potentials.
Generally speaking, this allowed us to illustrate that the most important
structural differences between the two models lie in the way the $\nu$-deformation
is implemented in the spin sector.
More precisely:
\begin{itemize}
\item[(i)] 
{\it Centrifugal radius and spin-spin sector}. 
We have pointed out that in the orbital part of the \SEOBNRvq{} Hamiltonian it is 
possible to identify a {\it centrifugal radius} function $\bar{r}_c$, similarly to $r_c$ within \TEOBResumS{}. 
This function incorporates, in resummed form, some of the even-in-spin contribution, 
as in the case of a nonspinning particle on Kerr. 
However, $\bar{r}_c$ and $r_c$ are very different functions, notably because of
the choice of the effective spin quantity. In particular, in \TEOBResumS{} the use
of $\tilde{a}_0$ allows one to automatically incorporate within $r_c$
the LO quadratic-in-spin (as well as quartic-in-spin) terms. 
This is not the case for \SEOBNRvq{}, that uses $\hat{S}$, so that a
compensation term in the effective Hamiltonian, $\hat{H}_{SS}^{\rm eff}$
has to be introduced. Another important difference comes from the fact
that the resummation choices of \SEOBNRvq{} include in $\bar{r}_c$ the
$\nu$-dependent terms of $\Delta_u$ which are not present in \TEOBResumS{}.
\item[(ii)] 
{\it Spin-orbit sector}. 
We attempted to provide a one to one comparison between the spin-orbit sectors 
of the two models, rewriting the corresponding part of the \SEOBNRvq{} Hamiltonian like the \TEOBResumS{} one. 
We identified the two gyro-gravitomagnetic functions $(\bar{G}_S,\bar{G}_{S_*})$ in the former that correspond 
to $(G_S,G_{S_*})$ in the latter. 
These functions differ both in the gauge choice and in the analytical content.
We have explicitly showed that in \SEOBNRvq{}, the spin-orbit Hamiltonian can
be obtained starting from the expression of Ref.~\cite{Bini:2015xua} 
 	and $\nu$-{\it deforming} it in some way, replacing the Kerr functions $(r_c^K,A^K,B^K,Q^K)$ with
 	$(\bar{r}_c,\AA,\BB,\QQ)$, that incorporate additional $\nu$-dependent effects.
From this point of view, we want to stress that most of the spinning-particle information that is
encoded in $G_{S_*}^K$ is missing in \TEOBResumS{}, that is thus analytically {\it less complete} than \SEOBNRvq{}.
	
One should however be aware that nothing prevents us from the possibility of injecting the same information
in an alternative Hamiltonian that, however, maintains the same global structure as the current one. In particular,
the features that we want to preserve are: (i) the use of $r_c$ with $\tilde{a}_0$ for spin-spin interaction 
and (ii) the use of factorized (and then resummed) $(G_S,G_{S_*})$ function. In particular, one would like
to keep for $G_{S_*}$ a factorized expression of the form
\begin{equation}
G_{S_*}=G_{S_*}^0\hat{G}_{S_*},
\end{equation}
where now $G_{S_*}^0$ reduces to $G_{S_*}^K$, Eq.~\eqref{kerrGss}, when $\nu=0$ and {\it not} just to the first
term of the PN expansion, $3/2\, u^3$. To achieve this, one cannot work 
in the DJS gauge, but in a different gauge such that the standard PN-expanded $G_{S_*}$ coincides with the 
Taylor expansion of $G_{S_*}^K$ when $\nu=0$. One finds that this gauge is defined by the condition that all 
the $\nu$-dependent terms that depend on the radial momentum disappear. 
The corresponding choice of the gauge parameters is reported at the end of Appendix~\ref{app:gauge}.
A new spin-orbit sector that 
fully incorporates the spinning particle information can be obtained as follows: (i) one factorizes out
from $g^{\rm eff}_{S_*}$ the $r^3 G_{S*}^K$ terms up to NNLO; (ii) $G_{S_*}^0$ is taken to have the
same functional form of $G_{S_*}^K$ where, however, the various Kerr functions $(r_c^K,A^K,B^K,Q^K)$ 
are replaced by the EOB ones, $(r_c,A,B,Q)$, with their complete $\nu$-dependence. 
Similarly, the 
Kerr spin is replaced by the $\tilde{a}_0$ effective spin variable. The functions $A$ and $B$ are then 
resummed using the usual \TEOBResumS{} prescriptions; finally, the new functions 
$(\hat{G}_S,\hat{G}_{S_*})$, that explicitly depend on $\nu$, and are both in the form $1+\dots$,
are also resummed using their inverse Taylor representation, analogously to what is done in the DJS gauge.
We found that incorporating the ($\nu$-deformed) spinning-particle information within this new
flavor of \TEOBResumS{} fixes one of the long standing issues of the model in DJS gauge, i.e. the
fact that the LSO does not exist for large, positive spins $\geq 0.7$, as recalled in the text and
as pointed out in Ref.~\cite{Balmelli:2015zsa}.
For this study, we also kept $r_c$ at LO, i.e. setting $\delta a^2 = 0$ in Eq.\eqref{rcTEOB}, so to use the same amount of PN information as \SEOBNRvq{}.
Figure~\ref{fig:EbLSO_part} shows the binding energy at the LSO obtained with this new model: one sees that the LSO always exists also for quasi-extremal, positive spins.
We could also verify that, once implemented in the time-domain
code to provide the full transition from early inspiral to plunge, merger and ringdown, the Hamiltonian
in the new gauge maintains the same robustness and flexibility that was typical of the DJS gauge one.
We also found that, analogously to this case, an effective spin-orbit parameter is necessary to get
a good phasing agreement with NR simulations. A detailed investigation of these aspects is beyond 
the scope of this work and will be discussed elsewhere.
\end{itemize}
As last remarks, to make the comparison between the \SEOBNRvq{} and \TEOBResumS{} Hamiltonians
more quantitative, we performed the following tasks.
\begin{itemize}
\item[{\it (a)}] We compared the gauge-invariant relations between energy, angular momentum and orbital frequency,
both for adiabatic and nonadiabatic dynamics. In doing so, we compared and contrasted the linear-in-spin
and quadratic-in-spin contributions of the two Hamiltonians. We found relevant qualitative and quantitative
differences, especially in the spin-spin sector.
\item[{\it (b)}] We proved the robustness of the \TEOBResumS{} analytical framework by replacing the Pad\'e resummed
$A$-potential by the log-resummed $A$-potential used in \SEOBNRvq{}. 
By a new NR-calibration of the $K$-function
we constructed a new, {\it nonspinning} EOB model that is faithful ($\sim 10^{-3}$ level) with state-of-the-art
NR simulations of the SXS catalog~\cite{SXS:catalog} and obtained with the BAM code, up to mass ratio $q=18$.
\item[{\it (c)}] Finally, we have illustrated that the post-adiabatic approximation introduced in Ref.~\cite{Nagar:2018gnk}
to efficiently generate long-inspiral waveforms can be implemented also using a \SEOBNRvq{}-like Hamiltonian.
Our result can thus be useful to improve the performance of the actual \SEOBNRvq{} in its {\tt LALInference}
implementation, possibly avoiding the current need of building surrogate waveform models.
\end{itemize}

\acknowledgments
D.~B., P.~R. and G.~R. thank IHES for hospitality during the development of this work. We are grateful to R.~Gamba for cross checking part of the analytical calculations. We thank A.~Buonanno for comments on the manuscript.

\newpage
\appendix

\section{Re-writing of the \SEOBNRvq{} Hamiltonian for generic spin orientations}
\label{App:seob_general}

In the main text we re-wrote the \SEOBNRvq{} Hamiltonian using the \TEOBResumS{}
formalism, once the former was restricted to equatorial orbits.
This Appendix completes the discussion taking into account generic spin orientations.

\subsection{Explicit calculation of $\hat{H}_{\rm NS}$}
In the generic case (with any value of $\theta$), the metric components of Eqs.~\eqref{eq:metric1}--\eqref{eq:metric2} read 
\begin{align}
g^{tt} &= -\dfrac{\Lambda_t}{\Delta_t \Sigma}, \\
g^{rr} &= \dfrac{\Delta_r}{\Sigma}, \\
g^{\theta \theta} &= \dfrac{1}{\Sigma}, \\
g^{\varphi \varphi}& = \dfrac{1}{\Lambda_t} \bigg( -\dfrac{\widetilde{\omega}^2_{fd}}{\Delta_t \Sigma} + \dfrac{\Sigma}{\sin^2 \theta} \bigg), \\
g^{t \varphi} &= -\dfrac{\widetilde{\omega}_{fd}}{\Delta_t \Sigma},
\end{align}
where 
\begin{align}
\label{eqapp:deltat}
\Delta_t &= r^2 \Delta_u, \\
\Delta_r &= \Delta_t D^{-1}, \\
\Lambda_t &= \left(r^2+\hat{S}^2 \right)^2-\hat{S}^2 \Delta_t \sin^2 \theta, \\
\Sigma &= r^2 + \hat{S}^2 \cos^2 \theta, \\
\widetilde{\omega}_{fd} &= 2 \, \hat{S} \, r,
\label{eqapp:sigma}
\end{align}
in which we already used the gauge freedom and imposed $\omega^0_{fd} = \omega^1_{fd} = 0$.

The ``non-spinning'' Hamiltonian $\hat{H}_{\rm NS}$ can then be computed using Eqs.~\eqref{eq:Hns1} and substituting the newly defined functions.

At the same time, our rewriting of $\hat{H}_{\rm NS}$, Eq.~\eqref{eq:Hns}, still holds, when taking into account that the \SEOBNRvq{} metric potentials and centrifugal radius are again defined as 
\begin{align}
\bar{r}_c^2 &\equiv \dfrac{\Lambda_t}{\Sigma}, \\
\AA &\equiv \dfrac{\Delta_t \Sigma}{\Lambda_t},\\
\BB &\equiv \dfrac{\Sigma}{\Delta_r},\\
\QQ &\equiv 1 + p_\varphi^2 \bar{u}_c^2 + \dfrac{p_{r^{*}}^2}{\AA}, \\
\bar{G}^0_S &\equiv \dfrac{\tilde{\omega}_{fd}}{\Lambda_t \hat{S}}.
\end{align} 

However, their explicit form is different and can be calculated using Eqs.~\eqref{eqapp:deltat}--\eqref{eqapp:sigma}. For example, the relation between the A and D function reads
\begin{equation}
\DD = \frac{\bar{r}_c^2}{\Sigma}\, \mathbb{A}\, \mathbb{B}.
\end{equation}

It is easily checked that we recover the equatorial orbits case for $\theta = \pi/2$.

\subsection{Explicit calculation of $\hat{H}_{\rm SO}$}
In order to write the generic form of $\hat{H}_{\rm SO}$, we first need to define some convention.
We indicate with $\bm{r}$ and $\bm{p}$ the dimensionless position and momentum vectors respectively. 
The spin vectors corresponding to the variables of Eqs.~\eqref{eq:S_BB}--\eqref{eq:Sstar_BB} are denoted by $\vec{\mathbb S}$ and $\vec{\mathbb S}_*$.
Using this notation, the general form of $\hat{H}_{\rm SO}$, given by Eq.~(4.18) of Ref.~\cite{Barausse:2009xi}, reads
\begin{widetext}
\begin{align}
\label{eqapp:HSO}
\hat{H}_{\rm SO} =&~ \pphih \Skerr \dfrac{e^{2 \tilde{\nu} - \mut} \left(e^{\mut + \tilde{\nu}}-\widetilde{B} \right)}{\widetilde{B}^2 \qrq\, \xi^2} + \dfrac{e^{\tilde{\nu}-2\mut}}{\widetilde{B}^2 \left(1+\qrq \right) \qrq\, \xi^2} \biggl\{ \widetilde{B}^2 \widetilde{J} \bigg[ \left(\tilde{\mu}\,'-\frac{1}{\widetilde{J}}\right) \pthh \left( 1+\qrq \right) + \notag \\
&- \de_{\cos \theta}\left(\tilde{\mu}\right) \prh \xi^2 - \qrq \left( \tilde{\nu}\,' \pthh + \de_{\cos \theta}\left(\tilde{\mu}-\tilde{\nu}\right) \prh \xi^2 \right) \bigg]\Sxi + \notag \\
&+ e^{\mut + \tilde{\nu}} \pphih \left( 2 \qrq + 1 \right) \bigg[ \widetilde{J}\, \tilde{\nu}\,' \Sv - \de_{\cos \theta}\left(\tilde{\nu}\right) \xi^2 \Sn \bigg] \widetilde{B} + \notag \\
&- \widetilde{J} \left(\widetilde{B}\,' - \frac{\widetilde{B}}{\widetilde{J}}\right) e^{\mut + \tilde{\nu}} \pphih (1+\qrq) \Sv \biggl\},
\end{align}
\end{widetext}
where the used unit vectors are $\bm{n} = \bm{r}/r$, $\bm{s} = \vec{\mathbb S} / |\vec{\mathbb S}|$, $\bm{\xi} = \bm{s} \times \bm{n}$ and $\bm{v} = \bm{n} \times \bm{\xi}$.

The functions that enter the ``spin-orbit'' Hamiltonian still formally read
\begin{align}
e^{2\tilde{\mu}} =&~ \Sigma, \hspace{1.4cm} e^{2\tilde{\nu}} = \frac{\Delta_t \Sigma}{\Lambda_t}, \\
\widetilde{B} =&~ \sqrt{\Delta_t}, \hspace{1.15cm} \widetilde{J} = \sqrt{\Delta_r},
\end{align}
but are related to the \TEOBResumS{}-like functions by
\begin{align}
e^{2\tilde{\mu}} =&~ \Sigma, \hspace{1.4cm} e^{2\tilde{\nu}} = \AA, \\
\widetilde{B} =&~ \sqrt{\AA}\, \bar{r}_c, \hspace{1cm} \widetilde{J} = \frac{\sqrt{\Sigma}}{\sqrt{\BB}}.
\end{align}

Substituting these formulas into Eq.~\eqref{eqapp:HSO}, we get
\begin{widetext}
\begin{align}
\hat{H}_{\rm SO} =&~ \frac{\sqrt{\AA}}{\qrq\, \xi^2} \frac{\sqrt{\Sigma} - \bar{r}_c}{\bar{r}_c^2 \sqrt{\Sigma}} \pphih \Skerr +\frac{\sqrt{\AA}}{\sqrt{\BB}\, \bar{r}_c\, \xi^2}\left[\frac{\AA'}{2\left(1+\qrq\right) \AA} - \frac{\left(\bar{r}_c\right)'}{\qrq\, \bar{r}_c} + \frac{\sqrt{\BB}}{\qrq \sqrt{\Sigma}}\right] \pphih \Sv + \notag \\
&+ \frac{1}{2\sqrt{\AA}\sqrt{\BB}\sqrt{\Sigma}} 
\Bigg\{\left[ \frac{\de_{\cos \theta}\AA}{1+\qrq}
- \frac{\AA}{\qrq} \frac{\de_{\cos \theta}\Sigma}{\Sigma} \right]\prh - \frac{1}{\xi^2}\left[\frac{\AA'}{\left(1+\qrq\right)} - \frac{\AA}{\qrq} \frac{\left(\Sigma\right)'-2\sqrt{\BB}\sqrt{\Sigma}}{\Sigma}\right] \pthh \Bigg\} \Sxi + \notag \\
&- \frac{1+2\qrq}{2 \qrq \left(1+\qrq\right)} \frac{\de_{\cos \theta}\AA}{\sqrt{\Sigma}\sqrt{\AA}\, \bar{r}_c} \pphih \Sn.
\end{align}
\end{widetext}

In the spin-aligned case, we find $\Sxi = \Sn = 0$ and $\Sv = \Skerr = \hat{\mathbb{S}}_*$. 
We also define the radial and angular momentum as $\prh = p_r$ and $\pphih = p_\varphi$ respectively, while $\pthh = 0$.
Finally, we fix $\theta = \pi/2$, such that $\xi^2 = 1$.

\subsection{Explicit calculation of $\hat{H}_{\rm SS}$} 
Let us turn now to the function $\hat{H}_{\rm SS}^{\rm eff}$.
Following Eq.~(4.19) of Ref.~\cite{Barausse:2009xi} we write
\begin{widetext}
	\begin{align}
	\label{eqapp:HSS}
	\hat{H}_{\rm SS} =&~ \omega \Skerr + \dfrac{e^{-3 \mut-\tilde{\nu}} \widetilde{J}}{2 \widetilde{B} \xi^2}
	\dfrac{\omega\,'}{\qrq \left(1+\qrq\right)} \biggl\{ -e^{\tilde{\nu} + \mut} \pthh \pphih \Sxi \widetilde{B} + e^{2 \left(\tilde{\nu} + \mut\right)} \pphih^2 \Sv + \notag \\
	&+e^{2 \mut} \left( 1 + \qrq \right) \qrq \Sv \xi^2 \widetilde{B}^2 + \widetilde{J} \prh \left[ \pthh \Sn - \widetilde{J} \prh \Sv \right] \xi^2 \widetilde{B}^2 \biggl\} + \notag \\
	&+ \dfrac{e^{-3 \mut-\tilde{\nu}} \de_{\cos{\theta}}\left(\omega\right) }{2 \widetilde{B} \qrq (1+\qrq)} \biggl\{ -e^{2\left(\mut + \tilde{\nu}\right)} \pphih^2 \Sn + e^{\mut + \tilde{\nu}} \tilde{J} \widetilde{B} \prh \pphih \Sxi + \notag \\
	&+ \widetilde{B}^2 \left[\Sn \pthh^2
	- \widetilde{J} \prh \pthh \Sv - e^{2 \mut} \left( 1+\qrq \right) \qrq\, \xi^2 \Sn\right] \biggl\},
	\end{align}
\end{widetext}
where again
\begin{equation}
\omega \equiv \dfrac{\tilde{\omega}_{fd}}{\Lambda_t} = \bar{G}^0_S\, \hat{S}.
\end{equation}

Eq.~\eqref{eqapp:HSS} then becomes
\begin{widetext}
\begin{align}
\hat{H}_{\rm SS} =&~ \bar{G}_S^0\, \hat{S}\, \Skerr + \dfrac{\bar{r}_c }{2\, \xi^2 \sqrt{\BB}} \left(\bar{G}_S^0\right)' \hat{S}\, \Bigg\{\xi^2\Sv + \frac{1}{\qrq\left(1+\qrq\right)} \times \notag \\
&\times \left[\left(\frac{\pphih^2}{\bar{r}_c^2 } - \frac{\xi^2 \prh^2}{\BB}\right) \Sv - \frac{\pphih \pthh}{\sqrt{\Sigma} \, \bar{r}_c} \Sxi + \frac{\xi^2\prh \pthh}{\sqrt{\BB}\sqrt{\Sigma}} \Sn \right] \Bigg\} + \notag \\
&- \dfrac{\bar{r}_c}{2 \sqrt{\Sigma}}\, \de_{\cos \theta}\left(\bar{G}_S^0\right) \hat{S}\, \Bigg\{\xi^2\Sn + \frac{1}{\qrq\left(1+\qrq\right)} \times \notag \\
&\times \left[\left(\frac{\pphih^2}{\bar{r}_c^2} - \frac{\pthh^2}{\Sigma}\right) \Sn - \frac{\pphih \prh}{\sqrt{\BB} \, \bar{r}_c} \Sxi + \frac{\prh \pthh}{\sqrt{\BB}\sqrt{\Sigma}} \Sv \right] \Bigg\}
.
\end{align}
\end{widetext}

Finally, the total $\hat{H}_{\rm SS}^{\rm eff}$, without NR calibration, is defined as
\begin{equation}
\hat{H}_{\rm SS}^{\rm eff} = \hat{H}_{\rm SS} - \dfrac{1}{2} u^3 \left( \delta_{i j} - n_{i} n_{j} \right) \hat{{\mathbb S}}_{*}^i \hat{{\mathbb S}}_{*}^j.
\end{equation} 

To go back to the spin-aligned case, we impose the conditions described in the previous section.

\section{PN coefficients in $G_S$ and $G_{S_*}$}
\label{app:GScoeff}
We write here explicitly the coefficients that enter Eqs.~\eqref{eq:GS_T} and \eqref{eq:GSs_T}. They are
\begin{align}
\label{gs-coef}
c_{10} &= \frac{5}{16} \nu, \hspace{2.4cm} c^*_{10} = \frac{3}{4} + \frac{1}{2}\nu, \nonumber \\
c_{20} &= \frac{51}{8}\nu + \frac{41}{256}\nu^2, \hspace{0.95cm} c^*_{20} = \frac{27}{16} + \frac{29}{4} \nu + \frac{3}{8} \nu^2, \nonumber \\
c_{30} &= \nu c_3, \hspace{2.5cm} c^*_{30} = \frac{135}{32} + \nu c_{3}, \nonumber \\
c_{12} &= 12 \nu - \frac{49}{128}\nu^2, \hspace{1.05cm} c^*_{12} = 4 + 11 \nu - \frac78 \nu^2, \nonumber \\
c_{02} &= \frac{27}{16} \nu,\hspace{2.4cm} c^*_{02} = \frac{5}{84} + \frac{3}{2}\nu, \nonumber \\
c_{04} &= -\frac{5}{16} \nu + \frac{169}{256} \nu^2, \hspace{0.7cm} c^*_{04} = \frac{5}{48} + \frac{25}{12}\nu + \frac{3}{8}\nu^2,\nonumber \\
&\hspace{3.65cm} c^*_{40} = \frac{2835}{256}, 
\end{align}

\section{Coefficients of the $\Delta_u$ function}
\label{app:Alog}
We list here the explicit expression of the $(\Delta_0,\Delta_i)$
coefficients used in the log-resummation of the potential in \SEOBNRvq{}
in Eq.~\eqref{Du_resum}. They read
\begin{widetext}
	\begin{align}
	\Delta_0=&~K (K \nu -2), \\
	\Delta_1 =&-2 (\Delta_0+K) (K \nu -1),\\
	\Delta_2 =&~ \frac{1}{2} \left(\Delta_1 (\Delta_1-4 K
	\nu +4)-2 a^2\Delta_0 (K \nu -1)^2\right),\\
	\Delta_3=&-a^2\Delta_1 (K \nu -1)^2-\frac{\Delta_1^3}{3}+\Delta_1^2 (K \nu -1)+\Delta_1\Delta_2-2 (K \nu -1) (\Delta_2-K
	\nu +1),\\
	\Delta_4 =&~\frac{1}{96} \bigg[8 \left(6 a^2 \left(\Delta_1^2-2 \Delta_2\right) (K \nu -1)^2+3 \Delta_1^4+\Delta_1^3 (8-8 K \nu )-12 \Delta_1^2\Delta_2+12\Delta_1 (2
	\Delta_2 K \nu -2 \Delta_2+\Delta_3)\right)\nonumber\\
	&+48\Delta_2^2-64 (K \nu -1) (3 \Delta_3-47 K \nu +47)-123 \pi ^2 (K\nu -1)^2\bigg],\\
	\Delta_5 =&~ \frac{(K \nu -1)^2}{\nu } \bigg[ \frac{64}{5} \nu \log (u)
	+ \nu \left(-\frac{1}{3} a^2
	\left(\Delta_1^3-3\Delta_1
	\Delta_2+3 \Delta _3\right)+\frac{\Delta_1^4-4 \Delta _1^2\Delta_2+4\Delta_1\Delta_3+2\Delta_2^2-4 \Delta_4}{2 K \nu -2}\right. \nonumber \\
	&\left.-\frac{\Delta_1^5-5\Delta_1^3
		\Delta_2+5\Delta_1^2 \Delta_3+5\Delta_1\Delta_2^2-5
		\Delta_2 \Delta_3-5\Delta_4\Delta_1}{5 (K \nu -1)^2}+\frac{2275 \pi^2}{512}+\frac{128 \gamma }{5}-\frac{4237}{60}+\frac{256
		\log (2)}{5}\right) + \nonumber \\
	& +\left(\frac{41}{32}\pi^2 - \frac{221}{6}\right)\nu^2\bigg].
	\end{align}
\end{widetext}

\section{Re-writing of the spin mapping of $\hat{{\mathbb S}}_*$}
\label{app:delta_*}

In this Appendix we show the explicit form of the coefficients that enter the spin mapping of Eq.~\eqref{eq:Sstar_BB}.

Comparing Eqs.~\eqref{eq:delta_s1} and \eqref{eq:delta_s2} with Eqs.~(51) and (52) of Ref.~\cite{Barausse:2011ys}, we get 
\begin{align}
c_u =&~ \frac{1}{6}\left(-4 b_0 + 7 \nu \right) \hat{S} - \frac{2}{3}\left(a_0 + \nu \right) \hat{S}_* , \\
c_Q =&~ \frac{1}{3}\left(2 b_0 + \nu\right) + \frac{1}{12}\left(8 a_0 + 3 \nu \right) \hat{S}_*, \\
c_{pr^2} =& -\frac{1}{2}\left(4 b_0 + 5\nu\right) - \left(2 a_0 + 3 \nu \right) \hat{S}_*,
\end{align}
and 
\begin{align}
c_{u^2} =&~ \frac{1}{36}\left(-56 b_0 -24 b_2 + 353\nu - 60 b_0 \nu - 27 \nu^2\right) \hat{S} + \notag \\
&+\frac{1}{9}\left(-14 a_0 -6 a_2 - 56\nu - 15 a_0 \nu - 21 \nu^2\right) \hat{S}_* , \\
c_{\QQ^2} =&~ \frac{1}{72}\left(-4 b_0 +48 b_1 - 23\nu - 12 b_0 \nu - 3 \nu^2\right) \hat{S} + \notag \\
&+\frac{1}{144}\left(-8 a_0 +96 a_1 - 45\nu - 24 a_0 \nu\right) \hat{S}_* , \\
c_{u \QQ} =&~ \frac{1}{36}\big(-14 b_0 -24 b_1 +24 b_2 -103\nu + \notag \\
&+ 66 b_0 \nu + 60 \nu^2\big) \hat{S} +\frac{1}{36}\big(-14 a_0 -24 a_1 + \notag \\
&+ 24 a_2 - 109\nu + 66 a_0 \nu + 51 \nu^2\big) \hat{S}_* , \\
c_{p_r^4} =&~ \frac{5}{3}\left(-2 b_3 + 3 b_0 \nu + 3 \nu^2\right) \hat{S} + \notag \\
&+\frac{5}{24}\left(-16 a_3 + 24 a_0 \nu + 27 \nu^2\right) \hat{S}_* , \\
c_{u p_r^2} =&~ \frac{1}{12}\big(-24 b_0 -16 b_1 - 32b_2 -24 b_3 + 43\nu \notag \\
&- 24 b_0 \nu - 54 \nu^2\big) \hat{S} +\frac{1}{24}\big(-48 a_0 - 36a_1 + \notag \\
&-64 a_2 -48 a_3 - 16\nu - 48 a_0 \nu - 147 \nu^2\big) \hat{S}_* , \\
c_{p_r^2 \QQ} =&~ \frac{1}{12}\left(2 b_0 -24 b_1 + 24 b_3 + 16\nu - 30 b_0 \nu - 21 \nu^2\right) \hat{S} + \notag \\
&+\frac{1}{24}\left(4 a_0 -48 a_1 + 48 a_3 + 6\nu - 60 a_0 \nu - 39 \nu^2\right) \hat{S}_* .
\end{align}

If we now decide to extract the spin dependence, we can write 
\begin{align}
\Delta_{\sigma^{*}}^{(1)} = c^{(1)}_S \hat{S} + c^{(1)}_{S_*} \hat{S}_{*}, \\
\Delta_{\sigma^{*}}^{(2)} = c^{(2)}_S \hat{S} + c^{(2)}_{S_*} \hat{S}_{*},
\end{align}
where the coefficients $\left(c_S^{(i)},c_{S*}^{(i)}\right)$ now are independent of spins, but depend on dynamical variables. These, 
with their complete gauge flexibility, read
\begin{align}
c^{(1)}_{S} &= \dfrac{2}{3} u \left( a_0 + \nu \right)
+ \dfrac{1}{12} \left( 8 a_0 + 3 \nu \right) \left( \QQ - 1 \right) + \notag \\
&- (2 a_0 + 3 \nu) \dfrac{p_r^2}{\BB}, \\
c^{(1)}_{S_{*}} &= \dfrac{1}{6} u \big( -4 b_0 + 7 \nu \big) + \dfrac{1}{3} \big( \QQ - 1 \big) \big( 2 b_0 + \nu \big) + \notag \\
&-\dfrac{1}{2} \dfrac{p_r^2}{\BB} \big( 4 b_0 + 5 \nu \big),
\end{align}
and
\begin{align}	
c^{(2)}_{S} &= \dfrac{1}{9} u^2 \left( -14 a_0 - 6 a_2 - 56 \nu -15 a_0 \nu - 21 \nu^2 \right) + \notag \\
&+ \dfrac{5}{24} \dfrac{p_r^4}{\BB^2} \left( -16 a_3 + 24 a_0 \nu + 27 \nu^2 \right) + \notag \\
&+ \dfrac{1}{144} \left( \QQ - 1 \right)^2 \left( -8 a_0 + 96 a_1 - 45 \nu - 24 a_0 \nu \right) + \notag \\
&+ \dfrac{1}{36} u \left( \QQ - 1 \right) \big( -14 a_0 - 24 a_1 + 24 a_2 - 109 \nu + \notag \\
&+ 66 a_0 \nu + 51 \nu^2 \big) + \dfrac{1}{24} \left( \QQ - 1 \right) \dfrac{p_r^2}{\BB} \big( 4 a_0 - 48 a_1 + 48 a_3 + \notag \\
&+ 6 \nu - 60 a_0 \nu - 39 \nu^2 \big) + \dfrac{1}{24} u \dfrac{p_r^2}{\BB} \big( - 48 a_0 - 32 a_1 + \notag \\
&- 64 a_2 - 48 a_3 - 16 \nu - 48 a_0 \nu - 147 \nu^2 \big) \ , \\ 
c^{(2)}_{S_{*}} &= \dfrac{1}{36} u^2 \big( -56 b_0 - 24 b_2 + 353 \nu - 60 b_0 \nu - 27 \nu^2 \big) + \notag \\
&+ \dfrac{5}{3} \dfrac{p_r^4}{\BB^2} \big( -2 b_3 + 3 b_0 \nu + 3 \nu^2 \big) + \notag \\
&+ \dfrac{1}{72} \big(\QQ-1\big)^2 \big( -4 b_0 + 48 b_1 - 23 \nu - 12 b_0 \nu - 3 \nu^2 \big) + \notag \\
&+ \dfrac{1}{36} u \big( \QQ - 1 \big) \big( -14 b_0 - 24 b_1 + 24 b_2 - 103 \nu + 66 b_0 \nu + \notag \\
&+60 \nu^2 \big) + \dfrac{1}{12} \big( \QQ - 1 \big) \dfrac{p_r^2}{\BB} \big( 2 b_0 - 24 b_1 + 24 b_3 + 16 \nu + \notag \\
&- 30 b_0 \nu - 21 \nu^2 \big) + \dfrac{1}{12} u \dfrac{p_r^2}{\BB} \big( -24 b_0 - 16 b_1 - 32 b_2 + \notag \\
&- 24 b_3 + 47 \nu - 24 b_0 \nu - 54 \nu^2 \big).
\end{align}

Applying the \SEOBNRvq{} gauge choice ($a_i=b_i=0$), the former coefficients become
\begin{align}
c_S^{(1)}&= \dfrac{\nu}{4}(\QQ-1)+\dfrac{2}{3}u\nu,\\
c_S^{(2)}&=-\frac{5}{16} \nu (Q-1)^2+\frac{1}{36} \left(51 \nu ^2-109 \nu \right) (Q-1) u\nonumber\\
&+\frac{1}{9} \left(-21 \nu ^2-56 \nu \right)u^2,
\end{align}
and 
\begin{align}
c_{S_*}^{(1)}&=\frac{1}{3} \nu (Q-1)+\frac{7 \nu u}{6}, \\
c_{S_*}^{(2)}&=\frac{1}{72} \left(-3 \nu ^2-23 \nu \right) (Q-1)^2 + \notag\\
&+\frac{1}{36} \left(60 \nu^2-103 \nu \right) (Q-1) u
+\frac{1}{36} \left(353 \nu -27 \nu ^2\right) u^2.
\end{align}

\section{Gauge fixings}
\label{app:gauge}
We here report explicitly the connection between the spin gauges used 
in the spin-orbit sectors of the two models. 
We recall that \TEOBResumS{} makes use of the DJS gauge, while \SEOBNRvq{} sets all gauge parameters to zero.

Let us list here the explicit gauge choices on the PN-expanded gyro-gravitomagnetic
ratios~\cite{Damour:2008qf} $g_S^{\rm eff}\equiv r^3 G_S$ and $g_{S_*}^{\rm eff}\equiv r^3 G_{S_*}$.
At NNLO level they read
\begin{align}
g_S^{\rm eff} = 2 + \dfrac{1}{c^2} g_S^{\rm eff_{NLO}} + \dfrac{1}{c^4} g_S^{\rm eff_{NNLO}}, \notag \\
g_{S_*}^{\rm eff} = \dfrac{3}{2} + \dfrac{1}{c^2} g_{S_*}^{\rm eff_{NLO}} + \dfrac{1}{c^4} g_{S_*}^{\rm eff_{NNLO}},
\end{align}
where the NLO terms read
\begin{align}
g_S^{\rm eff_{NLO}} &= \bigg( \left(\dfrac{3}{8} \nu + a \right) \bm{p}^2 + \notag \\ &- \left(\dfrac{9}{2} \nu + 3 a \right) (\bm{p} \cdot \bm{n})^2 - u (\nu + a) \bigg), \\
g_{S_*}^{\rm eff_{NLO}} &= \left(-\dfrac{5}{8} + \dfrac{\nu}{2} + b \right) \bm{p}^2 + \notag \\ &- \left(\dfrac{15}{4} \nu + 3 b \right) (\bm{p} \cdot \bm{n})^2 - u \left(\dfrac{1}{2} + \dfrac{5}{4} \nu + b \right),
\end{align}
and the NNLO ones are expressed as
\begin{align}
g_S^{\rm eff_{NNLO}} &= -u^2 \left(9 \nu + \dfrac{3}{2} \nu^2 + a + \alpha \right) + \notag \\ &+ u \bigg[(\bm{p} \cdot \bm{n})^2 \bigg(\dfrac{35}{4} \nu - \dfrac{3}{16} \nu^2 + 6 a - 4 \alpha - 3 \beta - 2 \gamma \bigg)+ \notag \\ &+ \bm{p}^2 \bigg(-\dfrac{17}{4} \nu + \dfrac{11}{8} \nu^2 - \dfrac{3}{2} a + \alpha - \gamma \bigg) \bigg] + \notag \\ &+ \bigg(\dfrac{9}{4} \nu - \dfrac{39}{16} \nu^2 + \dfrac{3}{2} a + 3 \beta- 3 \gamma \bigg) \bm{p}^2 (\bm{p} \cdot \bm{n})^2 + \notag \\ &+ \left(\dfrac{135}{16} \nu^2 - 5 \beta \right) (\bm{p} \cdot \bm{n})^4+ \left(-\dfrac{5}{8} \nu - \dfrac{a}{2} + \gamma \right) \bm{p}^4, \\
g_{S_*}^{\rm eff_{NNLO}} &= -u^2 \left(\dfrac{1}{2} + \dfrac{55}{8} \nu + \dfrac{13}{8} \nu^2 + 
b + \delta \right) + \notag \\ 
&+ u \bigg[\bm{p}^2 \left( \dfrac{1}{4} - \dfrac{59}{16} \nu + \dfrac{3}{2} \nu^2 - \dfrac{3}{2} b + \delta - \eta \right) + \notag \\ 
&+ (\bm{p} \cdot \bm{n})^2 \left(\dfrac{5}{4} + \dfrac{109}{8} \nu + \dfrac{3}{4} \nu^2 + 6 b - 
4 \delta - 3 \zeta - 2 \eta \right) \bigg] + \notag \\
&+ \bigg(\dfrac{57}{16} \nu - \dfrac{21}{8} \nu^2 + \dfrac{3}{2} b + 3 \zeta - 3 \eta \bigg) \bm{p}^2 (\bm{p} \cdot \bm{n})^2 +\notag \\ &+ \left(\dfrac{15}{2} \nu^2 - 
5 \zeta \right) (\bm{p} \cdot \bm{n})^4 + \notag \\ 
&+ \left(\dfrac{7}{16} - \dfrac{11}{16} \nu - \dfrac{\nu^2}{16} - \dfrac{b}{2} + \eta \right) \bm{p}^4.
\end{align}
The DJS gauge used within \TEOBResumS{} is defined by
\begin{align}
a=&-\dfrac{3}{8} \nu, \hspace{2.95cm} b=\dfrac{5}{8} -\dfrac{\nu}{2}, \notag \\
\alpha=&~\dfrac{11}{8} \nu \left(3-\nu \right), \hspace{1.8cm} \beta=\dfrac{\nu}{16} \left(13\nu-2 \right), \notag \\
\gamma=&~\dfrac{7}{16}\nu, \hspace{3.05cm} \delta= \dfrac{1}{16} \left( 9+54 \nu -23\nu^2 \right), \notag \\
\zeta=&~\dfrac{1}{16} \left( -7-8 \nu +15\nu^2 \right), \hspace{0.4cm} \eta=\dfrac{1}{16} \left( -2+7 \nu +\nu^2 \right). 
\end{align}
By contrast, the gauge chosen for \SEOBNRvq{} is defined by imposing $a_i=b_i=0$ with $i=(0,1,2,3)$, as pointed out in Ref.~\cite{Taracchini:2012ig}.
The gauge parameters $(a_i,b_i)$ (see Eqs.~(51) and (52) therein) are related to the former ones by
\begin{align}
&a_0=~a, \qquad\qquad\qquad\qquad\! a_1=~\gamma + \dfrac{\nu}{4} a, \notag \\
&a_2=~\alpha - a \left( 1+ \dfrac{\nu}{2} \right), \qquad\! a_3=~\beta + \dfrac{3}{2} a \nu, \notag \\
&b_0=~b,\qquad\qquad \qquad\qquad b_1=~\eta + \dfrac{\nu}{4} b, \notag \\
&b_2=~\delta - b \left( 1 + \dfrac{\nu}{2} \right), \qquad \, \, b_3=~\zeta + \dfrac{3}{2} \nu b .
\end{align}

\subsection{Gauge-fixing for the new \TEOBResumS{} $G_{S_*}$}
In order to include the complete Hamiltonian of a spinning particle on a Kerr background within \TEOBResumS{}, we cannot use the DJS gauge.
Instead we need to work in a gauge such that the PN-expanded $G_{S_*}$ coincides with the Taylor expansion of $G_{S_*}^K$ when $\nu \to 0$.
This gauge is defined by the condition that all 
the $\nu$-dependent terms that depend on $({\bf n}\cdot {\bf p})$ disappear. 
The corresponding choice of the
gauge parameters in $(g_S^{\rm eff},g_{S_*}^{\rm eff})$ is then
\begin{align}
a&=-\dfrac{3}{2}\nu, \hspace{2.1cm} b = -\dfrac{5}{4}\nu, \notag \\
\alpha&=-\dfrac{\nu}{16}\left(1+28\nu\right),\hspace{0.5cm} \beta=\dfrac{27}{16}\nu^2, \notag \\
\gamma&=\dfrac{7}{8}\nu^2, \hspace{2.2cm}
\delta = \dfrac{5}{4}\nu(1-\nu), \notag\\
\zeta &=\dfrac{3}{2}\nu^2, \hspace{2.2cm}
\eta = \dfrac{\nu}{16}\left(9+10\nu\right).
\end{align}

\bibliography{refs20191127.bib}

\begin{thebibliography}{42}%
\makeatletter
\providecommand \@ifxundefined [1]{%
 \@ifx{#1\undefined}
}%
\providecommand \@ifnum [1]{%
 \ifnum #1\expandafter \@firstoftwo
 \else \expandafter \@secondoftwo
 \fi
}%
\providecommand \@ifx [1]{%
 \ifx #1\expandafter \@firstoftwo
 \else \expandafter \@secondoftwo
 \fi
}%
\providecommand \natexlab [1]{#1}%
\providecommand \enquote  [1]{``#1''}%
\providecommand \bibnamefont  [1]{#1}%
\providecommand \bibfnamefont [1]{#1}%
\providecommand \citenamefont [1]{#1}%
\providecommand \href@noop [0]{\@secondoftwo}%
\providecommand \href [0]{\begingroup \@sanitize@url \@href}%
\providecommand \@href[1]{\@@startlink{#1}\@@href}%
\providecommand \@@href[1]{\endgroup#1\@@endlink}%
\providecommand \@sanitize@url [0]{\catcode `\\12\catcode `\$12\catcode
  `\&12\catcode `\#12\catcode `\^12\catcode `\_12\catcode `\%12\relax}%
\providecommand \@@startlink[1]{}%
\providecommand \@@endlink[0]{}%
\providecommand \url  [0]{\begingroup\@sanitize@url \@url }%
\providecommand \@url [1]{\endgroup\@href {#1}{\urlprefix }}%
\providecommand \urlprefix  [0]{URL }%
\providecommand \Eprint [0]{\href }%
\providecommand \doibase [0]{http://dx.doi.org/}%
\providecommand \selectlanguage [0]{\@gobble}%
\providecommand \bibinfo  [0]{\@secondoftwo}%
\providecommand \bibfield  [0]{\@secondoftwo}%
\providecommand \translation [1]{[#1]}%
\providecommand \BibitemOpen [0]{}%
\providecommand \bibitemStop [0]{}%
\providecommand \bibitemNoStop [0]{.\EOS\space}%
\providecommand \EOS [0]{\spacefactor3000\relax}%
\providecommand \BibitemShut  [1]{\csname bibitem#1\endcsname}%
\let\auto@bib@innerbib\@empty
\bibitem [{\citenamefont {Abbott}\ \emph {et~al.}(2018)\citenamefont {Abbott}
  \emph {et~al.}}]{LIGOScientific:2018mvr}%
  \BibitemOpen
  \bibfield  {author} {\bibinfo {author} {\bibfnamefont {B.~P.}\ \bibnamefont
  {Abbott}} \emph {et~al.} (\bibinfo {collaboration} {LIGO Scientific,
  Virgo}),\ }\href@noop {} {\  (\bibinfo {year} {2018})},\ \Eprint
  {http://arxiv.org/abs/1811.12907} {arXiv:1811.12907 [astro-ph.HE]}
  \BibitemShut {NoStop}%
\bibitem [{\citenamefont {Buonanno}\ and\ \citenamefont
  {Damour}(1999)}]{Buonanno:1998gg}%
  \BibitemOpen
  \bibfield  {author} {\bibinfo {author} {\bibfnamefont {A.}~\bibnamefont
  {Buonanno}}\ and\ \bibinfo {author} {\bibfnamefont {T.}~\bibnamefont
  {Damour}},\ }\href {\doibase 10.1103/PhysRevD.59.084006} {\bibfield
  {journal} {\bibinfo  {journal} {Phys. Rev.}\ }\textbf {\bibinfo {volume}
  {D59}},\ \bibinfo {pages} {084006} (\bibinfo {year} {1999})},\ \Eprint
  {http://arxiv.org/abs/gr-qc/9811091} {arXiv:gr-qc/9811091} \BibitemShut
  {NoStop}%
\bibitem [{\citenamefont {Buonanno}\ and\ \citenamefont
  {Damour}(2000)}]{Buonanno:2000ef}%
  \BibitemOpen
  \bibfield  {author} {\bibinfo {author} {\bibfnamefont {A.}~\bibnamefont
  {Buonanno}}\ and\ \bibinfo {author} {\bibfnamefont {T.}~\bibnamefont
  {Damour}},\ }\href {\doibase 10.1103/PhysRevD.62.064015} {\bibfield
  {journal} {\bibinfo  {journal} {Phys. Rev.}\ }\textbf {\bibinfo {volume}
  {D62}},\ \bibinfo {pages} {064015} (\bibinfo {year} {2000})},\ \Eprint
  {http://arxiv.org/abs/gr-qc/0001013} {arXiv:gr-qc/0001013} \BibitemShut
  {NoStop}%
\bibitem [{\citenamefont {Damour}\ \emph {et~al.}(2000)\citenamefont {Damour},
  \citenamefont {Jaranowski},\ and\ \citenamefont {Schaefer}}]{Damour:2000we}%
  \BibitemOpen
  \bibfield  {author} {\bibinfo {author} {\bibfnamefont {T.}~\bibnamefont
  {Damour}}, \bibinfo {author} {\bibfnamefont {P.}~\bibnamefont {Jaranowski}},
  \ and\ \bibinfo {author} {\bibfnamefont {G.}~\bibnamefont {Schaefer}},\
  }\href {\doibase 10.1103/PhysRevD.62.084011} {\bibfield  {journal} {\bibinfo
  {journal} {Phys. Rev.}\ }\textbf {\bibinfo {volume} {D62}},\ \bibinfo {pages}
  {084011} (\bibinfo {year} {2000})},\ \Eprint
  {http://arxiv.org/abs/gr-qc/0005034} {arXiv:gr-qc/0005034 [gr-qc]}
  \BibitemShut {NoStop}%
\bibitem [{\citenamefont {Damour}\ \emph {et~al.}(2009)\citenamefont {Damour},
  \citenamefont {Iyer},\ and\ \citenamefont {Nagar}}]{Damour:2008gu}%
  \BibitemOpen
  \bibfield  {author} {\bibinfo {author} {\bibfnamefont {T.}~\bibnamefont
  {Damour}}, \bibinfo {author} {\bibfnamefont {B.~R.}\ \bibnamefont {Iyer}}, \
  and\ \bibinfo {author} {\bibfnamefont {A.}~\bibnamefont {Nagar}},\ }\href
  {\doibase 10.1103/PhysRevD.79.064004} {\bibfield  {journal} {\bibinfo
  {journal} {Phys. Rev.}\ }\textbf {\bibinfo {volume} {D79}},\ \bibinfo {pages}
  {064004} (\bibinfo {year} {2009})},\ \Eprint {http://arxiv.org/abs/0811.2069}
  {arXiv:0811.2069 [gr-qc]} \BibitemShut {NoStop}%
\bibitem [{\citenamefont {Blanchet}(2014)}]{Blanchet:2013haa}%
  \BibitemOpen
  \bibfield  {author} {\bibinfo {author} {\bibfnamefont {L.}~\bibnamefont
  {Blanchet}},\ }\href {\doibase 10.12942/lrr-2014-2} {\bibfield  {journal}
  {\bibinfo  {journal} {Living Rev. Relativity}\ }\textbf {\bibinfo {volume}
  {17}},\ \bibinfo {pages} {2} (\bibinfo {year} {2014})},\ \Eprint
  {http://arxiv.org/abs/1310.1528} {arXiv:1310.1528 [gr-qc]} \BibitemShut
  {NoStop}%
\bibitem [{\citenamefont {Schaefer}\ and\ \citenamefont
  {Jaranowski}(2018)}]{Schafer:2018kuf}%
  \BibitemOpen
  \bibfield  {author} {\bibinfo {author} {\bibfnamefont {G.}~\bibnamefont
  {Schaefer}}\ and\ \bibinfo {author} {\bibfnamefont {P.}~\bibnamefont
  {Jaranowski}},\ }\href {\doibase 10.1007/s41114-018-0016-5} {\bibfield
  {journal} {\bibinfo  {journal} {Living Rev. Rel.}\ }\textbf {\bibinfo
  {volume} {21}},\ \bibinfo {pages} {7} (\bibinfo {year} {2018})},\ \Eprint
  {http://arxiv.org/abs/1805.07240} {arXiv:1805.07240 [gr-qc]} \BibitemShut
  {NoStop}%
\bibitem [{\citenamefont {Boh{\'e}}\ \emph {et~al.}(2017)\citenamefont
  {Boh{\'e}} \emph {et~al.}}]{Bohe:2016gbl}%
  \BibitemOpen
  \bibfield  {author} {\bibinfo {author} {\bibfnamefont {A.}~\bibnamefont
  {Boh{\'e}}} \emph {et~al.},\ }\href {\doibase 10.1103/PhysRevD.95.044028}
  {\bibfield  {journal} {\bibinfo  {journal} {Phys. Rev.}\ }\textbf {\bibinfo
  {volume} {D95}},\ \bibinfo {pages} {044028} (\bibinfo {year} {2017})},\
  \Eprint {http://arxiv.org/abs/1611.03703} {arXiv:1611.03703 [gr-qc]}
  \BibitemShut {NoStop}%
\bibitem [{\citenamefont {Cotesta}\ \emph {et~al.}(2018)\citenamefont
  {Cotesta}, \citenamefont {Buonanno}, \citenamefont {Boh\'e}, \citenamefont
  {Taracchini}, \citenamefont {Hinder},\ and\ \citenamefont
  {Ossokine}}]{Cotesta:2018fcv}%
  \BibitemOpen
  \bibfield  {author} {\bibinfo {author} {\bibfnamefont {R.}~\bibnamefont
  {Cotesta}}, \bibinfo {author} {\bibfnamefont {A.}~\bibnamefont {Buonanno}},
  \bibinfo {author} {\bibfnamefont {A.}~\bibnamefont {Boh\'e}}, \bibinfo
  {author} {\bibfnamefont {A.}~\bibnamefont {Taracchini}}, \bibinfo {author}
  {\bibfnamefont {I.}~\bibnamefont {Hinder}}, \ and\ \bibinfo {author}
  {\bibfnamefont {S.}~\bibnamefont {Ossokine}},\ }\href {\doibase
  10.1103/PhysRevD.98.084028} {\bibfield  {journal} {\bibinfo  {journal} {Phys.
  Rev.}\ }\textbf {\bibinfo {volume} {D98}},\ \bibinfo {pages} {084028}
  (\bibinfo {year} {2018})},\ \Eprint {http://arxiv.org/abs/1803.10701}
  {arXiv:1803.10701 [gr-qc]} \BibitemShut {NoStop}%
\bibitem [{\citenamefont {Nagar}\ \emph {et~al.}(2018)\citenamefont {Nagar}
  \emph {et~al.}}]{Nagar:2018zoe}%
  \BibitemOpen
  \bibfield  {author} {\bibinfo {author} {\bibfnamefont {A.}~\bibnamefont
  {Nagar}} \emph {et~al.},\ }\href {\doibase 10.1103/PhysRevD.98.104052}
  {\bibfield  {journal} {\bibinfo  {journal} {Phys. Rev.}\ }\textbf {\bibinfo
  {volume} {D98}},\ \bibinfo {pages} {104052} (\bibinfo {year} {2018})},\
  \Eprint {http://arxiv.org/abs/1806.01772} {arXiv:1806.01772 [gr-qc]}
  \BibitemShut {NoStop}%
\bibitem [{\citenamefont {Nagar}\ \emph
  {et~al.}(2019{\natexlab{a}})\citenamefont {Nagar}, \citenamefont {Messina},
  \citenamefont {Rettegno}, \citenamefont {Bini}, \citenamefont {Damour},
  \citenamefont {Geralico}, \citenamefont {Akcay},\ and\ \citenamefont
  {Bernuzzi}}]{Nagar:2018plt}%
  \BibitemOpen
  \bibfield  {author} {\bibinfo {author} {\bibfnamefont {A.}~\bibnamefont
  {Nagar}}, \bibinfo {author} {\bibfnamefont {F.}~\bibnamefont {Messina}},
  \bibinfo {author} {\bibfnamefont {P.}~\bibnamefont {Rettegno}}, \bibinfo
  {author} {\bibfnamefont {D.}~\bibnamefont {Bini}}, \bibinfo {author}
  {\bibfnamefont {T.}~\bibnamefont {Damour}}, \bibinfo {author} {\bibfnamefont
  {A.}~\bibnamefont {Geralico}}, \bibinfo {author} {\bibfnamefont
  {S.}~\bibnamefont {Akcay}}, \ and\ \bibinfo {author} {\bibfnamefont
  {S.}~\bibnamefont {Bernuzzi}},\ }\href {\doibase 10.1103/PhysRevD.99.044007}
  {\bibfield  {journal} {\bibinfo  {journal} {Phys. Rev.}\ }\textbf {\bibinfo
  {volume} {D99}},\ \bibinfo {pages} {044007} (\bibinfo {year}
  {2019}{\natexlab{a}})},\ \Eprint {http://arxiv.org/abs/1812.07923}
  {arXiv:1812.07923 [gr-qc]} \BibitemShut {NoStop}%
\bibitem [{\citenamefont {Pan}\ \emph {et~al.}(2014)\citenamefont {Pan},
  \citenamefont {Buonanno}, \citenamefont {Taracchini}, \citenamefont {Kidder},
  \citenamefont {Mroue} \emph {et~al.}}]{Pan:2013rra}%
  \BibitemOpen
  \bibfield  {author} {\bibinfo {author} {\bibfnamefont {Y.}~\bibnamefont
  {Pan}}, \bibinfo {author} {\bibfnamefont {A.}~\bibnamefont {Buonanno}},
  \bibinfo {author} {\bibfnamefont {A.}~\bibnamefont {Taracchini}}, \bibinfo
  {author} {\bibfnamefont {L.~E.}\ \bibnamefont {Kidder}}, \bibinfo {author}
  {\bibfnamefont {A.~H.}\ \bibnamefont {Mroue}},  \emph {et~al.},\ }\href
  {\doibase 10.1103/PhysRevD.89.084006} {\bibfield  {journal} {\bibinfo
  {journal} {Phys.Rev.}\ }\textbf {\bibinfo {volume} {D89}},\ \bibinfo {pages}
  {084006} (\bibinfo {year} {2014})},\ \Eprint {http://arxiv.org/abs/1307.6232}
  {arXiv:1307.6232 [gr-qc]} \BibitemShut {NoStop}%
\bibitem [{\citenamefont {Babak}\ \emph {et~al.}(2017)\citenamefont {Babak},
  \citenamefont {Taracchini},\ and\ \citenamefont {Buonanno}}]{Babak:2016tgq}%
  \BibitemOpen
  \bibfield  {author} {\bibinfo {author} {\bibfnamefont {S.}~\bibnamefont
  {Babak}}, \bibinfo {author} {\bibfnamefont {A.}~\bibnamefont {Taracchini}}, \
  and\ \bibinfo {author} {\bibfnamefont {A.}~\bibnamefont {Buonanno}},\ }\href
  {\doibase 10.1103/PhysRevD.95.024010} {\bibfield  {journal} {\bibinfo
  {journal} {Phys. Rev.}\ }\textbf {\bibinfo {volume} {D95}},\ \bibinfo {pages}
  {024010} (\bibinfo {year} {2017})},\ \Eprint
  {http://arxiv.org/abs/1607.05661} {arXiv:1607.05661 [gr-qc]} \BibitemShut
  {NoStop}%
\bibitem [{\citenamefont {Ossokine}\ \emph {et~al.}()\citenamefont {Ossokine}
  \emph {et~al.}}]{Ossokine:2019}%
  \BibitemOpen
  \bibfield  {author} {\bibinfo {author} {\bibfnamefont {S.}~\bibnamefont
  {Ossokine}} \emph {et~al.},\ }\href@noop {} {\bibinfo  {journal} {in
  preparation, 2019}\ }\BibitemShut {NoStop}%
\bibitem [{\citenamefont {Breschi}\ \emph {et~al.}(2019)\citenamefont
  {Breschi}, \citenamefont {Bernuzzi}, \citenamefont {Zappa}, \citenamefont
  {Agathos}, \citenamefont {Perego}, \citenamefont {Radice},\ and\
  \citenamefont {Nagar}}]{Breschi:2019srl}%
  \BibitemOpen
\bibfield  {journal} {  }\bibfield  {author} {\bibinfo {author} {\bibfnamefont
  {M.}~\bibnamefont {Breschi}}, \bibinfo {author} {\bibfnamefont
  {S.}~\bibnamefont {Bernuzzi}}, \bibinfo {author} {\bibfnamefont
  {F.}~\bibnamefont {Zappa}}, \bibinfo {author} {\bibfnamefont
  {M.}~\bibnamefont {Agathos}}, \bibinfo {author} {\bibfnamefont
  {A.}~\bibnamefont {Perego}}, \bibinfo {author} {\bibfnamefont
  {D.}~\bibnamefont {Radice}}, \ and\ \bibinfo {author} {\bibfnamefont
  {A.}~\bibnamefont {Nagar}},\ }\href {\doibase 10.1103/PhysRevD.100.104029}
  {\bibfield  {journal} {\bibinfo  {journal} {Phys. Rev.}\ }\textbf {\bibinfo
  {volume} {D100}},\ \bibinfo {pages} {104029} (\bibinfo {year} {2019})},\
  \Eprint {http://arxiv.org/abs/1908.11418} {arXiv:1908.11418 [gr-qc]}
  \BibitemShut {NoStop}%
\bibitem [{\citenamefont {Hinderer}\ \emph {et~al.}(2016)\citenamefont
  {Hinderer} \emph {et~al.}}]{Hinderer:2016eia}%
  \BibitemOpen
  \bibfield  {author} {\bibinfo {author} {\bibfnamefont {T.}~\bibnamefont
  {Hinderer}} \emph {et~al.},\ }\href {\doibase 10.1103/PhysRevLett.116.181101}
  {\bibfield  {journal} {\bibinfo  {journal} {Phys. Rev. Lett.}\ }\textbf
  {\bibinfo {volume} {116}},\ \bibinfo {pages} {181101} (\bibinfo {year}
  {2016})},\ \Eprint {http://arxiv.org/abs/1602.00599} {arXiv:1602.00599
  [gr-qc]} \BibitemShut {NoStop}%
\bibitem [{\citenamefont {Steinhoff}\ \emph {et~al.}(2016)\citenamefont
  {Steinhoff}, \citenamefont {Hinderer}, \citenamefont {Buonanno},\ and\
  \citenamefont {Taracchini}}]{Steinhoff:2016rfi}%
  \BibitemOpen
  \bibfield  {author} {\bibinfo {author} {\bibfnamefont {J.}~\bibnamefont
  {Steinhoff}}, \bibinfo {author} {\bibfnamefont {T.}~\bibnamefont {Hinderer}},
  \bibinfo {author} {\bibfnamefont {A.}~\bibnamefont {Buonanno}}, \ and\
  \bibinfo {author} {\bibfnamefont {A.}~\bibnamefont {Taracchini}},\ }\href
  {\doibase 10.1103/PhysRevD.94.104028} {\bibfield  {journal} {\bibinfo
  {journal} {Phys. Rev.}\ }\textbf {\bibinfo {volume} {D94}},\ \bibinfo {pages}
  {104028} (\bibinfo {year} {2016})},\ \Eprint
  {http://arxiv.org/abs/1608.01907} {arXiv:1608.01907 [gr-qc]} \BibitemShut
  {NoStop}%
\bibitem [{\citenamefont {Lackey}\ \emph {et~al.}(2018)\citenamefont {Lackey},
  \citenamefont {Pürrer}, \citenamefont {Taracchini},\ and\ \citenamefont
  {Marsat}}]{Lackey:2018zvw}%
  \BibitemOpen
  \bibfield  {author} {\bibinfo {author} {\bibfnamefont {B.~D.}\ \bibnamefont
  {Lackey}}, \bibinfo {author} {\bibfnamefont {M.}~\bibnamefont {Pürrer}},
  \bibinfo {author} {\bibfnamefont {A.}~\bibnamefont {Taracchini}}, \ and\
  \bibinfo {author} {\bibfnamefont {S.}~\bibnamefont {Marsat}},\ }\href@noop {}
  {\  (\bibinfo {year} {2018})},\ \Eprint {http://arxiv.org/abs/1812.08643}
  {arXiv:1812.08643 [gr-qc]} \BibitemShut {NoStop}%
\bibitem [{\citenamefont {Barausse}\ and\ \citenamefont
  {Buonanno}(2010)}]{Barausse:2009xi}%
  \BibitemOpen
  \bibfield  {author} {\bibinfo {author} {\bibfnamefont {E.}~\bibnamefont
  {Barausse}}\ and\ \bibinfo {author} {\bibfnamefont {A.}~\bibnamefont
  {Buonanno}},\ }\href {\doibase 10.1103/PhysRevD.81.084024} {\bibfield
  {journal} {\bibinfo  {journal} {Phys.Rev.}\ }\textbf {\bibinfo {volume}
  {D81}},\ \bibinfo {pages} {084024} (\bibinfo {year} {2010})},\ \Eprint
  {http://arxiv.org/abs/0912.3517} {arXiv:0912.3517 [gr-qc]} \BibitemShut
  {NoStop}%
\bibitem [{\citenamefont {Barausse}\ \emph {et~al.}(2009)\citenamefont
  {Barausse}, \citenamefont {Racine},\ and\ \citenamefont
  {Buonanno}}]{Barausse:2009aa}%
  \BibitemOpen
  \bibfield  {author} {\bibinfo {author} {\bibfnamefont {E.}~\bibnamefont
  {Barausse}}, \bibinfo {author} {\bibfnamefont {E.}~\bibnamefont {Racine}}, \
  and\ \bibinfo {author} {\bibfnamefont {A.}~\bibnamefont {Buonanno}},\ }\href
  {\doibase 10.1103/PhysRevD.80.104025} {\bibfield  {journal} {\bibinfo
  {journal} {Phys. Rev.}\ }\textbf {\bibinfo {volume} {D80}},\ \bibinfo {pages}
  {104025} (\bibinfo {year} {2009})},\ \Eprint {http://arxiv.org/abs/0907.4745}
  {arXiv:0907.4745 [gr-qc]} \BibitemShut {NoStop}%
\bibitem [{\citenamefont {Barausse}\ and\ \citenamefont
  {Buonanno}(2011)}]{Barausse:2011ys}%
  \BibitemOpen
  \bibfield  {author} {\bibinfo {author} {\bibfnamefont {E.}~\bibnamefont
  {Barausse}}\ and\ \bibinfo {author} {\bibfnamefont {A.}~\bibnamefont
  {Buonanno}},\ }\href {\doibase 10.1103/PhysRevD.84.104027} {\bibfield
  {journal} {\bibinfo  {journal} {Phys.Rev.}\ }\textbf {\bibinfo {volume}
  {D84}},\ \bibinfo {pages} {104027} (\bibinfo {year} {2011})},\ \Eprint
  {http://arxiv.org/abs/1107.2904} {arXiv:1107.2904 [gr-qc]} \BibitemShut
  {NoStop}%
\bibitem [{\citenamefont {Taracchini}\ \emph {et~al.}(2012)\citenamefont
  {Taracchini}, \citenamefont {Pan}, \citenamefont {Buonanno}, \citenamefont
  {Barausse}, \citenamefont {Boyle} \emph {et~al.}}]{Taracchini:2012ig}%
  \BibitemOpen
  \bibfield  {author} {\bibinfo {author} {\bibfnamefont {A.}~\bibnamefont
  {Taracchini}}, \bibinfo {author} {\bibfnamefont {Y.}~\bibnamefont {Pan}},
  \bibinfo {author} {\bibfnamefont {A.}~\bibnamefont {Buonanno}}, \bibinfo
  {author} {\bibfnamefont {E.}~\bibnamefont {Barausse}}, \bibinfo {author}
  {\bibfnamefont {M.}~\bibnamefont {Boyle}},  \emph {et~al.},\ }\href {\doibase
  10.1103/PhysRevD.86.024011} {\bibfield  {journal} {\bibinfo  {journal}
  {Phys.Rev.}\ }\textbf {\bibinfo {volume} {D86}},\ \bibinfo {pages} {024011}
  (\bibinfo {year} {2012})},\ \Eprint {http://arxiv.org/abs/1202.0790}
  {arXiv:1202.0790 [gr-qc]} \BibitemShut {NoStop}%
\bibitem [{\citenamefont {Taracchini}\ \emph {et~al.}(2014)\citenamefont
  {Taracchini}, \citenamefont {Buonanno}, \citenamefont {Pan}, \citenamefont
  {Hinderer}, \citenamefont {Boyle} \emph {et~al.}}]{Taracchini:2013rva}%
  \BibitemOpen
  \bibfield  {author} {\bibinfo {author} {\bibfnamefont {A.}~\bibnamefont
  {Taracchini}}, \bibinfo {author} {\bibfnamefont {A.}~\bibnamefont
  {Buonanno}}, \bibinfo {author} {\bibfnamefont {Y.}~\bibnamefont {Pan}},
  \bibinfo {author} {\bibfnamefont {T.}~\bibnamefont {Hinderer}}, \bibinfo
  {author} {\bibfnamefont {M.}~\bibnamefont {Boyle}},  \emph {et~al.},\ }\href
  {\doibase 10.1103/PhysRevD.89.061502} {\bibfield  {journal} {\bibinfo
  {journal} {Phys.Rev.}\ }\textbf {\bibinfo {volume} {D89}},\ \bibinfo {pages}
  {061502} (\bibinfo {year} {2014})},\ \Eprint {http://arxiv.org/abs/1311.2544}
  {arXiv:1311.2544 [gr-qc]} \BibitemShut {NoStop}%
\bibitem [{\citenamefont {Hinderer}\ \emph {et~al.}(2013)\citenamefont
  {Hinderer} \emph {et~al.}}]{Hinderer:2013uwa}%
  \BibitemOpen
  \bibfield  {author} {\bibinfo {author} {\bibfnamefont {T.}~\bibnamefont
  {Hinderer}} \emph {et~al.},\ }\href {\doibase 10.1103/PhysRevD.88.084005}
  {\bibfield  {journal} {\bibinfo  {journal} {Phys. Rev.}\ }\textbf {\bibinfo
  {volume} {D88}},\ \bibinfo {pages} {084005} (\bibinfo {year} {2013})},\
  \Eprint {http://arxiv.org/abs/1309.0544} {arXiv:1309.0544 [gr-qc]}
  \BibitemShut {NoStop}%
\bibitem [{\citenamefont {Damour}\ and\ \citenamefont
  {Nagar}(2014)}]{Damour:2014sva}%
  \BibitemOpen
  \bibfield  {author} {\bibinfo {author} {\bibfnamefont {T.}~\bibnamefont
  {Damour}}\ and\ \bibinfo {author} {\bibfnamefont {A.}~\bibnamefont {Nagar}},\
  }\href {\doibase 10.1103/PhysRevD.90.044018} {\bibfield  {journal} {\bibinfo
  {journal} {Phys.Rev.}\ }\textbf {\bibinfo {volume} {D90}},\ \bibinfo {pages}
  {044018} (\bibinfo {year} {2014})},\ \Eprint {http://arxiv.org/abs/1406.6913}
  {arXiv:1406.6913 [gr-qc]} \BibitemShut {NoStop}%
\bibitem [{\citenamefont {Nagar}\ \emph {et~al.}(2016)\citenamefont {Nagar},
  \citenamefont {Damour}, \citenamefont {Reisswig},\ and\ \citenamefont
  {Pollney}}]{Nagar:2015xqa}%
  \BibitemOpen
  \bibfield  {author} {\bibinfo {author} {\bibfnamefont {A.}~\bibnamefont
  {Nagar}}, \bibinfo {author} {\bibfnamefont {T.}~\bibnamefont {Damour}},
  \bibinfo {author} {\bibfnamefont {C.}~\bibnamefont {Reisswig}}, \ and\
  \bibinfo {author} {\bibfnamefont {D.}~\bibnamefont {Pollney}},\ }\href
  {\doibase 10.1103/PhysRevD.93.044046} {\bibfield  {journal} {\bibinfo
  {journal} {Phys. Rev.}\ }\textbf {\bibinfo {volume} {D93}},\ \bibinfo {pages}
  {044046} (\bibinfo {year} {2016})},\ \Eprint
  {http://arxiv.org/abs/1506.08457} {arXiv:1506.08457 [gr-qc]} \BibitemShut
  {NoStop}%
\bibitem [{\citenamefont {Nagar}\ \emph {et~al.}(2017)\citenamefont {Nagar},
  \citenamefont {Riemenschneider},\ and\ \citenamefont
  {Pratten}}]{Nagar:2017jdw}%
  \BibitemOpen
  \bibfield  {author} {\bibinfo {author} {\bibfnamefont {A.}~\bibnamefont
  {Nagar}}, \bibinfo {author} {\bibfnamefont {G.}~\bibnamefont
  {Riemenschneider}}, \ and\ \bibinfo {author} {\bibfnamefont {G.}~\bibnamefont
  {Pratten}},\ }\href {\doibase 10.1103/PhysRevD.96.084045} {\bibfield
  {journal} {\bibinfo  {journal} {Phys. Rev.}\ }\textbf {\bibinfo {volume}
  {D96}},\ \bibinfo {pages} {084045} (\bibinfo {year} {2017})},\ \Eprint
  {http://arxiv.org/abs/1703.06814} {arXiv:1703.06814 [gr-qc]} \BibitemShut
  {NoStop}%
\bibitem [{\citenamefont {Nagar}\ and\ \citenamefont
  {Rettegno}(2019)}]{Nagar:2018gnk}%
  \BibitemOpen
  \bibfield  {author} {\bibinfo {author} {\bibfnamefont {A.}~\bibnamefont
  {Nagar}}\ and\ \bibinfo {author} {\bibfnamefont {P.}~\bibnamefont
  {Rettegno}},\ }\href {\doibase 10.1103/PhysRevD.99.021501} {\bibfield
  {journal} {\bibinfo  {journal} {Phys. Rev.}\ }\textbf {\bibinfo {volume}
  {D99}},\ \bibinfo {pages} {021501} (\bibinfo {year} {2019})},\ \Eprint
  {http://arxiv.org/abs/1805.03891} {arXiv:1805.03891 [gr-qc]} \BibitemShut
  {NoStop}%
\bibitem [{\citenamefont {Akcay}\ \emph {et~al.}(2019)\citenamefont {Akcay},
  \citenamefont {Bernuzzi}, \citenamefont {Messina}, \citenamefont {Nagar},
  \citenamefont {Ortiz},\ and\ \citenamefont {Rettegno}}]{Akcay:2018yyh}%
  \BibitemOpen
  \bibfield  {author} {\bibinfo {author} {\bibfnamefont {S.}~\bibnamefont
  {Akcay}}, \bibinfo {author} {\bibfnamefont {S.}~\bibnamefont {Bernuzzi}},
  \bibinfo {author} {\bibfnamefont {F.}~\bibnamefont {Messina}}, \bibinfo
  {author} {\bibfnamefont {A.}~\bibnamefont {Nagar}}, \bibinfo {author}
  {\bibfnamefont {N.}~\bibnamefont {Ortiz}}, \ and\ \bibinfo {author}
  {\bibfnamefont {P.}~\bibnamefont {Rettegno}},\ }\href {\doibase
  10.1103/PhysRevD.99.044051} {\bibfield  {journal} {\bibinfo  {journal} {Phys.
  Rev.}\ }\textbf {\bibinfo {volume} {D99}},\ \bibinfo {pages} {044051}
  (\bibinfo {year} {2019})},\ \Eprint {http://arxiv.org/abs/1812.02744}
  {arXiv:1812.02744 [gr-qc]} \BibitemShut {NoStop}%
\bibitem [{\citenamefont {Damour}\ \emph {et~al.}(2008)\citenamefont {Damour},
  \citenamefont {Jaranowski},\ and\ \citenamefont
  {Sch{\"a}fer}}]{Damour:2008qf}%
  \BibitemOpen
  \bibfield  {author} {\bibinfo {author} {\bibfnamefont {T.}~\bibnamefont
  {Damour}}, \bibinfo {author} {\bibfnamefont {P.}~\bibnamefont {Jaranowski}},
  \ and\ \bibinfo {author} {\bibfnamefont {G.}~\bibnamefont {Sch{\"a}fer}},\
  }\href {\doibase 10.1103/PhysRevD.78.024009} {\bibfield  {journal} {\bibinfo
  {journal} {Phys.Rev.}\ }\textbf {\bibinfo {volume} {D78}},\ \bibinfo {pages}
  {024009} (\bibinfo {year} {2008})},\ \Eprint {http://arxiv.org/abs/0803.0915}
  {arXiv:0803.0915 [gr-qc]} \BibitemShut {NoStop}%
\bibitem [{\citenamefont {Damour}(2001)}]{Damour:2001tu}%
  \BibitemOpen
  \bibfield  {author} {\bibinfo {author} {\bibfnamefont {T.}~\bibnamefont
  {Damour}},\ }\href {\doibase 10.1103/PhysRevD.64.124013} {\bibfield
  {journal} {\bibinfo  {journal} {Phys. Rev.}\ }\textbf {\bibinfo {volume}
  {D64}},\ \bibinfo {pages} {124013} (\bibinfo {year} {2001})},\ \Eprint
  {http://arxiv.org/abs/gr-qc/0103018} {arXiv:gr-qc/0103018} \BibitemShut
  {NoStop}%
\bibitem [{\citenamefont {Bini}\ \emph {et~al.}(2015)\citenamefont {Bini},
  \citenamefont {Damour},\ and\ \citenamefont {Geralico}}]{Bini:2015xua}%
  \BibitemOpen
  \bibfield  {author} {\bibinfo {author} {\bibfnamefont {D.}~\bibnamefont
  {Bini}}, \bibinfo {author} {\bibfnamefont {T.}~\bibnamefont {Damour}}, \ and\
  \bibinfo {author} {\bibfnamefont {A.}~\bibnamefont {Geralico}},\ }\href
  {\doibase 10.1103/PhysRevD.93.109902, 10.1103/PhysRevD.92.124058} {\bibfield
  {journal} {\bibinfo  {journal} {Phys. Rev.}\ }\textbf {\bibinfo {volume}
  {D92}},\ \bibinfo {pages} {124058} (\bibinfo {year} {2015})},\ \bibinfo
  {note} {[Erratum: Phys. Rev.D93,no.10,109902(2016)]},\ \Eprint
  {http://arxiv.org/abs/1510.06230} {arXiv:1510.06230 [gr-qc]} \BibitemShut
  {NoStop}%
\bibitem [{\citenamefont {Akcay}\ \emph {et~al.}(2012)\citenamefont {Akcay},
  \citenamefont {Barack}, \citenamefont {Damour},\ and\ \citenamefont
  {Sago}}]{Akcay:2012ea}%
  \BibitemOpen
  \bibfield  {author} {\bibinfo {author} {\bibfnamefont {S.}~\bibnamefont
  {Akcay}}, \bibinfo {author} {\bibfnamefont {L.}~\bibnamefont {Barack}},
  \bibinfo {author} {\bibfnamefont {T.}~\bibnamefont {Damour}}, \ and\ \bibinfo
  {author} {\bibfnamefont {N.}~\bibnamefont {Sago}},\ }\href {\doibase
  10.1103/PhysRevD.86.104041} {\bibfield  {journal} {\bibinfo  {journal} {Phys.
  Rev.}\ }\textbf {\bibinfo {volume} {D86}},\ \bibinfo {pages} {104041}
  (\bibinfo {year} {2012})},\ \Eprint {http://arxiv.org/abs/1209.0964}
  {arXiv:1209.0964 [gr-qc]} \BibitemShut {NoStop}%
\bibitem [{\citenamefont {Nagar}(2011)}]{Nagar:2011fx}%
  \BibitemOpen
  \bibfield  {author} {\bibinfo {author} {\bibfnamefont {A.}~\bibnamefont
  {Nagar}},\ }\href {\doibase 10.1103/PhysRevD.84.084028} {\bibfield  {journal}
  {\bibinfo  {journal} {Phys.Rev.}\ }\textbf {\bibinfo {volume} {D84}},\
  \bibinfo {pages} {084028} (\bibinfo {year} {2011})},\ \Eprint
  {http://arxiv.org/abs/1106.4349} {arXiv:1106.4349 [gr-qc]} \BibitemShut
  {NoStop}%
\bibitem [{\citenamefont {Nagar}\ \emph
  {et~al.}(2019{\natexlab{b}})\citenamefont {Nagar}, \citenamefont {Pratten},
  \citenamefont {Riemenschneider},\ and\ \citenamefont
  {Gamba}}]{Nagar:2019wds}%
  \BibitemOpen
  \bibfield  {author} {\bibinfo {author} {\bibfnamefont {A.}~\bibnamefont
  {Nagar}}, \bibinfo {author} {\bibfnamefont {G.}~\bibnamefont {Pratten}},
  \bibinfo {author} {\bibfnamefont {G.}~\bibnamefont {Riemenschneider}}, \ and\
  \bibinfo {author} {\bibfnamefont {R.}~\bibnamefont {Gamba}},\ }\href@noop {}
  {\  (\bibinfo {year} {2019}{\natexlab{b}})},\ \Eprint
  {http://arxiv.org/abs/1904.09550} {arXiv:1904.09550 [gr-qc]} \BibitemShut
  {NoStop}%
\bibitem [{\citenamefont {Field}\ \emph {et~al.}(2014)\citenamefont {Field},
  \citenamefont {Galley}, \citenamefont {Hesthaven}, \citenamefont {Kaye},\
  and\ \citenamefont {Tiglio}}]{Field:2013cfa}%
  \BibitemOpen
  \bibfield  {author} {\bibinfo {author} {\bibfnamefont {S.~E.}\ \bibnamefont
  {Field}}, \bibinfo {author} {\bibfnamefont {C.~R.}\ \bibnamefont {Galley}},
  \bibinfo {author} {\bibfnamefont {J.~S.}\ \bibnamefont {Hesthaven}}, \bibinfo
  {author} {\bibfnamefont {J.}~\bibnamefont {Kaye}}, \ and\ \bibinfo {author}
  {\bibfnamefont {M.}~\bibnamefont {Tiglio}},\ }\href {\doibase
  10.1103/PhysRevX.4.031006} {\bibfield  {journal} {\bibinfo  {journal}
  {Phys.Rev.}\ }\textbf {\bibinfo {volume} {X4}},\ \bibinfo {pages} {031006}
  (\bibinfo {year} {2014})},\ \Eprint {http://arxiv.org/abs/1308.3565}
  {arXiv:1308.3565 [gr-qc]} \BibitemShut {NoStop}%
\bibitem [{\citenamefont {P{\"u}rrer}(2014)}]{Purrer:2014fza}%
  \BibitemOpen
  \bibfield  {author} {\bibinfo {author} {\bibfnamefont {M.}~\bibnamefont
  {P{\"u}rrer}},\ }\href {\doibase 10.1088/0264-9381/31/19/195010} {\bibfield
  {journal} {\bibinfo  {journal} {Class. Quant. Grav.}\ }\textbf {\bibinfo
  {volume} {31}},\ \bibinfo {pages} {195010} (\bibinfo {year} {2014})},\
  \Eprint {http://arxiv.org/abs/1402.4146} {arXiv:1402.4146 [gr-qc]}
  \BibitemShut {NoStop}%
\bibitem [{\citenamefont {P{\"u}rrer}(2016)}]{Purrer:2015tud}%
  \BibitemOpen
  \bibfield  {author} {\bibinfo {author} {\bibfnamefont {M.}~\bibnamefont
  {P{\"u}rrer}},\ }\href {\doibase 10.1103/PhysRevD.93.064041} {\bibfield
  {journal} {\bibinfo  {journal} {Phys. Rev.}\ }\textbf {\bibinfo {volume}
  {D93}},\ \bibinfo {pages} {064041} (\bibinfo {year} {2016})},\ \Eprint
  {http://arxiv.org/abs/1512.02248} {arXiv:1512.02248 [gr-qc]} \BibitemShut
  {NoStop}%
\bibitem [{\citenamefont {Galley}\ and\ \citenamefont
  {Schmidt}(2016)}]{Galley:2016mvy}%
  \BibitemOpen
  \bibfield  {author} {\bibinfo {author} {\bibfnamefont {C.~R.}\ \bibnamefont
  {Galley}}\ and\ \bibinfo {author} {\bibfnamefont {P.}~\bibnamefont
  {Schmidt}},\ }\href@noop {} {\  (\bibinfo {year} {2016})},\ \Eprint
  {http://arxiv.org/abs/1611.07529} {arXiv:1611.07529 [gr-qc]} \BibitemShut
  {NoStop}%
\bibitem [{\citenamefont {Lackey}\ \emph {et~al.}(2017)\citenamefont {Lackey},
  \citenamefont {Bernuzzi}, \citenamefont {Galley}, \citenamefont {Meidam},\
  and\ \citenamefont {Van Den~Broeck}}]{Lackey:2016krb}%
  \BibitemOpen
  \bibfield  {author} {\bibinfo {author} {\bibfnamefont {B.~D.}\ \bibnamefont
  {Lackey}}, \bibinfo {author} {\bibfnamefont {S.}~\bibnamefont {Bernuzzi}},
  \bibinfo {author} {\bibfnamefont {C.~R.}\ \bibnamefont {Galley}}, \bibinfo
  {author} {\bibfnamefont {J.}~\bibnamefont {Meidam}}, \ and\ \bibinfo {author}
  {\bibfnamefont {C.}~\bibnamefont {Van Den~Broeck}},\ }\href {\doibase
  10.1103/PhysRevD.95.104036} {\bibfield  {journal} {\bibinfo  {journal} {Phys.
  Rev.}\ }\textbf {\bibinfo {volume} {D95}},\ \bibinfo {pages} {104036}
  (\bibinfo {year} {2017})},\ \Eprint {http://arxiv.org/abs/1610.04742}
  {arXiv:1610.04742 [gr-qc]} \BibitemShut {NoStop}%
\bibitem [{\citenamefont {Balmelli}\ and\ \citenamefont
  {Damour}(2015)}]{Balmelli:2015zsa}%
  \BibitemOpen
  \bibfield  {author} {\bibinfo {author} {\bibfnamefont {S.}~\bibnamefont
  {Balmelli}}\ and\ \bibinfo {author} {\bibfnamefont {T.}~\bibnamefont
  {Damour}},\ }\href {\doibase 10.1103/PhysRevD.92.124022} {\bibfield
  {journal} {\bibinfo  {journal} {Phys. Rev.}\ }\textbf {\bibinfo {volume}
  {D92}},\ \bibinfo {pages} {124022} (\bibinfo {year} {2015})},\ \Eprint
  {http://arxiv.org/abs/1509.08135} {arXiv:1509.08135 [gr-qc]} \BibitemShut
  {NoStop}%
\bibitem [{SXS()}]{SXS:catalog}%
  \BibitemOpen
  \href@noop {} {\enquote {\bibinfo {title} {{SXS Gravitational Waveform
  Database}},}\ }\bibinfo {howpublished}
  {\url{https://data.black-holes.org/waveforms/index.html}}\BibitemShut
  {NoStop}%
\end{thebibliography}%

\end{document}